\documentclass{appolb}
\usepackage{graphicx}
\usepackage{amssymb,amsmath}

\begin{document}

\title{Relativity without tears}
\author{ Z.~K.~Silagadze
\address{Budker Institute of Nuclear Physics and Novosibirsk State
University, 630 090, Novosibirsk, Russia }
}

\maketitle

\begin{abstract}
Special relativity is no longer a new revolutionary theory but a firmly 
established cornerstone of modern physics. The teaching of special 
relativity, however, still follows its presentation as it unfolded 
historically, trying to convince the audience of this teaching that 
Newtonian physics is natural but incorrect and special relativity is its 
paradoxical but correct amendment. I argue in this article in favor of 
logical instead of historical trend in teaching of relativity and that 
special relativity is neither paradoxical nor correct (in the absolute sense
of the nineteenth century) but the most natural and expected description of 
the real space-time around us valid for all practical purposes. This last 
circumstance constitutes a profound mystery of modern physics better known 
as the cosmological constant problem.
\end{abstract}
\PACS{03.30.+p}

\section*{Preface}
{\it 
``To the few who love me and whom I love -- to those who feel rather 
than to those who think -- to the dreamers and those who put faith in
dreams as in the only realities -- I offer this Book of Truths, not in its 
character of Truth-Teller, but for the Beauty that abounds in its Truth; 
constituting it true. To these I present the composition as an Art-Product 
alone; let us say as a Romance; or, if I be not urging too lofty a claim, 
as a Poem.

What I here propound is true: -- therefore it cannot die: -- or if by any 
means it be now trodden down so that it die, it will rise again "to the
Life Everlasting."

Nevertheless, it is a poem only that I wish this work to be judged after I am 
dead''} \cite{Poe}.

\section{Introduction}
The emergence of special relativity constitutes one of the major revolutions 
in physics. Since then this revolution is repeated over and over in physics
students' minds. ``Often the result is to destroy completely the confidence
of the student in perfectly sound and useful concepts already acquired'' 
\cite{1}. What is the problem the students stumble upon?
 
maybe the following passage from \cite{1-B} might give you an insight:
``At first, relativity was considered shocking, anti-establishment and highly 
mysterious, and all presentations intended for the population at large were 
meant to emphasize these shocking and mysterious aspects, which is hardly 
conducive to easy teaching and good understanding. They tended to emphasize 
the revolutionary aspects of the theory whereas, surely, it would be good 
teaching to emphasize the continuity with earlier thought''. 

The standard presentation of special relativity follows the royal way Einstein
laid down in his breakthrough paper \cite{2} and is based on two postulates. 
The first postulate is the Principle of Relativity that the laws of physics 
are the same in all inertial reference frames. This is precisely what the 
students are taught in their undergraduate mechanics course and should hardly 
create any conceptual problems because the Principle of Relativity ``appeals 
to our common sense with irresistible force'' \cite{3}.

There is a subtle difference in the formulations of the second postulate as 
given by Einstein and as presented in modern textbooks \cite{4}. In Einstein's 
paper \cite{2} the postulate states that ``any ray of light moves in the 
`stationary' system of coordinates with the determined velocity $c$, 
whether the ray be emitted by a stationary or by a moving body''. Taken 
separately, this statement is also both orthodox and obvious in the context of 
the pre-1905 physics with luminiferous {\ae}ther as its conceptual basis. 
Indeed, after a light wave is launched in the {\ae}ther it propagates with the 
speed which is completely independent of the state of motion of the light 
source, determined solely by the elastic properties of the {\ae}ther.

It is the combination of these two postulates, considered impossible in the
pre-1905 physics, that shattered the very foundations of contemporary 
physics  and changed forever our perspective of space-time.

Many modern textbooks make explicit what is so cunningly hidden in Einstein's
seemingly harmless postulates. They state the second postulate as follows: 
``(in vacuum) light travels rectilinearly at speed $c$ in every
direction in every inertial frame'' \cite{5,5-G}. And here is buried the root 
of confusion students experience while studying special relativity. Because 
for Newtonian intuition ``to take as a postulate that the speed of light is 
constant relative to changes in reference frame is to assume an apparent 
absurdity. It goes against common sense. No wonder, thinks a student, that we 
can derive other absurdities, such as time dilation and length contraction, 
from the premises'' \cite{4}.

Laying aside the question of culture shock students experience when 
confronting the second postulate, there are several reasons, from the modern 
perspective, not to base relativity on the second postulate.

First of all, it anchors special relativity in the realm of 
electromagnetism while now we know that relativity embraces all natural 
phenomena. Strong and weak interactions, unknown to Einstein in 1905, are 
also Lorentz invariant.

The second postulate assumes light propagation in vacuum. But which vacuum?
Vacuum in modern physics is quite different from just empty space and 
looks, in a sense, like an incarnation of {\ae}ther, which ``at present, 
renamed and thinly disguised, dominates the accepted laws of physics'' 
\cite{6}. Of course this new ``{\ae}ther'',  being Lorentz invariant and 
quantum mechanical, has nothing in common with the {\ae}ther of the nineteenth
century. Anyway, in some cases it reveals itself as an nontrivial medium 
capable to alter the propagation properties of light. 

For example, between conducting plates \cite{7,8} or in a background 
gravitational field \cite{9,10} light can propagate with speeds greater 
than $c$. Yet the Lorentz invariance remains intact at a fundamental level
\cite{11}. Simply the boundary conditions or the background field single out 
a preferred rest frame and the ground state (quantum vacuum) becomes not
Lorentz-invariant. The presence of such not Lorentz-invariant ``{\ae}ther'' 
can be detected in contrast to the situation in infinite space with no 
boundaries \cite{11}.

Whether light propagates with invariant velocity $c$ is subject of photon 
being massless. This masslessness of the photon by itself originates from 
the particular pattern of the electroweak symmetry breaking. However, there is
no compelling theoretical reason for the photon to be strictly massless: a 
tiny photon mass would not destroy the renormalizability of quantum 
electrodynamics and hence the beautiful agreement between its predictions and
experiment \cite{12}. Moreover, it was shown that the photon mass can be 
generated by inflation \cite{13}. As the current universe is entering another
phase of inflation, according to the supernovae results \cite{14}, the photon
should have a miniscule mass \cite{13}  of about $10^{-42}~\mathrm{GeV}/c^2$,
far below of the present experimental limits. Anyway this miniscule mass makes
the photon not the best choice the special relativity to base on.

Surprisingly, it was known for a long time that in fact one does not need the 
second postulate to develop the special relativity. The Relativity Principle
alone along with some ``self-evident'' premises such as the homogeneity of
space and time and the isotropy of space would suffice. 

To my knowledge, von Ignatowsky \cite{15,15-T} was the first to discover this 
remarkable fact as early as in 1910, followed by Frank and Rothe \cite{16}.
``However, like numerous others that followed these [papers] have gone largely 
unnoticed'' \cite{17}. An impressive list of those largely neglected and
forgotten papers can be compiled from citations in \cite{11,18,19}.

The idea has got a better chance after it was rediscovered in the modern 
context  \cite{20,21,22}. At least,  it  attracts interest up to date 
\cite{23,23-R,24,25,26}. ``At this point we can sharpen and somewhat displace
some typical questions of historical character: Why, although it was logically
and epistemologically perfectly possible, did not Einstein see the connection 
between his two postulates? Was his embarrassment a kind of a subconscious 
inkling of this connection? Why such early papers as that of Ignatowsky did 
not catch more the interest of physicists, historians and philosophers of 
science? Why such a connection is rarely mentioned in pedagogical 
presentations of Relativity?'' \cite{27}

I think answers to these questions can be found partly in the concrete 
historical circumstances of the emergence and development of relativity
and partly in the ``intellectual inertia''\cite{28} of society.

As was already mentioned, in Einstein's original form the second postulate 
was not shocking at all for contemporary physicists. The focus was displaced
to the weird but logically inevitable consequences that followed when the 
second postulate was combined with the Relativity Principle. The one-postulate
derivations of Ignatowsky et al. involved a somewhat higher level of 
mathematics (group theory, more intricate and rather less familiar analysis;
See for example \cite{21,22}, or earlier considerations by Lalan \cite{29}).
I suspect, this was considered as an unnecessary complication, making the 
approach ``unavailable for a general education physics course'' \cite{22}.

The focus has shifted since then from reference frames and clock 
synchronization to symmetries and space-time structure, and the situation is
different today. For contemporary students the luminiferous {\ae}ther is just
an historical anachronism  and can not serve as the epistemological basis
for the second postulate. Einstein's brilliant magic when he, ``having taken 
from the idea of light waves in the {\ae}ther the one aspect that he needed''
\cite{4}, declared later in his paper that "the introduction of a 
`luminiferous {\ae}ther' will prove to be superfluous'' \cite{2}, does not 
work any more. Therefore Ignatowsky's approach is much more appealing today
than it was in 1910, because it leads to Lorentz transformations, which are
at the heart of special relativity, ``without ever having to face the 
distracting sense of paradox that bedevils more conventional attempts from the
very first steps'' \cite{22}.

Below I will try to show that, combining ideas from \cite{20,21,22,30},
it is possible to make the one-postulate derivation of Lorentz 
transformations mathematically as simple as was Einstein's original 
presentation.

In fact, much richer algebraic and geometric structures are lurking 
behind special relativity \cite{30-G}. Some of them will be considered in 
subsequent chapters.
\clearpage

\section{Relativity without light}
Let an inertial frame of reference $S^\prime$ move along the $x$-axis with 
velocity $V$ relative to the ``stationary'' frame $S$. 
\begin{figure}[htb]
 \begin{center}
    \mbox{\includegraphics[scale=0.55]{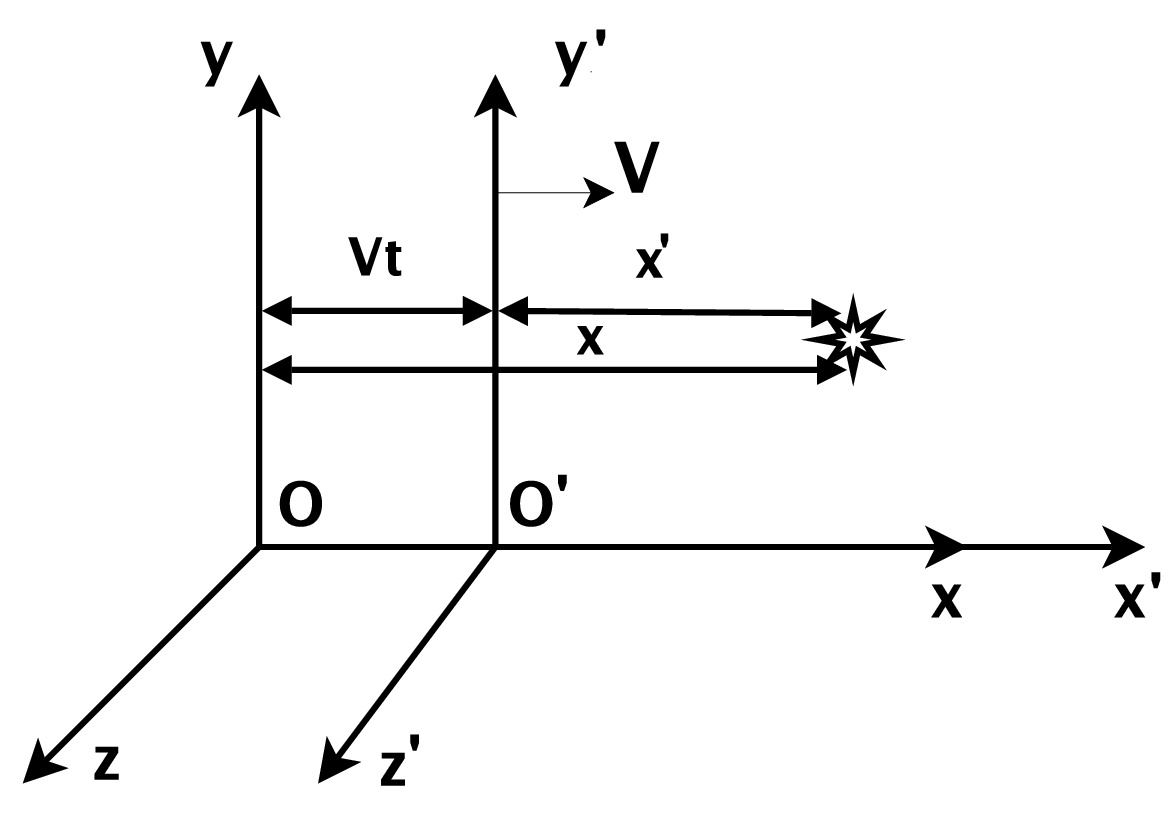}}
  \end{center}
\caption {Two frames of reference in relative motion. xxxx xxx xxx}
\label{gal}
\end{figure}

A simple glance at Fig.\ref{gal} is sufficient to write down the Galilean 
transformation that relates the $x$-coordinates of some event (for example, 
an explosion) in the frames $S$ and $S^\prime$,
\begin{equation}
x=Vt+x^\prime.
\label{eq1}
\end{equation}

But in (\ref{eq1}) we have implicitly assumed that meter sticks do not change 
their lengths when gently set in uniform motion. Although intuitively 
appealing, this is not quite obvious. For example, according to Maxwell's
equations, charges at rest interact only through the Coulomb field while in
motion they experience also the magnetic interaction. Besides, when a
source moves very fast its electric field is no longer spherically
symmetrical. It is therefore not unreasonable to expect that a meter stick
set in rapid motion will change shape in so far as electromagnetic forces
are important in ensuring the internal equilibrium of matter \cite{1}. Anyway
we just admit this more general possibility and change (\ref{eq1}) to
\begin{equation}
x=Vt+k(V^2)\,x^\prime,
\label{eq2}
\end{equation}  
where the scale factor $k(V^2)$ accounts for the possible change in the length
of of the meter stick. 

The Relativity Principle and the isotropy of space are implicit in (\ref{eq2}) 
because we assumed that the scale factor depends only on the magnitude
of the relative velocity $V$.

Equation (\ref{eq2}) allows us to express the primed coordinates through the 
unprimed ones
\begin{equation}
x^\prime=\frac{1}{k(V^2)}\left (x-Vt\right ).
\label{eq3}
\end{equation}
Then the Relativity Principle tells us that the same relation holds if 
unprimed coordinates are expressed through the primed ones, with $V$ replaced
by $-V$, because the velocity of $S$ with respect to $S^\prime$ is $-V$.
Therefore,
$$x=\frac{1}{k(V^2)}\left (x^\prime+Vt^\prime\right )=
\frac{1}{k(V^2)}\left [\frac{1}{k(V^2)}\left (x-Vt\right )+ Vt^\prime
\right ].$$
Solving for $t^\prime$, we get
\begin{equation}
t^\prime =\frac{1}{k(V^2)}\left [t-\frac{1-k^2(V^2)}{V}\,x\right ].
\label{eq4}
\end{equation}
We see immediately that time is not absolute if the scale factor $k(V^2)$
is not identically one.

From (\ref{eq3}) and (\ref{eq4}) we can infer the velocity addition rule
$$v^\prime_x=\frac{dx^\prime}{dt^\prime}=\frac{dx-Vdt}{dt-\frac{1-k^2}{V}
\,dx}=\frac{v_x-V}{1-\frac{1-k^2}{V}\,v_x}.$$
In what follows it will be convenient to write down the velocity addition 
rule with unprimed velocity expressed through the primed one,
\begin{equation}
v_x=\frac{v^\prime_x+V}{1+\frac{1-k^2}{V}\,v^\prime_x}\equiv F(v^\prime_x,V).
\label{eq5}
\end{equation}
If we change the signs of both velocities $v^\prime_x$ and $V$ it is obvious 
that the sign of the resulting velocity $v_x$ will also change.
Therefore $F$ must be an odd function of its arguments 
\begin{equation}
F(-x,-y)=-F(x,y). 
\label{eq6}
\end{equation}
Consider now three bodies $A$, $B$ and $C$ in relative motion. Let $V_{AB}$
denote the velocity of $A$ with respect to $B$. Then
\begin{equation}
V_{BA}=-V_{AB}.
\label{eq7}
\end{equation}
This is the reciprocity principle already used above. I think it is obvious 
enough to require no proof in an introductory course of special 
relativity. However, in fact it can be deduced from the Relativity Principle,
space-time homogeneity and space isotropy \cite{19,31}. Some subtleties of
such proof are discussed in \cite{32}.

Now, using (\ref{eq6}) and (\ref{eq7}), we get \cite{22}
$$F(V_{CB},V_{BA})=V_{CA}=-V_{AC} $$ 
$$=-F(V_{AB},V_{BC})=-F(-V_{BA},-V_{CB})=
F(V_{BA},V_{CB}). $$
Therefore $F$ is a symmetric function of its arguments. Then $F(v^\prime_x,V)
=F(V,v^\prime_x)$ immediately yields, according to (\ref{eq5}),
$$\frac{1-k^2(V^2)}{V}\,v^\prime_x=\frac{1-k^2(v^{\prime\,2}_x)}
{v^\prime_x}\,V,$$
or
\begin{equation}
\frac{1-k^2(V^2)}{V^2}=\frac{1-k^2(v^{\prime\,2}_x)}{v^{\prime\,2}_x}
\equiv K,
\label{eq8}
\end{equation}
where at the last step we made explicit that the only way to satisfy 
equation (\ref{eq8}) for all values of $V$ and $v^\prime_x$ is to assume that 
its left- and right-hand sides are equal to some constant $K$. Then the 
velocity addition rule will take the form
\begin{equation}
v_x=\frac{v^\prime_x+V}{1+Kv^\prime_x V}.
\label{eq9}
\end{equation}

If $K=0$, one recovers the Galilean transformations and velocity addition 
rule $v_x=v^\prime_x+V$. 

If $K<0$, one can take $K=-\frac{1}{c^2}$ and introduce a dimensionless 
parameter $\beta=\frac{V}{c}$. Then
\begin{eqnarray} & &
x^\prime=\frac{1}{\sqrt{1+\beta^2}}(x-Vt), \nonumber \\ & &
t^\prime=\frac{1}{\sqrt{1+\beta^2}}\left (t+\frac{V}{c^2}\,x\right ), 
\label{eq10}
\end{eqnarray}
while the velocity addition rule takes the form
\begin{equation}
v_x=\frac{v^\prime_x+V}{1-\frac{v^\prime_x V}{c^2}}.
\label{eq11}
\end{equation}
If $v^\prime_x=V=\frac{c}{2}$, then (\ref{eq11}) gives $v_x=\frac{4}{3}c$.
Therefore velocities greater than $c$ are easily obtained in this case.
But if $v^\prime_x=V=\frac{4}{3}c$, then
$$v_x=\frac{\frac{8}{3}c}{1-\frac{16}{9}}=-\frac{24}{7}c<0.$$  
Therefore two positive velocities can sum up into a negative velocity!
This is not the only oddity of the case $K<0$. For example, if $v^\prime_x V
=c^2$, then $v^\prime_x$ and $V$ will sum up into an infinite velocity.

If we perform two successive transformations $(x,t)\to(x^\prime,t^\prime)
\to(x^{\prime\prime},t^{\prime\prime})$, according to (\ref{eq10}) with 
$\beta=4/3$, we end up with
$$x^{\prime\prime}=-\frac{7}{25}x-\frac{24}{25}ct.$$
This can not be expressed as the result of a transformation $(x,t)\to
(x^{\prime\prime},t^{\prime\prime})$ of type (\ref{eq10}) because the
coefficient $1/\sqrt{1+\beta^2}$ of $x$ in (\ref{eq10}) is always
positive. Hence the breakdown of the group law.

However, the real reason why we should  discard the case $K<0$ is the absence
of causal structure. The transformations (\ref{eq10}) can be recast
in the form
\begin{eqnarray} &&
x_0^\prime=cos{\theta}\,x_0+\sin{\theta}\, x, \nonumber \\ &&
x^\prime=cos{\theta}\,x-\sin{\theta}\, x_0,
\label{eq12} 
\end{eqnarray}
where $x_0=ct$ and $\cos{\theta}=\frac{1}{\sqrt{1+\beta^2}}$. This is the 
usual rotation and hence the only invariant quantity related to the event
is $x_0^2+x^2$. This Euclidean norm does not allow us to define an invariant
time order between events, much like of the ordinary three-dimensional
space where one can not say which points precede a given one.

Finally, if $K=\frac{1}{c^2}>0$, one gets the Lorentz transformations
\begin{eqnarray} & &
x^\prime=\frac{1}{\sqrt{1-\beta^2}}(x-Vt), 
\nonumber \\ & &
t^\prime=\frac{1}{\sqrt{1-\beta^2}}\left (t-\frac{V}{c^2}\,x\right ), 
\label{eq13}
\end{eqnarray}
with the velocity addition rule
\begin{equation}
v_x=\frac{v^\prime_x+V}{1+\frac{v^\prime_x V}{c^2}}.
\label{eq14}
\end{equation}
Now $c$ is an invariant velocity, as (\ref{eq14}) shows. However, in the 
above derivation nothing as yet indicates that $c$ is the velocity of light.
Only when we invoke the Maxwell equations it becomes clear that the velocity
of electromagnetic waves implied by these equations must coincide
with $c$ if we want the Maxwell equations to be invariant under Lorentz
transformations (\ref{eq13}).

The choice between the Galilean ($K=0$) and Lorentzian ($K>0$) cases is not
only an experimental choice. Some arguments can be given supporting the idea
that this choice can be made even on logical grounds.

First of all, special relativity is more friendly to determinism than the 
classical Newtonian mechanics \cite{33}. An example of indeterministic
behavior of a seemingly benign Newtonian system is given by Xia's five-body
supertask \cite{34,35}.

\begin{figure}[htb]
 \begin{center}
    \mbox{\includegraphics{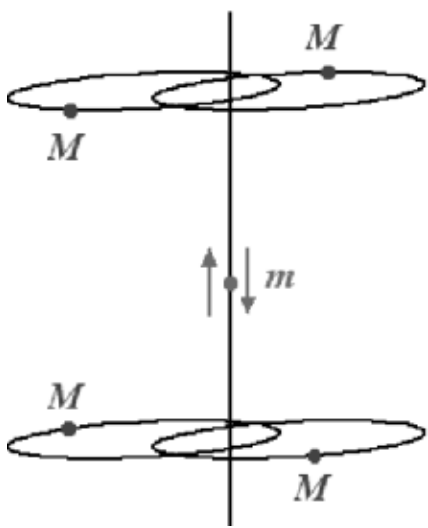}}
  \end{center}
\caption {Xia's five-body supertask.}
\label{xia}
\end{figure}
Two symmetrical highly eccentric gravitationally bound binaries are placed
at a large distance from each other, while a fifth body of much smaller mass
oscillates between the planes of these binaries (see Fig.\ref{xia}). 
It can be shown \cite{34} that there exists a set of initial conditions under 
which the binaries from this construction will escape to spatial infinity
in a finite time, while the fifth body will oscillate back and forth with ever
increasing velocity.

The time reverse of Xia's supertask is an example of ``space invaders'' 
\cite{33} -- particles appearing from spatial infinity without any causal
reason. This indeterminacy of idealized Newtonian world is immediately killed
by special relativity because without unbounded velocities there are no space 
invaders.

That causality and Lorentz symmetry are tightly bound is a remarkable fact.
Another less known royal way to relativity can be traced back to Alfred 
Robb's synthetic approach \cite{36} which emphasizes the primary role
of causal order relation in determining the space-time geometry. Later this 
line of reasoning was further developed by Reichenbach \cite{37} from the 
side of philosophy and Alexandrov \cite{38} and Zeeman  \cite{39} from the 
side of mathematics. The famous Alexandrov-Zeeman theorem states that, in 
a sense, causality implies Lorentz symmetry.

Group theory provides another argument to prefer the Lorentzian world over
the Galilean one \cite{40} because mathematically the Lorentz group $G_c$
has much simpler and elegant structure than its singular limit $G_\infty$
which is the symmetry group of the Galilean world. This argument goes back to 
Minkowski. In his own words, cited in \cite{40}, from  his famous Cologne 
lecture ``{\it Raum und Zeit}'':
  
``Since $G_c$ is mathematically more intelligible than $G_\infty$, it looks 
as though the thought might have struck some mathematician, fancy free, that 
after all, as a matter of fact, natural phenomena do not possess an invariance 
with the group $G_\infty$, but rather with the group $G_c$, $c$ being finite 
and determinate, but in ordinary units of measure extremely great. Such 
a premonition would have been an extraordinary triumph for pure mathematics.
Well, mathematics, though it can now display only staircase-wit, has the 
satisfaction of being wise after the event, and is able, thanks to its happy 
antecedents, with its senses sharpened by an unhampered outlook to far 
horizons, to grasp forthwith the far-reaching consequences of such a 
metamorphosis of our concept of nature.''

\section{Relativistic energy and momentum}
Although Minkowski's contribution and his notion of space-time is the most
crucial event placing relativity in the modern context, students reluctantly
swallow the four-vector formalism due to their Newtonian background.
Therefore, some elementary derivation of the relativistic expressions for 
energy and momentum, stressing not the radical break but continuity with 
concepts already acquired by students \cite{1}, is desirable in a general 
education physics course.

Usually that is done by using the concept of relativistic mass -- another
unfortunate historical heritage stemming from an inappropriate 
generalization of the Newtonian relationship between momentum and mass 
$\vec{p}=m\vec{v}$ to the relativistic domain.

Crystal-clear arguments were given \cite{41,42} against the use of 
relativistic mass as a concept foreign to the logic of special relativity. 
However, the ``intellectual inertia'' still prevails, the most famous
(wrong!) formula associated to special relativity for the general public is
still $E=mc^2$, and students are still exposed to the outmoded notion of 
velocity dependent mass (unfortunately, this is maybe the only concept of
special relativity they absorb with ease because of its Newtonian roots).

In fact, even to ensure continuity with Newtonian concepts there is no need 
to preserve historical artifacts in teaching special relativity a hundred 
years after its creation. Below a modification of Davidon's derivation 
\cite{18,43} of relativistic energy and momentum is presented which 
quite gently harmonizes with students' Newtonian background.

Suppose a ball cools down by emitting photons isotropically in its rest frame
$S^\prime$. Let the total energy emitted during some time be 
${\cal E^\prime}$. Because of isotropy, the total momentum carried away  by
radiation is zero and the ball will stay motionless. The Principle of 
Relativity then implies that in every other inertial frame the velocity of the
ball is also unchanged.

Let us see how things look in the reference frame $S$ in which the ball 
moves with velocity $V$ along the $x$-axis. A bunch of photons emitted 
within the solid angle $d\Omega^\prime=2\pi\sin{\theta^\prime}
d\theta^\prime$ around the polar angle $\theta^\prime$ in the frame $S^\prime$
has total energy $d{\cal E^\prime}={\cal E^\prime}\,\frac{d\Omega^\prime}
{4\pi}$ in this frame. Due to the Doppler effect, the corresponding energy
in the frame $S$ will be
$$d{\cal E}=\gamma d{\cal E^\prime}(1+\beta\cos{\theta^\prime})=
-\gamma {\cal E^\prime}\,\frac{1}{2}\,(1+\beta\cos{\theta^\prime})
\,d\cos{\theta^\prime}.$$
(Note that the Doppler formula for the photon frequency shift follows from
the Lorentz transformations quite easily \cite{5}. We have also assumed the 
relation $E=\hbar \omega$ for the photon energy from elementary quantum 
theory).

Therefore, the total energy emitted in the frame $S$ is
$${\cal E}=-\gamma {\cal E^\prime}\,\frac{1}{2}\,\int\limits_0^\pi 
(1+\beta\cos{\theta^\prime})\,d\cos{\theta^\prime}=\gamma {\cal E^\prime}.$$
In the frame $S$, the radiation is no longer isotropic and therefore it takes
away some momentum. Let us calculate how much. The emission angle of the 
bunch of light in the frame $S$ can be found from the 
aberration formula (which follows from the velocity addition rule 
(\ref{eq14}))
$$\cos{\theta}=\frac{\cos{\theta^\prime}+\beta}{1+\beta\cos{\theta^\prime}}.$$
But each photon of energy $\hbar \omega$ carries the momentum $\hbar \omega/c$
(another elementary fact from the quantum theory). Therefore the x-component
of the momentum of the light-bunch is
$$d\,{\cal P}_x=\frac{d{\cal E}}{c}\cos{\theta}=-\frac{\gamma {\cal E^\prime}}
{2c}(\cos{\theta^\prime}+\beta)\,d\cos{\theta^\prime}.$$
Integrating, we get the total momentum taken away by radiation in the 
$x$-direction
$${\cal P}_x=-\frac{\gamma {\cal E^\prime}}{2c}\int\limits_0^\pi
(\cos{\theta^\prime}+\beta)\,d\cos{\theta^\prime}=\gamma\beta\,
\frac{{\cal E^\prime}}{c}.$$
Therefore, for the energy and momentum of the ball in the frame $S$ we should
have
\begin{equation}
\Delta E=-\gamma{\cal E^\prime},\;\;\; \Delta p=-\gamma\beta\,\
\frac{{\cal E^\prime}}{c}.
\label{eq15} 
\end{equation}
It is natural to assume that the momentum and velocity of the ball are 
parallel to each other,
\begin{equation}
\vec{p}=N(V)\vec{V},
\label{eq16} 
\end{equation}
where $N(V)$ is some unknown function (let us call it the inertia of the ball
\cite{18,43}). Then 
$$\Delta p=V\Delta N,$$
because $\Delta{V}=0$ as explained above. But (\ref{eq15}) implies 
that $\Delta p=\frac{\beta\Delta E}{c}$. Therefore, $\frac{\beta
\Delta E}{c}=V\Delta N$ and
\begin{equation}
\Delta E=c^2\Delta N.
\label{eq17} 
\end{equation}
Although this result was obtained in the particular circumstances, we now 
assume that it is universally valid. That is, we assume that every change in 
the energy of a body implies the corresponding change in its inertia 
according to (\ref{eq17}). If the body is subject to a force $\vec{F}$ then
$$\frac{dE}{dt}=\vec{F}\cdot\vec{V}=\vec{V}\cdot\frac{d\vec{p}}{dt}=
\vec{V}\cdot \left (\vec{V}\frac{dN}{dt}+N\frac{d\vec{V}}{dt}\right )=
V^2\frac{dN}{dt}+\frac{N}{2}\frac{dV^2}{dt}.$$
Using (\ref{eq17}) we get
$$c^2\frac{dN}{dt}=V^2\frac{dN}{dt}+\frac{N}{2}\frac{dV^2}{dt},$$
or
$$\frac{dN}{N}=\frac{dV^2}{2(c^2-V^2)}.$$
Integrating, we get
\begin{equation}
N=\frac{N_0}{\sqrt{1-\beta^2}},
\label{eq18} 
\end{equation}
where $N_0$ is an integration constant. Therefore, 
\begin{equation}
\vec{p}=\frac{N_0\vec{V}}{\sqrt{1-\beta^2}}=\frac{m\vec{V}}{\sqrt{1-\beta^2}}
\label{eq19} 
\end{equation}
because by considering the nonrelativistic limit $\beta\ll 1$ we conclude 
that $N_0$ is just the mass of the body. It follows from (\ref{eq17}) and 
(\ref{eq18}) that
\begin{equation}
E=Nc^2=\frac{mc^2}{\sqrt{1-\beta^2}}
\label{eq20} 
\end{equation}
up to irrelevant overall constant. This derivation of relativistic energy and
momentum clarifies the real meaning of the notorious $E=mc^2$. It is the 
inertia of the body, defined as the coefficient of the velocity in the 
expression of the momentum $\vec{p}=N\vec{V}$, to which the energy of the 
body is proportional, while the mass of the body is an invariant, 
frame-independent quantity -- much like the Newtonian concept of mass.

\section{Relativity without reference frames}
You say that relativity without reference frames does not make even a 
linguistic sense? Yes, of course, if the words are understood literally. 
But ``the name `{\it theory of relativity}' is an unfortunate choice. Its 
essence is not the relativity of space and time but rather the independence 
of the laws of nature from the point of view of the observer" \cite{44}. 
Of course, I do not advocate changing the name, so celebrated, of the theory.
But it should be instructive to bear in mind the conventionality of the name 
and the shift of focus from the length contraction and time dilation (which 
are rather obvious effects in Minkowski's four-dimensional geometric 
formulation) to space-time structure and symmetries.

Above we have discarded the $K<0$ possibility as unphysical. But should we 
really? The main argument was that the corresponding Euclidean space-time 
does not support causal structure and, therefore, it is in fact some kind of 
a timeless nirvana. The real turbulent universe around us is certainly not
of this type. But we cannot exclude that it may contain inclusions of
Euclidean domains, maybe formed at the centers of black holes as a result of 
quantum signature change during gravitational collapse \cite{45}. Moreover, 
the Hartle--Hawking's 'No Boundary Proposal' in quantum cosmology assumes 
that Euclidean configurations play an important role in the initial 
wavefunction of the Universe \cite{46}. Loosely speaking, it is suggested that
the whole Universe was  initially Euclidean and then by quantum tunneling  
a transition to the usual Lorentzian space-time occurred.

But once we accept the Euclidean space-time as physically viable we have to
change the principal accents of the formalism, depriving reference frames of 
their central role, because in timeless Euclidean nirvana there are no 
observers and hence no reference frames. The Principle of Relativity then
should be replaced by the more general concept of symmetry transformations 
which leave the space-time structure invariant. In doing so, a question 
naturally arises as to what is the concept of space-time geometry which 
embraces even space-times inhospitable to intelligent life. And at this point 
it's a good thing to acquaint students with the Erlangen program of 
Felix Klein.

\section{What is geometry?}
The word {\it geometry} is derived from Greek and means {\it earth 
measurement}. It is not surprising, therefore, that beginning from Gauss and 
Riemann the length concept is considered to be central in geometry. The 
distance between two infinitesimally close points in Riemannian geometry is 
given by some positive definite, or at least non-degenerate, quadratic 
differential form 
$$ds^2=\sum\limits_{i,j=1}^n g_{ij}(x)\,dx^i dx^j.$$
Other geometric concepts like the angle between two intersecting curves
can be defined in terms of the metric $g_{ij}$.

The main feature of the Riemannian geometry is that it is local and 
well suited for field theory in physics \cite{47}, general relativity being
the most notable example. However, some interesting and important questions
are left outside the scope of Riemannian geometry. 

For example, there are two basic geometric measurements: the determination of 
the distance between two points and the determination of the angle between two 
intersecting lines. The corresponding measures are substantially different 
\cite{48}. The distance between two points is an algebraic function of their
coordinates and therefore it is uniquely defined (up to a sign). In contrast,
the angle between two intersecting lines is determined as a transcendental 
(trigonometric) function of coordinates and is defined only up to $2\pi n$,
for an arbitrary integer $n$. As a result, every interval can easily be 
divided into an arbitrary number of equal parts, while the division of an 
arbitrary angle into three equal angles by using only a compass and a  
straightedge  is an ancient problem proved impossible by Wantzel in 1836.

This difference between length and angle measurements is just implied
in Riemannian geometry without any explanation and hardly disturbs our
intuition until we recognize its Euclidean roots. The existence of
non-Euclidean geometries when calls for a more careful examination of the 
question how the corresponding measures arise. For this purpose we should take 
another road to geometry, the Kleinian road where the equality of figures is 
a basic geometrical concept \cite{49}. 

In Euclidean geometry the equality of two figures means that they can be 
superimposed with an appropriate rigid motion. The axiom that if $A$ equals 
$B$ and $B$ equals $C$, then $A$ equals $C$, in fact indicates \cite{49} that 
rigid motions form a group. A far-reaching generalization of this almost 
trivial observation was given by Felix Klein in his famous inaugural lecture 
\cite{50} that he prepared as professor at Erlangen in 1872 but never actually 
gave. In this lecture Klein outlined his view of geometry which later became
known as the Erlangen program. Actually the Erlangen program is just the basic 
principle of Galois theory applied to geometry \cite{51}. According this
principle discovered by Galois around 1832 one can classify the ways 
a little thing (a point or figure) can sit in a bigger thing (space) by 
keeping track of the symmetries of the bigger thing that leave the little 
thing unchanged \cite{51}. 

Every particular  geometry determines the group of transformations (motions) 
which preserve this geometry. Klein calls this group the principal group and 
for him the geometry is just the study of invariants of the principal group 
because genuine geometric properties of a figure are only those which remain
unchanged under transformations belonging to the  principal group.

The converse is also true: a group $G$ acting on a space $X$ determines
some geometry \cite{52}. In classical geometries all points of $X$
look alike (the space is homogeneous). In the group-theoretical language this
means that $G$ acts on $X$ transitively; that is, for every pair of points 
$x$ and $y$ of $X$ there exists a symmetry transformation (motion) $g\in G$
which takes $x$ into $y$: $y=g(x)$. Let $H$ be the stabilizer of a point $x$
--- the set of all transformations in $G$ which leave $x$ invariant. Then $X$
can be identified in fact with the coset space $G/H$. Indeed, we have a 
one-to-one map of $X$ onto $G/H$ which takes each point $g(x)\in X$ to the 
equivalence class $[g]\in G/H$. The symmetries $x\mapsto s(x),\;s\in G$ of 
the space $X$ are then represented by the transformations $[g]\to [sg]$ 
of the coset space $G/H$. 

Hence, interestingly, the Erlangen program provides a very Kantian view of 
what space (space-time) is. Kant in his {\it Critique of Pure Reason} denies 
the objective reality of space and time, which for him are only forms in which 
objects appear to us due to the hardwired features of our consciousness 
(intuition) and not the properties of objects themselves. 

The Kantian character of relativity theory was advocated by Kurt G\"odel 
\cite{53,54}. In G\"odel's view the relativity of simultaneity deprives time 
of its objective meaning and ``In short, it seems that one obtains an 
unequivocal proof for the view of those philosophers who, like Parmenides, 
Kant, and the modern idealists, deny the objectivity of change and consider 
change as an illusion or an appearance due to our special mode of perception''
\cite{54}.

One can argue that the Minkowski space-time represents a kind of reality
that comes for the change to the Newtonian absolute time. In Minkowski's own 
words: ``Henceforth space by itself, and time by itself are doomed to fade 
away into mere shadows, and only a kind of union of the two will preserve an 
independent reality'' \cite{55}.

However, the Erlangen program, if pushed to its logical extreme, indicates 
that the Minkowski space-time is just a useful parameterization of the coset
space $G/H$. The symmetry group $G$ (the Poincar\'e group) and its subgroup 
$H$ (the Lorentz group) as the stabilizer of ``points'' (space-time events) 
are all that really does matter. 

However, a  too abstract group-theoretical approach is not always instructive
because even if space and time are really illusions they proved to be very 
useful concepts ``due to our special mode of perception''. Therefore we leave
the interesting question of the reality of space-time to philosophers; below
we follow Klein's more classical presentation \cite{48} because it emphasizes 
not the radical break but continuity with concepts already acquired by 
students \cite{1}.

\section{Projective metrics}
Our special mode of perception, especially vision, determines our implicit
belief that everything is made of points as the most basic structural 
elements. For a blind man, however, who examines things by touching them, the
most basic geometric elements are, perhaps, planes \cite{56}. This duality
between various  basic geometrical elements is most naturally incorporated
into projective geometry \cite{57,58}, which makes it a good starting point
for studying of different kinds of linear and angular measures \cite{48,59}.
For simplicity we will consider only plane projective geometry to 
demonstrate basic principles, a generalization to higher dimensional spaces 
being rather straightforward.

Imagine a Euclidean plane $R^2$ embedded into the three dimensional Euclidean
space $R^3$ equipped with a Cartesian coordinate system. The coordinate 
system can be chosen in such a way that the equation of the plane becomes 
$z=a\neq 0$. Then every point in the plane $R^2$ together with the origin 
$(0,0,0)$ of the coordinate system determines a line in $R^3$. Therefore 
points in the plane $R^2$ can be considered as the remnants in this plane of 
the corresponding lines and can be uniquely  determined by the coordinates  
$(x,y,z)$ of any point on these lines except $(0,0,0)$. Two sets  
$(x_1,y_1,z_1)$ and $(x_2,y_2,z_2)$ of coordinates represent the same 
point of $R^2$ if and only if $\frac{x_1}{x_2}=\frac{y_1}{y_2}=
\frac{z_1}{z_2}$. Therefore the points of $R^2$ are in fact the equivalence 
classes of triples $[x]=(\lambda x_1,\lambda x_2,\lambda x_3)$ of real numbers
with $\lambda\neq 0$. If we now define the projective plane $P^2$ as the set 
of all such equivalence classes except $[0]=(0,0,0)$, then $P^2$ will not be 
just the Euclidean plane $R^2$ because it will contain points 
$(\lambda x_1,\lambda x_2,0)$ corresponding to the lines in $R^3$ which are 
parallel to the plane $R^2$ in the Euclidean sense. These points are, of 
course, points at infinity from the Euclidean perspective, lying on the circle
of the infinite radius. Alternatively we can consider this circle as the line 
at infinity in $P^2$. Other lines in $P^2$ can be identified with remnants of 
the planes in $R^3$ incident to the origin. Such planes are uniquely 
determined by their normal vectors $(u_1,u_2,u_3)$, or more precisely by 
equivalence classes $[u]=(\lambda u_1,\lambda u_2,\lambda u_3),\;\lambda
\neq 0$. Therefore, we can consider $(u_1,u_2,u_3)$ as the (projective) 
coordinates of a line in $P^2$.

Now we can forget the Euclidean scaffolding of our model of the projective 
plane $P^2$. Then all points as well as all lines in $P^2$ will look alike.
Every pair of points $x$ and $y$ are incident to a unique line $x\times y$, 
and every pair of lines $\xi$ and $\eta$ are incident to the unique point 
$\xi \times \eta$ of their intersection. Here we assume that points and lines 
are given by their three projective coordinates and $\times$ denotes the usual 
vector product of 3-dimensional vectors.  

It is important to note  that projective geometry, and the projective 
coordinates of points and lines in it, can be defined synthetically by a 
small set of axioms without any reference to the Cartesian coordinates and the 
Euclidean concept of length \cite{60,61}. At that, every four points 
$P_1,P_2,P_3$ and $E$ such that no three of them are collinear uniquely 
determine a projective coordinate system \cite{58}. In this coordinate system 
the basic points $P_1,P_2,P_3,E$ have coordinates
$$P_1=(1,0,0),\;\; P_2=(0,1,0),\;\; P_3=(0,0,1)\;\;\mathrm{and}\;\;
E=(1,1,1).$$
The projective coordinates of a given point in two different coordinate 
systems are related by a projective transformation
\begin{equation}
 x_i^\prime = \sum \limits_{i=1}^3 A_{ij}x_j, \;\;
\mathrm{det}\,(A_{ij})\neq 0.
\label{P2trans}
\end{equation}
Projective transformations form a group and projective geometry studies 
invariants of this group. In particular, incidence relations are invariant
under projective transformations and constitute the most basic geometric
notion in projective geometry. 

A point $x$ is incident to a line $\xi$ if and only if the scalar 
product $x\cdot \xi$ is equal to zero. Remarkable symmetry of this condition 
reflects the duality property of the projective plane: the notions `line' 
and `point' can be used interchangeably in the plane projective geometry. 
Every figure in $P^2$ can be considered on equal footing as made from points 
or as made from lines. This duality between points and lines is surprising. 
Our intuition does not grasp it, as our geometric terminology  witnesses
\cite{58}.

Another important projective invariant is the cross-ratio of four collinear 
points. If points $x,y,z,z^\prime$ are collinear, so that $z=\lambda x+
\mu y$ and $z^\prime=\lambda^\prime x+\mu^\prime y$, then the cross-ratio
of these points equals
\begin{equation} 
R(x,y,z,z^\prime)=\frac{\mu\lambda^\prime}{\lambda\mu^\prime}.
\label{crr}
\end{equation}

The notions `angle' and `distance' are not projective geometry notions
because angles and distances are not invariant under projective 
transformations. Therefore, to define these notions we should select a 
subgroup of the projective group that will leave invariant appropriately 
defined angles and distances. Selecting different subgroups, we
get different geometries. In the spirit of the Erlangen program all these
geometries are equally feasible. It remains to find a ``natural'' definition 
of angles and distances for a given subgroup, not wildly different from what
we intuitively expect from these kinds of measurements.

Both measurements, be it the measurement of an angle or distance, share 
common features \cite{48}. An unknown distance is compared to 
some standard length regarded as the unit length. This comparison in fact 
involves translations of the unit length. That is, it is assumed that 
(Euclidean) length is translation invariant. The same is true for
angular measurements, but now we have rotations instead of translations. And
here comes the difference between angular and linear measurements. From
the projective perspective translations along a line are projective 
transformations with one fixed point -- a point at infinity, while rotations 
in a flat pencil of lines have no fixed lines.

This observation opens a way for a generalization \cite{48}. To define 
a projective metric on a line, first of all we should select a projective 
transformation $A$ which will play the role of the unit translation. If $x$ 
is some point on a projective line then $x, A(x), A^2(x),A^3(x) \ldots$ will
give mark-points of the distance scale. Therefore, we will have as many 
different measures of the distance on a projective line as there are 
substantially different projective transformations of this line. The Euclidean
example suggests that we can classify projective transformations by the number
of their fixed points.
     
On the projective line $P^1$ every three distinct points define the unique
coordinate system in which the basic points have coordinates $(1,0)$, $(0,1)$
and $(1,1)$. Any other point is given by an equivalence class $[x]=
(\lambda x_1,\lambda x_2)$, $\lambda\neq 0$, of two real numbers $x_1,x_2$.
Projective transformations of the line $P^1$ have the form
\begin{equation}
\begin{array}{c} 
x_1^\prime=A_{11}x_1+A_{12}x_2 \\
x_2^\prime=A_{21}x_1+A_{22}x_2
\end{array}, \;\;\;\; \mathrm{det}\,(A_{ij})\neq 0.
\label{P1trans}
\end{equation}
If $(x_1,x_2)$ is a fixed point of this transformation, then we should have
\begin{equation}
\begin{array}{c}
\lambda x_1=A_{11}x_1+A_{12}x_2 \; ,\\
\lambda x_2=A_{21}x_1+A_{22}x_2 \; ,
\end{array}
\label{P1fxp}
\end{equation}
for some $\lambda\neq 0$. The point  $(x_1,x_2)$ is uniquely determined by 
the ratio $z=\frac{x_1}{x_2}$ (the non-homogeneous coordinate of this point),
for which we get from (\ref{P1fxp})
\begin{equation}
A_{21}z^2+(A_{22}-A_{11})z-A_{12}=0,
\label{P1fxpz}
\end{equation}
or in the homogeneous form
\begin{equation}
\Omega(x,x)=0,
\label{P1fxpw}
\end{equation}
where we have introduced the quadratic form
$$\Omega(x,y)=\sum\limits_{i,j=1}^2 \Omega_{ij}x_iy_j,$$
with
$$\Omega_{11}=A_{21},
\;\; \Omega_{12}=\Omega_{21}=\frac{A_{22}-A_{11}}{2},\;\;
\Omega_{22}=-A_{12}.$$
If $\Delta=(A_{22}-A_{11})^2+4A_{12}A_{21}=4(\Omega_{12}^2-\Omega_{11}
\Omega_{22})>0$, then (\ref{P1fxpz}) has two different real solutions and the
transformation $A$ is called hyperbolic.
If $\Delta=0$, two different solutions degenerate into a single real solution
and  $A$ is called parabolic.
At last, if $\Delta<0$, then (\ref{P1fxpz}) has no real solutions at all and 
the transformation  $A$ is called elliptic.

Therefore, we have three different measures of the distance on the projective 
line $P^1$: hyperbolic, parabolic and elliptic. By duality the same is true
for a flat pencil of lines for which we have three different types of angular
measure.

Let us consider first the hyperbolic measure. Then the corresponding 
projective transformation $A$ has two fixed points $p_0$ and $p_\infty$. Let 
these points be the basic points of the projective coordinate system such 
that $p_0=(0,1)$ and $p_\infty=(1,0)$. Substituting these points into
(\ref{P1fxp}), we get $A_{12}=A_{21}=0$ and, therefore, the projective 
transformation $A$ takes the simple form
$$\begin{array}{c}
x_1^\prime = A_{11}x_1 \\ x_2^\prime = A_{22}x_2
\end{array} $$
in this coordinate system. Introducing again the non-homogeneous coordinate
$z=\frac{x_1}{x_2}$, we can represent the transformation $A$ in the form
$z^\prime =\lambda z$, where $\lambda = A_{11}/A_{22}$.

If we take some point $z=z_1$ as the beginning of the distance scale then the
mark-points of this scale will be $z_1,\lambda z_1,\lambda^2 z_1,\lambda^3 
z_1,\ldots$ and the distance between points $\lambda^n z_1$ and $z_1$ is 
equal to $n$.

To measure distances that constitute fractions of the unit distance, one 
should subdivide the unit intervals $[\lambda^n z_1,\lambda^{n+1} z_1]$ into
$N$ equal parts. This can be achieved by means of the projective 
transformation $z^\prime =\mu z$, leaving the basic points $p_0$ ($z=0$) and 
$p_\infty$ ($z=\infty$) invariant, such that $\mu^N z_1=\lambda z_1$, or
$\mu=\lambda^{\frac{1}{N}}$.  Then the points $z_1,\mu z_1,\mu^2 z_1,\ldots,
\mu^N z_1=\lambda z_1$ constitute the desired subdivision of the unit interval
$[z_1,\lambda z_1]$ into $N$ equal parts. Now the distance from the point 
$z_1$ to the point $\lambda^{n+\frac{m}{N}}z_1$ is equal to $n+\frac{m}{N}$.

Repeating subdivisions infinitely, we come to the conclusion \cite{48} that
the distance from $z_1$ to a point $z$ equals to the real number $\alpha$ 
such that $z=\lambda^\alpha z_1$. Therefore, in our particular coordinate 
system the hyperbolic distance is given by the formula 
$$d(z,z_1)=\frac{1}{\ln{\lambda}}\ln{\frac{z}{z_1}}.$$
As the distance on the line should be additive, $d(x,y)=d(x,z_1)+d(z_1,y)$, 
we get the general formula \cite{48} for the hyperbolic distance between 
points $x=(x_1,x_2)$ and $y=(y_1,y_2)$
\begin{equation}
d(x,y)=C\ln{\frac{x_1y_2}{x_2y_1}},
\label{hdst}
\end{equation}
where $C=\frac{1}{\ln{\lambda}}$ is some constant and the freedom to choose 
$C$ reflects the freedom to choose different units of the length measurement.

However, as $p_0=(0,1)$ and $p_\infty=(1,0)$, we have $x=x_2p_0+x_1p_\infty$,
$y=y_2p_0+y_1p_\infty$ and $$\frac{x_1y_2}{x_2y_1}=R(p_0,p_\infty,x,y)$$ 
is the cross-ratio formed by the points $x$ and $y$ with the fixed points 
$p_0$ and $p_\infty$ of the projective transformation $A$. Therefore we get 
the coordinate-independent form  
\begin{equation}
d(x,y)=C\ln{R(p_0,p_\infty,x,y)}
\label{hdstcr}
\end{equation} 
of the hyperbolic distance formula. Now we use this formula to express the 
hyperbolic distance in terms of homogeneous coordinates in a general 
coordinate system.

As the projective coordinates of a point are determined only up to a scale
factor, we can write
\begin{equation}
p_0=\lambda_0 x+y,\;\;\; p_\infty=\lambda_\infty x+y,
\label{p0pinfty}
\end{equation}
for some $\lambda_0$ and $\lambda_\infty$. Then, using \cite{58} 
$R(p_0,p_\infty,x,y)=R(x,y,p_0,p_\infty)$, we get
$$ R(p_0,p_\infty,x,y)=\frac{\lambda_\infty}{\lambda_0}.$$
But $\Omega(p_0,p_0)=\Omega(p_\infty,p_\infty)=0$ and (\ref{p0pinfty}) shows
that both $\lambda_0$ and $\lambda_\infty$ are solutions of the quadratic
equation
\begin{equation}
\lambda^2\,\Omega(x,x)+2\lambda\Omega(x,y)+\Omega(y,y)=0.
\label{lambdaeq}
\end{equation}
Therefore,
$$R(p_0,p_\infty,x,y)=\frac{\Omega(x,y)+\sqrt{\Omega^2(x,y)-\Omega(x,x)
\Omega(y,y)}}{\Omega(x,y)-\sqrt{\Omega^2(x,y)-\Omega(x,x)\Omega(y,y)}},$$
and the hyperbolic distance is
\begin{equation}
d(x,y)=C\ln{\frac{\Omega(x,y)+\sqrt{\Omega^2(x,y)-\Omega(x,x)\Omega(y,y)}}
{\Omega(x,y)-\sqrt{\Omega^2(x,y)-\Omega(x,x)\Omega(y,y)}}}.
\label{hdist}
\end{equation}
We have not mentioned one subtlety \cite{48}. The fixed points $p_0$ and 
$p_\infty$ divide the projective line $P^1$ into two intervals according to
the sign of the cross-ratio $R(p_0,p_\infty,x,y)$. Above we have assumed that
$x$ and $y$ are from the interval which corresponds to the positive sign of 
this cross-ratio. Therefore (\ref{hdist}) gives the distance only between 
points of this interval. As this formula implies, $p_0$ and $p_\infty$ 
are both at logarithmically infinite distance from every point of the interval 
under consideration. Therefore these points, to say nothing of the points 
beyond them, are unreachable for the inhabitants of the one-dimensional 
hyperbolic world, for whom the question of existence of the other interval is 
a metaphysical question.  

The analytic continuation of (\ref{hdist}) by means of the formula
$$\ln{x}=2i\arccos{\frac{x+1}{2\sqrt{x}}}$$
can be used to get the distance in the elliptic case, when (\ref{lambdaeq}) 
does not have real solutions, and, therefore, $\Omega^2(x,y)-\Omega(x,x)
\Omega(y,y)<0$ for all points $x$ and $y$. This gives
\begin{equation}
d(x,y)=2iC\arccos{\frac{\Omega(x,y)}{\sqrt{\Omega(x,x)\Omega(y,y)}}}.
\label{edist}
\end{equation}
Note that in the elliptic case there are no infinite points, all distances 
being finite and defined only up to $2\pi n,\;n\in \mathbb Z$. This is 
precisely the situation characteristic of the Euclidean angles. If we take 
$\Omega(x,x)=x_1^2+x_2^2$, so that $\Omega_{ij}=\delta_{ij}$, then 
(\ref{edist}) gives for the distance (angle) between lines $(u_1,u_2)$ and 
$(v_1,v_2)$
$$d(u,v)=2iC\arccos{\frac{u_1v_1+u_2v_2}
{\sqrt{(u_1^2+u_2^2)(v_1^2+v_2^2)}}}.$$
Obviously this is the usual Euclidean formula for the angle between two lines
with normals $(u_1,u_2)$ and $(v_1,v_2)$ provided that $C=-i/2$. 

It remains to consider the parabolic case, when $\Omega^2(x,y)-\Omega(x,x)
\Omega(y,y)$ is identically zero for all $x,y$, and, therefore, (\ref{hdist})
gives zero distance between every pair of points. Nevertheless it is possible 
to define a nontrivial parabolic distance if one considers how (\ref{hdist}) 
approaches zero \cite{48}.

Noting that
$$\Omega(x,x)\Omega(y,y)-\Omega^2(x,y)=(x_1y_2-y_1x_2)\frac{-\Delta}{4},$$
where $\Delta=4(\Omega_{12}^2-\Omega_{11}\Omega_{22})$ is the discriminant of
the quadratic form $\Omega$, we can rewrite (\ref{hdist}) for small $\Delta$ 
as follows: 
$$d(x,y)=2iC\arcsin{\sqrt{\frac{\Omega(x,x)\Omega(y,y)-\Omega^2(x,y)}
{\Omega(x,x)\Omega(y,y)}}}$$ $$\approx iC\sqrt{-\Delta}\frac{x_1y_2-y_1x_2}
{\sqrt{\Omega(x,x)\Omega(y,y)}}.$$
However, we can assume that the arbitrary constant $C$ goes to
infinity as $\Delta\to 0$ so that $iC\sqrt{-\Delta}=k$ remains finite and 
non zero. In the parabolic limit we have $\Omega(x,x)=(p_1x_1+p_2x_2)^2$, 
where $p_1=\sqrt{\Omega_{11}}$ and $p_2=\sqrt{\Omega_{22}}$. Therefore, we get
the following formula for the parabolic distance:
\begin{equation}
d(x,y)=k\frac{x_1y_2-y_1x_2}{(p_1x_1+p_2x_2)(p_1y_1+p_2y_2)}=
\frac{Q(x)}{P(x)}-\frac{Q(y)}{P(y)},
\label{pdist}
\end{equation}
where $P(x)=p_1x_1+p_2x_2=\sqrt{\Omega(x,x)}$ and $Q(x)$ is an arbitrary 
linear form not proportional to $P(x)$ (that is, $q_1p_2-q_2p_1=k\neq 0$).
In particular, if $P(x)=x_2$ and $Q(x)=x_1$, we get the usual Euclidean 
expression for the distance between points $x$ and $y$ whose non-homogeneous
coordinates are $\tilde x=\frac{x_1}{x_2}$ and  $\tilde y=\frac{y_1}{y_2}$:
$d(x,y)=\frac{x_1}{x_2}-\frac{y_1}{y_2}=\tilde x-\tilde y$.
 
\section{Nine Cayley-Klein geometries}
A point on the projective plane $P^2$  is determined by three real coordinates
$x_1,x_2,x_3$. Therefore, a natural generalization of equation (\ref{P1fxpw}),
which defines the fundamental points of the linear measure, is the equation of
a conic section
\begin{equation}
\Omega(x,x)=\sum\limits_{i,j=1}^3 \Omega_{ij}x_ix_j=0.
\label{P2Abs}
\end{equation}
The conic section (\ref{P2Abs}) will be called the fundamental conic, or the
Absolute in Cayley's terminology. Every line incident to an interior point of 
this conic section intersects it at two points, which can be used as the 
fundamental points of  a linear metric on the line. This allows us to
define a projective metric for interior points of the Absolute. For a pair 
$x$ and $y$ of such points the line $l=x\times y$ intersects the 
Absolute at some points $p_0$ and $p_\infty$.  As in the one-dimensional case
we can write $p_0=\lambda x+y, \;p_\infty=\lambda^\prime x+y$ and
applying the same reasoning we will end up with the same formula (\ref{hdist})
for the hyperbolic distance between interior points of the Absolute lying on 
the line $l$. Of course, the arbitrary constant $C$ occurring in this 
formula must be the same for all lines on which (\ref{hdist}) defines a 
metric.

Analogously, to define the angular measure, we can consider the Absolute as
made from (tangential) lines  
\begin{equation}
\tilde \Omega(u,u)=\sum\limits_{i,j=1}^3 \tilde \Omega_{ij} u_iu_j=0,
\label{P2Absl}
\end{equation} 
where  $(\tilde \Omega_{ij})$ is the matrix of cofactors \cite{58} of the 
matrix $(\Omega_{ij})$. Then for every flat pencil of lines at every interior 
point of the Absolute we can choose two lines tangent to the Absolute 
(belonging to the Absolute considered as made from lines) as the fundamental 
lines to define the cross-ratio and the associated hyperbolic angular measure.
If two fundamental lines degenerate into one line, we will have a parabolic
angular measure and if the Absolute is such that the given  pencil of lines
does not contain (real) tangents to the Absolute, the angular measure will be
elliptic. The arbitrary constant $C^\prime$ which will appear in the 
projective angular measure must be the same for all pencils of lines but
can be different from the constant $C$ which appears in the projective linear
measure.

Therefore, we have nine different combinations of types of the linear and 
angular measures on the plane, and hence, nine different plane geometries. 
These geometries are nowadays called Cayley-Klein geometries and they are 
listed in the  Table \ref{Cayley-Klein}.

\begin{table}[htb]
\begin{center}
\begin{tabular}{|c||c|c|c|}
\hline
measure of & \multicolumn{3}{|c|}{measure of lengths} \\
\cline{2-4} 
angles & Hyperbolic & Parabolic & Elliptic \\
\hline \hline
$\;$ Hyperbolic$\;$ & Doubly hyperbolic & \hspace*{2mm} Minkowski 
\hspace*{2mm} &  co-Hyperbolic \\
           & (de-Sitter) &  & (anti de-Sitter) \\
\hline
Parabolic & co-Minkowski & Galilean & 
\hspace*{1mm} co-Euclidean \hspace*{1mm} \\
          &              &           &   \\
\hline
Elliptic & Hyperbolic & Euclidean & Elliptic \\
         & (Lobachevsky) & & (Riemann) \\
\hline
\end{tabular}
\caption{The nine two-dimensional Cayley-Klein geometries}
\label{Cayley-Klein}
\end{center}
\end{table}

Let us take a closer look at these geometries and at relations among them.
First of all, some geometries are related by duality. The points of geometry
$X$ are the lines of its dual, or co-geometry $\tilde X$, and the lines of  
geometry $X$ are the points of $\tilde X$. What was the distance between points
in geometry $X$, in the dual geometry $\tilde X$ we should call the angle 
between lines, and what was called the angle between lines in $X$, becomes 
in $\tilde X$ the distance between points. Elliptic, Galilean and doubly 
hyperbolic geometries are self-dual, while all other geometries differ from 
their duals. 

There exists \cite{58} a projective coordinate system in which the equation 
of the conic $\Omega$ takes the simple form
$$\sum\limits_{i=1}^3 b_i x_i^2=0,\;\;\mathrm{or\; in\; the\; tangential\;
form}\;\;
\sum\limits_{i=1}^3 \frac{u_i^2}{b_i}=0.$$
The homogeneous coordinates are determined only up to a scale factor. 
Therefore, we can assume without loss of generality the following
equation \cite{62} for the conic $\Omega$:
\begin{equation}
\Omega(x,x)=\epsilon_1\epsilon_2 x_1^2-\epsilon_1 x_2^2+x_3^2=0,
\label{Omega}
\end{equation}
with some constants $\epsilon_1, \epsilon_2$, or in the tangential form,
\begin{equation}
\tilde\Omega(l,l)=l_1^2 -\epsilon_2 l_2^2+\epsilon_1\epsilon_2 l_3^2=0.
\label{OmegaT}
\end{equation}
Let us introduce the following non-homogeneous coordinates (in the Euclidean 
case they will turn out to be the usual Cartesian coordinates)
\begin{equation}
z=\frac{x_1}{\sqrt{\Omega(x,x)}+x_3},\;\;\;\;\; 
t=\frac{x_2}{\sqrt{\Omega(x,x)}+x_3}.
\label{nonhxt}
\end{equation}
If the quadratic form $\Omega$ is not positive definite, we will assume that
only those points for which $\Omega(x,x)>0$ are the points of the 
corresponding geometry. however, if $\Omega(x,x)>0$ then we can choose an 
arbitrary scale factor of the homogeneous coordinates so that 
\begin{equation}
 \Omega(x,x)=1
\label{Omega1}
\end{equation}
for the scaled homogeneous coordinates. From (\ref{nonhxt}) and 
(\ref{Omega1}) we get expressions of the corresponding values of the 
homogeneous coordinates in terms of $z$ and $t$:
\begin{equation}
x_1=\frac{2z}{1-\epsilon_1 (t^2-\epsilon_2 z^2)},\;\;\;\;
x_2=\frac{2t}{1-\epsilon_1 (t^2-\epsilon_2 z^2)},\;\;\;
x_3=\frac{1+\epsilon_1 (t^2-\epsilon_2 z^2)}
{1-\epsilon_1 (t^2-\epsilon_2 z^2)}.
\label{x1x2x3}
\end{equation} 
Taking $C=k/(2\sqrt{\epsilon_1})$, we get for the distance between
points $x_0(t_0,z_0)$ and  $x(t,z)$ from (\ref{edist})
$$\cosh{[\frac{\sqrt{\epsilon_1}}{k}\,d(x_0,x)]}=\Omega(x_0,x),$$
which by using $\sinh^2{(x/2)}=(\cosh{x}-1)/2$ can be 
transformed into
\begin{equation}
\sinh^2{\left (\frac{\sqrt{\epsilon_1}\,d(x_0,x)}{2k}\right )}=\epsilon_1
\frac{(t-t_0)^2-\epsilon_2 (z-z_0)^2}{[1-\epsilon_1 (t^2-\epsilon_2 z^2)]
[1-\epsilon_1 (t_0^2-\epsilon_2 z_0^2)]}.
\label{CKdist}
\end{equation}

It is well known that the points of the Euclidean plane can be represented by 
complex numbers. Remarkably, the points of all nine Cayley-Klein geometries
also can be represented by suitably generalized complex numbers \cite{63,64}. 

Complex numbers $a+ib$ are obtained by adding a special element $i$ to the 
real numbers. This special element is characterized by the property that it 
is a solution of the quadratic equation $i^2=-1$. However, there is nothing 
special to this quadratic equation. On equal footing we can assume a special
element $e$ to be a solution of the general  quadratic equation $Ae^2+Be+C=0$.
In fact this construction gives three different types of generalized complex
numbers $a+eb$ depending on the value of the discriminant: $\Delta=B^2-4AC<0$,
$\Delta=0$, or $\Delta>0$. In the first case we get the ordinary complex 
numbers $a+ib$ and one can assume without loss of generality that $i^2=-1$. 
If the discriminant is zero, we get the so called dual numbers $a+\epsilon b$
and one can assume that $\epsilon^2=0$. At last, for a positive discriminant
we get the double numbers $a+eb$ with $e^2=1$. Note that, for example, in the
double numbers $e$ is a special unit different from $1$ or $-1$. That is, the 
equation $x^2=1$ has four different solutions in double numbers. All that is 
explained in detail in Yaglom's book {\it Complex Numbers in Geometry} 
\cite{63}.  

Now let us introduce a special element $e$ such that $e^2=\epsilon_2$ and 
the corresponding generalized complex numbers $z=t+ex$. Then (\ref{CKdist}) 
yields for the distance between points $z_0=t_0+ex_0$ and $z$
\begin{equation}
\sinh^2{\left (\frac{\sqrt{\epsilon_1}\,d(z_0,z)}{2k}\right )}=\epsilon_1
\frac{(z-z_0)(\bar z - \bar z_0)}{[1-\epsilon_1 z\bar z]
[1-\epsilon_1 z_0\bar z_0]},
\label{CKdistz}
\end{equation}
where the conjugation operation is defined as usual: $\bar z = t-ex$ and 
$z\bar z=t^2-e^2x^2=t^2-\epsilon_2 x^2$.

Analogous considerations apply to the angular measure based on the tangential
conic (\ref{OmegaT}) and we get similar formulas for the angle between
two lines, with the roles of $\epsilon_1$ and $\epsilon_2$ interchanged. 

In fact for $\epsilon_1$ there are only three possibilities:
$\epsilon_1=1$, $\epsilon_1=0$ or $\epsilon_1=-1$. All other cases can be 
reduced to these three by changing the unit of the linear measure. The same 
is true for the parameter $\epsilon_2$ also: by changing the unit of the 
angular measure this parameter can be brought to $1$ or $-1$, if different 
from zero.

If $\epsilon_1=1$, then (\ref{CKdist}) shows that the linear measure is 
hyperbolic, the points of the corresponding Cayley-Klein geometry are 
represented by double numbers if $\epsilon_2=1$ (hyperbolic angular 
measure, the de Sitter geometry), by dual numbers if $\epsilon_2=0$ 
(parabolic angular measure, the co-Minkowski geometry), and by complex numbers
if $\epsilon_2=-1$ (elliptic angular measure, the Lobachevsky geometry). The 
corresponding distance formula is
$$\sinh^2{\frac{d(z_0,z)}{2k}}=
\frac{(z-z_0)(\bar z - \bar z_0)}{[1-z\bar z][1-z_0\bar z_0]}.$$
If $\epsilon_1=-1$ then we have elliptic linear measure and the distance 
formula is
$$\sin^2{\frac{d(z_0,z)}{2k}}=
\frac{(z-z_0)(\bar z - \bar z_0)}{[1+z\bar z][1+z_0\bar z_0]}.$$
At that, the  points of the anti de-Sitter geometry are represented by double
numbers, points of the co-Euclidean geometry -- by dual numbers, and points 
of the Elliptic geometry -- by complex numbers. For $\epsilon_1=\pm1$,
the choice $k=1$ corresponds to the usual definition of length in the
Lobachevsky and Elliptic (Riemann) geometries \cite{64}.

At last, taking the limit $\epsilon_1 \to 0$, we get the distance in the 
parabolic case
$$\frac{d^2(z_0,z)}{4k^2}=(z-z_0)(\bar z - \bar z_0).$$
In the case of the elliptic angular measure, $z$ and $z_0$ are complex 
numbers, and the last formula gives the usual Euclidean distance if we take 
$k=1/2$. The points of the Galilean geometry (parabolic angular measure) are 
represented by dual numbers and the points of the Minkowski geometry 
(hyperbolic angular measure) -- by double numbers \cite{64}.

We can make contact with Riemannian geometry by noting that for 
infinitesimally close $z$ and $z_0$, and for the unit scale factor $k=1$,
(\ref{CKdist}) takes the form
\begin{equation}
ds^2=\frac{4(dt^2-\epsilon_2 dx^2)}{[1-\epsilon_1(t^2-\epsilon_2 x^2)]^2}.
\label{CKds2}
\end{equation}
Let us define the generalized cosine and sine functions as follows \cite{65,
66}
\begin{equation}
C(x;\epsilon)=\left\{\begin{array}{c} \cos{(\sqrt{-\epsilon}x)},\;\; 
\mathrm{if} \;\, \epsilon<0 \\ 1, \hspace*{19mm} \mathrm{if} \hspace*{2mm}
\epsilon=0 \\ \cosh{(\sqrt{\epsilon}x)}\;\;\;\; \mathrm{if} \;\; \epsilon>0 
\end{array} \right . , \;\;\;\; S(x;\epsilon)=\left\{\begin{array}{c} 
\frac{\sin{(\sqrt{-\epsilon}x)}}{\sqrt{-\epsilon}},\;\, \mathrm{if} \;\, 
\epsilon<0 \\ x, \hspace*{12mm}\;\; \mathrm{if} \hspace*{2mm} \epsilon=0 \\ 
\frac{\sinh{(\sqrt{\epsilon}x)}}{\sqrt{\epsilon}},\;\; \mathrm{if} \;\; 
\epsilon>0 \end{array} \right . . 
\label{Gsincos}
\end{equation}
Then
\begin{equation}
C^2(x;\epsilon)-\epsilon S^2(x;\epsilon)=1,\;\;\;
\frac{dC(x;\epsilon)}{dx}=\epsilon S(x;\epsilon),\;\;\;
\frac{dS(x;\epsilon)}{dx}=C(x;\epsilon).
\label{CSrel}
\end{equation} 
If we define the ``polar'' coordinates $(r,\phi)$ (for points with
$0<t^2-\epsilon_2 x^2<1/\epsilon_1$) through relations
$$ t=r C(\phi;\epsilon_2),\;\;\;\; x=r S(\phi;\epsilon_2), $$
then, by using identities (\ref{CSrel}), we obtain from (\ref{CKds2})
\begin{equation}
ds^2=\frac{4(dr^2-\epsilon_2 r^2 d\phi^2)}{[1-\epsilon_1 r^2]^2}=
E(r)dr^2+G(r)d\phi^2,
\label{ds2rphi}
\end{equation}
where
\begin{equation}
E(r)=\frac{4}{[1-\epsilon_1 r^2]^2}\;\;\;\mathrm{and}\;\;\;
G(r)=\frac{-4\epsilon_2 r^2}{[1-\epsilon_1 r^2]^2}.
\label{EGfun}
\end{equation}
However, for the Riemannian metric (\ref{ds2rphi}) the corresponding Gaussian 
curvature $K$ satisfies the equation \cite{48}
$$4E^2G^2K=E\left (\frac{dG}{dr}\right)^2+G\frac{dE}{dr}\frac{dG}{dr}-
2EG\frac{d^2G}{dr^2}.$$
Substituting (\ref{EGfun}), we get
$$K=-\epsilon_1.$$
This clarifies the geometric meaning of the parameters $\epsilon_1$ and
(by duality)  $\epsilon_2$: the corresponding Cayley-Klein geometry has the 
constant curvature $-\epsilon_1$ and its dual geometry, the constant 
curvature $-\epsilon_2$.  

More details about the Cayley-Klein geometries can be found in \cite{64},
or from more abstract group-theoretical point of view, in 
\cite{65,66,67,68}. We will not follow this abstract trend here, but for the 
sake of future use give below the derivation of the Lie algebras of the 
Cayley-Klein symmetry groups. 

Rigid motions (symmetries) of the Cayley-Klein geometry are those projective
transformations of $P^2$ which leave the fundamental conic (\ref{Omega}) 
invariant. Let $S=e^{\alpha G}$ be such a transformation with $G$ as its 
generator (an element of the Lie algebra of the Cayley-Klein geometry symmetry
group). Writing (\ref{Omega}) in the matrix form $x^T\Omega\, x=0$,  for 
rigid motions we get $(x^\prime)^T\Omega\, x^\prime=x^T\Omega\, x$, where 
$x^\prime=Sx$. Therefore, the invariance of the Absolute $\Omega$ under the 
symmetry transformation $S$ implies that $S^T\Omega\, S=\Omega$. For the 
infinitesimal parameter $\alpha$ this condition reduces to
$$G^T\Omega+\Omega G=0.$$
Taking in this equation
$$\Omega=\left ( \begin{array}{ccc} \epsilon_1\epsilon_2 & 0 & 0 \\
0 & -\epsilon_1 & 0 \\ 0 & 0 & 1 \end{array}\right ),$$
we get (in the general case $\epsilon_1\epsilon_2\neq 0$)
$$G_{11}=G_{22}=G_{33}=0,\;\; G_{21}=\epsilon_2 G_{12},\;\;
G_{31}=-\epsilon_1\epsilon_2 G_{13},\;\;G_{32}=\epsilon_1 G_{23}.$$ 
Therefore, the Lie algebra of the symmetry group of the Cayley-Klein geometry 
has three linearly independent generators \cite{62}
$$ G_1=\left ( \begin{array}{ccc} 0 & 0 & 0 \\
0 & 0 & -1 \\ 0 & -\epsilon_1 & 0 \end{array}\right ),\;\;\;
G_2=\left ( \begin{array}{ccc} 0 & 0 & 1 \\
0 & 0 & 0 \\ -\epsilon_1\epsilon_2 & 0 & 0 \end{array}\right ), $$
\begin{equation}
G_3=\left ( \begin{array}{ccc} 0 & 1 & 0 \\
\epsilon_2 & 0 & 0 \\ 0 & 0 & 0 \end{array}\right ).
\label{GMat}
\end{equation}
Now it is not difficult to find the commutators 
\begin{equation}
[G_1,G_2]=\epsilon_1 G_3,\;\;\; [G_3,G_1]=-G_2,\;\;\; 
[G_2,G_3]=\epsilon_2 G_1.
\label{LieG}
\end{equation} 
In fact all relevant information is encoded in these commutation relations.
Let us consider, for example, the Minkowski geometry with $\epsilon_1=0,\, 
\epsilon_2=1$ and obtain the Lorentz transformations from its Lie algebra.

For the Minkowski geometry, the commuting generators $G_1=H$ and $G_2=P$ can 
be considered  as the generators of time and space translations, while $G_3=K$ 
will play the role of the Lorentz transformation generator. From (\ref{LieG}) 
we read the commutation relations 
\begin{equation}
[H,P]=0,\;\;\; [K,H]=-P,\;\;\; [K,P]=-H.
\label{LieP}
\end{equation} 
among these generators. This defines the Lie algebra of the two-dimensional 
Poincar\'{e} group $\cal{P}$. The two-dimensional Lorentz group $\cal{L}$  is 
generated by $K$ and consists of the transformations of the form $e^{\psi K}$,
where $\psi$ is the relevant group parameter. As we stated above, the 
Minkowski space-time can be identified with the coset space 
$\cal{M}=\cal{P}/\cal{L}$. Every element of $\cal{M}$ has the form 
$[g]=g\cal{L}$, where $g=e^{x_0H}e^{xP}=e^{x_0H+xP}$ is
the element of $\cal{P}$ characterized by two real parameters $x_0$ and $x$,
which we identify with the time and space coordinates of the ``point'' $[g]$.
The Lorentz transformation $s=e^{\psi K}$ acts on the point $[g]$ in the 
following way
$$[g]\to [sg]=[sgs^{-1}],$$
where the last equality follows from the fact that $s^{-1}\in \cal{L}$, and
therefore, $s^{-1}\cal{L}=\cal{L}$. Now we need to calculate 
$$e^{\psi K}e^{x_0H+xP}e^{-\psi K}.$$
First of all, note that (\ref{GMat}) implies that  $H^2=P^2=HP=0$,
and therefore, 
$$e^{x_0H+xP}=1+x_0H+xP.$$
To proceed, recall the Baker-Campbell-Hausdorff formula \cite{69}
\begin{equation}
e^ABe^{-A}=\sum\limits_{m=0}^\infty \frac{B_m}{m!},
\label{BCH}
\end{equation}
where $B_m$ is defined recursively by $B_m=[A,B_{m-1}]$ and $B_0=B$. Taking
$A=\psi K$, $B=x_0H+xP$, we get from (\ref{LieP})
$$B_1=[A,B]=-\psi (x_0P+xH),\;\;\; B_2=[A,B_1]=\psi^2 (x_0 H+x P)=\psi^2 B.$$
Therefore,
$$B_{2k+1}=-\psi^{2k+1} (x_0 P+x H),\;\;\; B_{2k}=\psi^{2k}(x_0 H+ x P)$$
and
$$\sum\limits_{m=0}^\infty \frac{B_m}{m!}=\sum\limits_{k=0}^\infty\left \{
\frac{\psi^{2k}}{(2k)!}(x_0 H+x P)-\frac{\psi^{2k+1}}{(2k+1)!}(x_0 P+x H)
\right \}.$$
Or, after summing up the infinite series, 
$$\sum\limits_{m=0}^\infty \frac{B_m}{m!}=
(\cosh{\psi}\;x_0-\sinh{\psi}\;x)H+(\cosh{\psi}\;x-\sinh{\psi}\;x_0)P.$$
Hence, if we write $sgs^{-1}=e^{x_0^\prime H}e^{x^\prime P}=1+x_0^\prime H+
x^\prime P$, that is, if we represent the point $[sg]$ by the transformed 
coordinates $x_0^\prime,\, x^\prime$, we get
\begin{eqnarray} & &
x_0^\prime=\cosh{\psi}\;x_0-\sinh{\psi}\;x,\nonumber \\ & &
x^\prime=\cosh{\psi}\;x-\sinh{\psi}\;x_0, 
\label{eq13new}
\end{eqnarray}
which is nothing but the Lorentz transformation (\ref{eq13}) provided $\psi$
is the rapidity defined by $\tanh{\psi}=\beta$, and $x_0=ct$. 

This derivation of Lorentz transformations demonstrates clearly that the 
natural (canonical) parameter associated with Lorentz transformations is 
rapidity, not velocity \cite{70,71,72}. Therefore, to follow the intrinsic 
instead of historical logic, we should ``introduce rapidity as soon as 
possible in the teaching of relativity, namely, at the start. There is no 
need to go through the expressions of Lorentz transformations using velocity,
and then to "discover" the elegant properties of rapidity as if they resulted
from some happy and unpredictable circumstance''\cite{70}. 

\section{Possible kinematics}
As we see, special relativity has its geometric roots in the Cayley-Klein 
geometries. However, special relativity is not a geometric theory, but a 
physical one. This means that it includes concepts like causality, reference 
frames, inertial motion, relativity principle, which, being basic physical
concepts, are foreign to geometry. We showed in the first chapters that if 
we stick to these concepts then the Minkowski geometry and its singular 
$c\to\infty$ Galilean cousin remain as the only possibilities.

But what about the other Cayley-Klein geometries? Do they also have physical 
meanings? To answer this question affirmatively we have to alter or modify the
physical premises of special relativity and, as the preceding brief discussion 
of relativity without reference frames indicates, we should first of all 
generalize the Relativity Principle, getting rid of the too restrictive 
framework of inertial reference frames, whose existence in general space-times
is neither obvious nor guaranteed.

The symmetry group of special relativity is the ten-parameter Poincar\'{e}
group. Ten basis elements of its Lie algebra are the following: the generator 
$H$ of time translations; three generators $P_i$ of space translations along 
the $i$-axis; three generators $J_i$ of spatial rotations; and three 
generators $K_i$ of pure Lorentz transformations,  which can be considered as 
the inertial transformations (boosts) along the $i$-axis. The commutation 
relations involving $J_i$ have the form
\begin{equation}
[J_i,H]=0,\;\;\;[J_i,J_j]=\epsilon_{ijk}J_k,\;\;\; 
[J_i,P_j]=\epsilon_{ijk}P_k,\;\;\; [J_i,K_j]=\epsilon_{ijk}K_k,
\label{Jcomrel}
\end{equation}
and they just tell us that $H$ is a scalar and $P_i,J_i,K_i$ are vectors. 
These commutation relations can not be altered without spoiling the isotropy 
of space if we still want to regard $H$ as a scalar and  $P_i,J_i,K_i$ as 
vectors. However, other commutation relations
\begin{equation}
[H,P_i]=0,\; [H,K_i]=P_i,\; [P_i,P_j]=0,\;
[K_i,K_j]=-\epsilon_{ijk}J_k,\;[P_i,K_j]=\delta_{ij}H
\label{KHPcomrel}
\end{equation}
are less rigid as they depend on the interpretation of inertial 
transformations which we want to change. 

Besides the continuous symmetries of the Poincar\'{e} group there are some 
discrete symmetries such as the space inversion (parity) and the 
time-reversal, which play important roles in physics. Under the time-reversal
\begin{equation}
H\mapsto -H,\;\;\; P\mapsto P
\label{HPinv}
\end{equation}
and the commutation relations (\ref{Jcomrel}) and (\ref{KHPcomrel}) indicate 
that (\ref{HPinv}) can be extended up to the involutive automorphism
\begin{equation}
\tau:\;\;\;\;H\mapsto -H, \;\;\;\;P\mapsto P,\;\;\;\;J\mapsto J,\;\;\;\; 
K\mapsto -K
\label{Tinv}
\end{equation}
of the Poincar\'{e} Lie Algebra. Analogously, the space inversion is 
represented by the involutive automorphism
\begin{equation}
\pi:\;\;\;\;H\mapsto H, \;\;\;\;P\mapsto -P,\;\;\;\;J\mapsto J,\;\;\;\; 
K\mapsto -K.
\label{Pinv}
\end{equation}
We will assume \cite{73} that the generalized Lie algebra we are looking 
for also possesses the automorphisms $\tau$ and $\pi$. From the physical 
point of view we are assuming that the observed $P$- and $T$-asymmetries of 
the weak interactions have no geometric origin and the space-time itself is 
mirror symmetric. Of course, it is tempting then to expect the world of 
elementary particles to be also mirror symmetric and the less known fact is 
that this mirror symmetry can be indeed ensured by introducing hypothetical 
mirror matter counterparts of the ordinary elementary particles \cite{74}.

After deforming the commutation relations (\ref{KHPcomrel}), the Poincar\'{e}
group is replaced by the so called kinematical group -- the generalized 
relativity group of nature. In a remarkable paper \cite{73} Bacry and 
L\'{e}vy-Leblond showed that under very general assumptions there are only
eleven possible kinematics. The assumptions they used are:
\begin{itemize}
\item The infinitesimal generators $H,P_i,J_i,K_i$ transform correctly under
rotations as required by the space isotropy. This leads to the commutation 
relations (\ref{Jcomrel}).
\item the parity $\pi$ and the time-reversal $\tau$ are automorphisms of the 
kinematical group.
\item Inertial transformations in any given direction form a noncompact 
subgroup. Otherwise, a sufficiently large boost would be no boost at all, 
like $4\pi$ rotations \cite{75}, contrary to the physical meaning we ascribe 
to boosts. The role of this condition is to exclude space-times with no causal 
order, that is, timeless universes like the Euclidean one. Although, as was 
mentioned earlier, the real space-time  quite might have such inclusions, but
as there is no time in these regions and hence no motion, it is logical that 
the corresponding symmetry groups are not called kinematical.
\end{itemize}
If we demand the parity and time-reversal invariance, the only possible 
deformations of the commutation relations (\ref{KHPcomrel}) will have the 
form
$$ [H,P_i]=\epsilon_1 K_i,\;\;\; [H,K_i]=\lambda P_i,\;\;\; 
[P_i,P_j]=\alpha \epsilon_{ijk}J_k,$$
\begin{equation}
[K_i,K_j]=\beta\epsilon_{ijk}J_k,\;\;\;
[P_i,K_j]=\epsilon_2\delta_{ij}H
\label{KHPdeformed}
\end{equation}
Now we have to ensure the Jacobi identities
$$[P_i,[P_j,K_k]]+[P_j,[K_k,P_i]]+[K_k,[P_i,P_j]]=0$$
and
$$[P_i,[K_j,K_k]]+[K_j,[K_k,P_i]]+[K_k,[P_i,K_j]]=0,$$
which are satisfied only if
\begin{equation}
\alpha-\epsilon_1\epsilon_2=0
\label{alpharel}
\end{equation}
and
\begin{equation}
\beta+\lambda\epsilon_2=0.
\label{betarel}
\end{equation}
It turns out \cite{73} (this can be checked by explicit calculations) that 
all other Jacobi identities are also satisfied if (\ref{alpharel}) and 
(\ref{betarel}) hold.

As we see, the structure of the generalized Lie algebra is completely 
determined by three real parameters $\epsilon_1,\epsilon_2$ and
$\lambda$. Note that the overall sign of the structure constants is irrelevant
as the sign change of all structure constants can be achieved simply by 
multiplying each infinitesimal generator by $-1$. Therefore we can assume
$\lambda\ge 0$ without loss of generality and by a scale change it can be 
brought either to $\lambda=1$ or  $\lambda=0$.

If $\lambda=1$, the commutation relations are  
$$ [H,P_i]=\epsilon_1 K_i,\;\;\; [H,K_i]=P_i,\;\;\; 
[P_i,P_j]= \epsilon_1\epsilon_2\,\epsilon_{ijk}J_k,$$
\begin{equation}
[K_i,K_j]=-\epsilon_2\,\epsilon_{ijk}J_k,\;\;\;
[P_i,K_j]=\epsilon_2\delta_{ij}H
\label{KHPlambda1}
\end{equation}
and comparing with (\ref{LieG}) we see that every three-dimensional subalgebra
$(H,P_1,K_1)$, $(H,P_2,K_2)$ and  $(H,P_3,K_3)$ realizes the Cayley-Klein 
geometry of the type $(\epsilon_1,\epsilon_2)$ if the identifications 
$G_1=H,\,G_2=P_i,\,G_3=K_i$ are made.

Let us note that $G_3^2=\epsilon_2 E_{12}$ and $G_3E_{12}=E_{12}G_3=G_3$, 
where
$$E_{12}=\left (\begin{array}{ccc} 1 & 0 & 0 \\ 0 & 1 & 0 \\ 0 & 0 & 0
\end{array}\right ). $$
Therefore,
$$e^{\psi K}=1+\left (\psi+\frac{\psi^3}{3!}\,\epsilon_2+\frac{\psi^5}{5!}\,
\epsilon_2^2+\frac{\psi^7}{7!}\,\epsilon_2^3\ldots\right )K$$ $$+
\left (\frac{\psi^2}{2!}\,\epsilon_2+\frac{\psi^4}{4!}\,\epsilon_2^2+
\frac{\psi^6}{6!}\,\epsilon_2^3+\ldots\right )E_{12}$$
and recalling the definition of the generalized sine and cosine functions
(\ref{Gsincos}) we get
\begin{equation}
e^{\psi K}=(1-E_{12})+S(\psi;\epsilon_2)K+C(\psi;\epsilon_2)E_{12}.
\label{epsiK}
\end{equation}
As we see, $e^{\psi K}$ is periodic if $\epsilon_2<0$; that is, in this case
the inertial transformations form a compact group and the corresponding
Cayley-Klein geometries with elliptic angular measure (Lobachevsky, Euclidean
and Elliptic) do not lead to kinematical groups. 

The remaining six cases include: the de Sitter kinematics ($DS$) with 
doubly-hyperbolic geometry $\epsilon_1>0,\,\epsilon_2>0$, the anti de 
Sitter kinematics ($ADS$) with the co-hyperbolic geometry $\epsilon_1<0,\,
\epsilon_2>0$, the Poincar\'{e} kinematics ($P$) with the Minkowski geometry
$\epsilon_1=0,\,\epsilon_2>0$, the Newton-Hook kinematics ($NH$) 
with the co-Minkowski geometry $\epsilon_1>0,\,\epsilon_2=0$, the anti 
Newton-Hook kinematics  ($ANH$) with the co-Euclidean geometry $\epsilon_1<0,
\,\epsilon_2=0$ and the Galilean kinematics ($G$) $\epsilon_1=0,\,
\epsilon_2=0$ whose geometry is described in detail in \cite{64}.

At last, if $\lambda=0$, the commutation relations are  
$$[H,P_i]=\epsilon_1 K_i,\;\;\; [H,K_i]=0,\;\;\; 
[P_i,P_j]= \epsilon_1\epsilon_2\,\epsilon_{ijk}J_k,$$
\begin{equation}
[K_i,K_j]=0,\;\;\;[P_i,K_j]=\epsilon_2\delta_{ij}H.
\label{KHPlambda0}
\end{equation}
Suppose that $\epsilon_1\ne 0$ and introduce another basis in the Lie algebra
(\ref{KHPlambda0}):
$$P^\prime_i=\epsilon_1 K_i,\;\;\; K_i^\prime=P_i, \;\;\;
H^\prime=H,\;\;\; J_I^\prime=J_i.$$
The commutation relations in the new bases take the form
$$ [H^\prime,P_i^\prime]=0,\;\;\; [H^\prime,K_i^\prime]=P^\prime,\;\;\; 
[P_i^\prime,P_j^\prime]=0,$$ $$
[K_i^\prime,K_j^\prime]=\epsilon_1\epsilon_2\,\epsilon_{ijk}J_k^\prime,\;\;\;
[P_i^\prime,K_j^\prime]=-\epsilon_1\epsilon_2\,\delta_{ij}H^\prime.$$
However, this is the same Lie algebra as (\ref{KHPlambda1}) for the 
Cayley-Klein parameters $\epsilon_1^\prime=0$ and $\epsilon_2^\prime=
-\epsilon_1\epsilon_2$. Nevertheless, the physics corresponding to the 
isomorphic algebras (\ref{KHPlambda0}) and  (\ref{KHPlambda1}) are completely 
different because we prescribe to the generators $H,\,P,\,K$ a well-defined 
concrete physical meaning and they cannot be transformed arbitrarily, except 
by scale changes \cite{73}.

Anyway, all these new possibilities also realize Cayley-Klein geometries. At
that, $K_i=G_2/\epsilon_1$ and by using $G_2^2=0$ (for  $\epsilon_1^\prime=0$)
we get $e^{\psi K_i}=1+\psi\,K_i$; that is, the subgroup generated by $K_i$
is always non-compact in this case. However, $(-\epsilon_1,-\epsilon_2)$ and
$(\epsilon_1,\epsilon_2)$ give the same Lie algebra as the first case just
corresponds to the change of  basis: $H\mapsto -H,\,P\mapsto -P,\,K\mapsto 
-K,\,J\mapsto J$. Therefore, one is left with three possibilities: the anti 
para-Poincar\'{e} kinematics ($AP^\prime$) $\epsilon_1=1,\,\epsilon_2=1$ 
with Euclidean geometry $\epsilon_1^\prime=0,\,\epsilon_2^\prime=-1$, the
para-Poincar\'{e} kinematics ($P^\prime$)  $\epsilon_1=-1,\,\epsilon_2=1$ 
with the Minkowski geometry $\epsilon_1^\prime=0,\,\epsilon_2^\prime=1$, and 
the para-Galilei kinematics ($G^\prime$) $\epsilon_1=1,\,\epsilon_2=0$ with 
Galilean geometry  $\epsilon_1^\prime=0,\,\epsilon_2^\prime=0$.

The case $\lambda=0,\;\epsilon_1=0$ adds two more possibilities: the Carroll 
kinematics ($C$) with $\epsilon_2=\pm 1$, first discovered in \cite{76}, and 
the static kinematics ($S$) with $\epsilon_2=0$ for which all commutators 
between $H,\,P,\,K$ vanish.

In the case of the Carroll kinematics, the Galilean geometry is realized in 
each $(H,P_i,K_i)$ subspaces with $G_1=P_i,\,G_2=H,\,G_3=K_i$. As we see, 
compared to the Galilean kinematics, the roles of time and space translations
are interchanged in the Carroll kinematics. This leads to the exotic situation
of absolute space but relative time. That is, an event has the same spatial
coordinates irrespective of the applied inertial transformation (change of 
the reference frame) \cite{76,77}. There are no interactions between spatially 
separated events, no true motion, and practically no causality. The evolution 
of isolated and immobile physical objects corresponds to the ultralocal 
approximation of strong gravity \cite{78}. The name of this strange kinematics
is after Lewis Carroll's tale {\it Through the Looking-Glass, and What Alice 
Found There} (1871), where the Red Queen points out to Alice: "A slow sort of 
country! Now, here, you see, it takes all the running you can do, to keep in 
the same place. If you want to get somewhere else, you must run at least twice 
as fast as that!"

It is remarkable that under very general assumptions all possible kinematics,
except the static one, have Cayley-Klein geometries at their background 
\cite{79,79-C}. By deforming the Absolute of the doubly-hyperbolic geometry 
(by changing parameters $\epsilon_1$ and $\epsilon_2$) one can get all other
Cayley-Klein geometries. It is not surprising, therefore, that all eleven
possible kinematics are in fact limiting cases of the de Sitter or anti 
de Sitter kinematics \cite{73} which are the most general relativity groups. 
As a result, there exists essentially only one way to generalize special 
relativity, namely, by endowing space-time with some constant curvature 
\cite{73}.
  
\section{Group contractions}
There is a subtlety in considering various limiting cases of the kinematical 
groups introduced in the previous chapter. We demonstrate this by considering
the non-relativistic limit of the Poincar\'{e} group. It is commonplace
that the Lorentz transformations (\ref{eq13}), written for the space and time
intervals $\Delta x$ and $\Delta t$,
\begin{eqnarray} & &
\Delta x^\prime=\frac{1}{\sqrt{1-\beta^2}}(\Delta x-V\Delta t),
\nonumber \\ & &
\Delta t^\prime=\frac{1}{\sqrt{1-\beta^2}}\left (\Delta t-\frac{V}{c^2}\,
\Delta x\right ), \nonumber
\end{eqnarray}
in the non-relativistic limit $\beta\to 0$ reduce to the Galilei 
transformations
\begin{eqnarray} & &
\Delta x^\prime=\Delta x-V\Delta t,
\nonumber \\ & &
\Delta t^\prime=\Delta t.
\label{Gtrans}
\end{eqnarray}
But to reach this conclusion, besides $\beta\ll 1$, we have implicitly 
assumed that
$$\beta \frac{\Delta x}{c\Delta t}\ll 1,\;\;\;\;\beta\frac{c\Delta t}
{\Delta x}\sim 1,$$
or
$$\beta\ll\frac{c\Delta t}{\Delta x},\;\;\;\; \frac{\Delta x}
{c\Delta t}\sim\beta\ll 1,$$
which is not necessarily true if $\Delta x \gg c\Delta t$. 

Therefore the Galilei transformations are not the non-relativistic limit of 
the Lorentz transformations but only one non-relativistic limit, which 
corresponds to the situation when space intervals are much smaller than time 
intervals \cite{76} (in units where $c=1$). There exists another 
non-relativistic limit
$$\beta\frac{c\Delta t}{\Delta x}\ll 1,\;\;\;\;
\beta \frac{\Delta x}{c\Delta t}\sim 1,$$
or
$$\beta\ll\frac{\Delta x}{c\Delta t},\;\;\;\; \frac{c\Delta t}{\Delta x}
\sim \beta\ll 1$$
and in this limit the Lorentz transformations reduce to the Carroll 
transformations \cite{76,77}
\begin{eqnarray} & &
\Delta x^\prime=\Delta x,
\nonumber \\ & &
\Delta t^\prime=\Delta t-\frac{V}{c^2}\Delta x.
\label{Ctrans}
\end{eqnarray}
The Carroll kinematics corresponds to the situation when space intervals are
much larger than time intervals, hence no causal relationship between events
are possible; events are isolated.

A systematic way to correctly treat limiting cases of symmetry groups was 
given by In\"{o}nu and Wigner \cite{80} and is called ``group contraction''.  
A symmetry group $G$ can be contracted towards its continuous subgroup $S$,
which remain intact under the contraction process. Denote by $J_i$ the 
generators of the subgroup $S$ and the  the remaining generators of $G$, by 
$I_i$. Therefore the Lie algebra of the group $G$ looks like
\begin{equation}
[J_i,J_j]=f_{ijk}^{(1)}J_k,\;\;\;[I_i,J_j]=f_{ijk}^{(2)}J_k+g_{ijk}^{(2)}I_k,
\;\;\;[I_i,I_j]=f_{ijk}^{(3)}J_k+g_{ijk}^{(3)}I_k.
\label{GSLie}
\end{equation}
Under the change of basis
\begin{equation}
J_i^\prime=J_i,\;\;\;\;I_i^\prime=\epsilon\,I_i,
\label{Icontr}
\end{equation}
the commutation relations transform to
\begin{equation}
[J_i^\prime,J_j^\prime]=f_{ijk}^{(1)}J_k^\prime,\;\;\;
[I_i^\prime,J_j^\prime]=\epsilon f_{ijk}^{(2)}J_k^\prime+g_{ijk}^{(2)}
I_k^\prime,\;\;\;[I_i^\prime,I_j^\prime]=\epsilon^2 f_{ijk}^{(3)}J_k^\prime+
\epsilon g_{ijk}^{(3)}I_k^\prime.
\label{GSLieeps}
\end{equation}
When $\epsilon\to 0$, the base change (\ref{Icontr}) becomes singular but the
commutation relations (\ref{GSLieeps}) still have a well-defined limit as 
$\epsilon\to 0$:  
\begin{equation}
[J_i^\prime,J_j^\prime]=f_{ijk}^{(1)}J_k^\prime,\;\;\;
[I_i^\prime,J_j^\prime]=g_{ijk}^{(2)}I_k^\prime,\;\;\;
[I_i^\prime,I_j^\prime]=0.
\label{GprimeLie}
\end{equation}
In general, the Lie algebra (\ref{GprimeLie}) is not isomorphic to the 
initial Lie algebra (\ref{GSLie}) and defines another symmetry group 
$G^\prime$, which is said to be the result of the contraction of the group 
$G$ towards its continuous subgroup $S$. The contracted generators 
$I_i^\prime$ form an abelian invariant subgroup in $G^\prime$, as 
(\ref{GprimeLie}) shows.

The contraction has a clear meaning in the group parameter space. If 
$I^\prime=\epsilon I$, then the corresponding group parameters should satisfy
$\alpha=\epsilon\alpha^\prime$ if we want $e^{\alpha I}$ and $e^{\alpha^\prime
I^\prime}$ to represent the same point (transformation) of the group. 
Therefore, when $\epsilon\to 0$ the parameter $\alpha$ becomes infinitesimal.
From the point of view of $G$, its contracted form $G^\prime$ covers only
infinitesimally small neighborhood of the subgroup $S$. This explains why the 
operation is called contraction.

\begin{figure}[htb]
 \begin{center}
    \mbox{\includegraphics[scale=0.38]{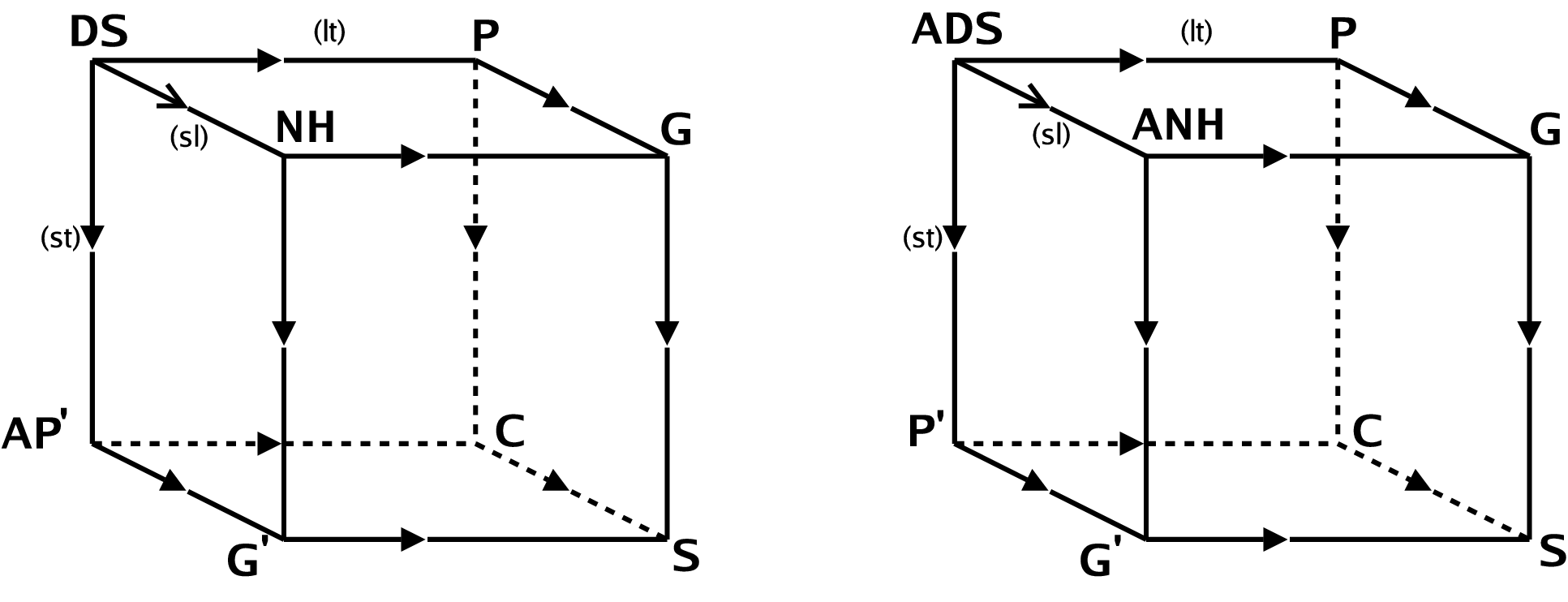}}
  \end{center}
\caption {The eleven kinematical groups and relations between them.}
\label{Kgroups}
\end{figure}

Let us return to kinematical groups. We do not want to spoil the space 
isotropy by contraction. Therefore, $S$ should be a rotation-invariant 
subgroup of $G$. Looking at the commutation relations (\ref{Jcomrel}) and
(\ref{KHPdeformed}), we find only four rotation-invariant subalgebras, 
generated respectively by $(J_i,H)$, $(J_i,P_i)$, $(J_i,K_i)$ and $(J_i)$,
which are common to all kinematical Lie algebras. Therefore, we can consider
four types of physical contractions of the kinematical groups \cite{73}:
\begin{itemize}
\item Contraction with respect to the rotation and time-translation subgroups 
generated by $(J_i,H)$. Under this contraction
\begin{equation}
P_i\to\epsilon\,P_i,\;\;\;\;K_i\to\epsilon\,K_i,
\label{JHcontr}
\end{equation}
and as $\epsilon\to 0$ the contracted algebra is obtained by the substitutions
\begin{equation}
\epsilon_1\to\epsilon_1 ,\;\;\;\;\lambda\to\lambda,\;\;\;\;\epsilon_2\to 0.
\label{JHcontreps}
\end{equation} 
As (\ref{JHcontr}) indicates, the corresponding limiting case is characterized
by small speeds (parameters of the inertial transformations) and small space
intervals. So this contraction can be called Speed-Space ($sl$) contraction 
\cite{73}. According to (\ref{JHcontreps}), the Speed-Space contraction 
describes a transition from the relative-time groups to the absolute-time
groups:
$$DS\to NH,\;ADS\to ANH,\; P\to G,\; AP^\prime \to G^\prime,\;
P^\prime \to G^\prime,\; C\to S.$$
\item Contraction with respect to the three-dimensional Euclidean group 
generated by $(J_i,P_i)$, which is the motion group of the 
three-dimensio\-nal Euclidean space. Under this contraction
$$H\to\epsilon\,H,\;\;\;\;K_i\to\epsilon\,K_i,$$
and the  limit $\epsilon\to 0$ produces the changes
$$\epsilon_1\to\epsilon_1 ,\;\;\;\;\lambda\to 0\;\;\;\;\epsilon_2\to 
\epsilon_2.$$
The physical meaning of this contraction is the limit when speeds are small 
and time intervals are small; hence the name  ``Speed-Time ($st$) 
contraction''. The speed-Time contraction leads to the absolute-space groups 
with essentially no causal relations between events, and hence, of reduced 
physical significance:
$$DS\to AP^\prime, \; ADS\to P^\prime, \; P\to C, \; NH\to G^\prime,
\; ANH\to G^\prime, \; G\to S.$$
The absolute space groups themselves remain intact under the Speed-Time 
contraction.
\item Contraction with respect to the Lorentz group generated by $(J_i,K_i)$.
This is the Space-Time ($lt$) contraction because under this contraction
$$H\to\epsilon\,H,\;\;\;\;P_i\to\epsilon\,P_i$$
and we get the limiting case of small space and time intervals. The contracted
groups are obtained by the changes 
$$\epsilon_1\to 0,\;\;\;\;\lambda\to \lambda, \;\;\;\;\epsilon_2\to 
\epsilon_2$$
and we get transitions from the global (cosmological) groups to the local 
groups:
$$DS\to P,\;\;\; ADS\to P,\;\;\; NH\to G,\;\;\; ANH\to G,$$ $$ AP^\prime 
\to C,\;\;\; P^\prime \to C,\;\;\; G^\prime \to S.$$
\item Contraction with respect to the rotation subgroup generated by $J_i$.
Under this Speed-Space-Time ($slt$) contraction
$$H\to\epsilon\,H,\;\;\;\;P_i\to\epsilon\,P_i,\;\;\;K_i\to\epsilon\,K_i,$$
which leads in the limit $\epsilon\to 0$ to
$$\epsilon_1\to 0,\;\;\;\;\lambda\to 0, \;\;\;\;\epsilon_2\to 0.$$ 
That is, all kinematical groups are contracted into the static group.
\end{itemize}
Schematically the relations between various kinematical groups are shown in 
Fig.\ref{Kgroups}. All these groups are limiting cases of the de Sitter or 
anti de Sitter groups.

As we have seen above, the natural parameter of the inertial transformations
in the Poincar\'{e} group is the rapidity, which is dimensionless. One can ask 
the similar question about the natural dimension of speeds in other kinematical
groups also. Note that the term speeds, as opposed to velocities, will be used 
to denote natural parameters of inertial transformations in general. 

It is natural to choose group parameters for which the Lie algebra 
structure constants are dimensionless \cite{81}. At that, as the Lie algebra 
generators have dimensions inverse to those of the corresponding group 
parameters, every non-zero commutation relation between $H,P_i$ and $K_i$ will
induce a non-trivial relation between their dimensions $[H],[P_i]$ and 
$[K_i]$.

In the case of static kinematics, all commutation relations vanish. Therefore,
there are no non-trivial dimensional relations, and the dimensions of time 
translations, space translations and speeds, which we denote respectively by
$T,L$ and $S$ (so that $[H]=T^{-1}$, $[P_i]=L^{-1}$, $[K_i]=S^{-1}$), are all
independent.

For the de Sitter and anti de Sitter groups, all structure constants are 
non-zero, which implies the following dimensional relations (note that angles, 
and hence $J_i$, are dimensionless)
$$T^{-1}L^{-1}=S^{-1},\;\;T^{-1}S^{-1}=L^{-1},\;\;L^{-2}=1,\;\;S^{-2}=1,\;\;
L^{-1}S^{-1}=T^{-1}.$$
This is possible only if $L=T=S=1$. Therefore the natural group parameters for
de Sitter and anti de Sitter kinematics are dimensionless. ``From our 
`Galilean' viewpoint, we could say that in the de Sitter universe there is a 
`characteristic' length, a `characteristic' time and a `characteristic'' speed
which may be used as natural units, and then lengths, times and speeds are
dimensionless'' \cite{81}.

In Poincar\'{e} kinematics we have three non-trivial dimensional relations
$$T^{-1}S^{-1}=L^{-1},\;\;\; S^{-2}=1,\;\;\;L^{-1}S^{-1}=T^{-1},$$
which imply that speeds are dimensionless and $L=T$. That is, space and time
are unified in one dimensional quantity while speeds are natural to measure 
in terms of a characteristic speed $c$.

In Galilei kinematics $T^{-1}S^{-1}=L^{-1}$ and we get $S=LT^{-1}$, which is 
the usual velocity. Space and time in this case are independent dimensional
quantities.

Dimensional structures of other kinematical groups are more exotic \cite{81}.
In para-Poincar\'{e} and anti para-Poincar\'{e} case $L=1$ and $S=T$. That is,
there is a characteristic length while time and speeds are unified in one 
dimensional quantity. In Carroll case $S=TL^{-1}$ and space and time are 
independent dimensional quantities. Newton-Hook and anti Newton-Hook 
space-times are characterized by $T=1$ and $S=L$; that is, there is a 
characteristic time while length and speeds are unified. At last, in 
para-Galilei case space and time are independent dimensional quantities but
$S=LT$.
 
Remarkably, all kinematical groups admit a four-dimensional space-time which
can be identified with the homogeneous space of the group, namely, with its
quotient by the six-dimensional subgroup generated by the rotations $J_i$ and
the inertial transformations $K_i$. At that, for the kinematical groups with
vanishing commutators $[K_i,H]=0$ and $[K_i,P_j]=0$, that is, for the
para-Galilei and static groups, inertial transformations do not act on
the space-time. For other groups the space-time transforms non-trivially under
inertial transformations. Let us find the corresponding transformations for
Newton-Hook groups, for example, in the case of (1+1)-dimensional space-time
to avoid unnecessary technical details.

A space-time point $(x_0,x)$ is the equivalence class of the group element
$e^{x_0H}e^{xP}$. After the inertial transformation $e^{\psi K}$, we get a new
point $(x_0^\prime,x^\prime)$, which, on the other hand, is the equivalence 
class of the group element $e^{\psi K}e^{x_0H}e^{xP}e^{-\psi K}$. For 
Newton-Hook groups $\epsilon_2=0$ and, as (\ref{GMat}) implies, the generators
$H=G_1,P=G_2,K=G_3$ satisfy
\begin{equation}
H^2=\epsilon_1E_{23},\;\;\;E_{23}H=H,\;\;\;P^2=0,\;\;\;K^2=0,\;\;\;
E_{23}K=0, $$ $$
E_{23}P=0,\;\;\;HP=0,\;\;\;HK=0,\;\;\;PK=0,
\label{EHPK}
\end{equation}
with
$$E_{23}=\left (\begin{array}{ccc} 0 & 0 & 0 \\ 
0 & 1 & 0 \\ 0 & 0 & 1 \end{array} \right ).$$
Using the first two of these relations, we get
$$e^{x_0H}=1+\left[C(x_0;\epsilon_1)-1\right]E_{23}+S(x_0;\epsilon_1)H,$$
and after applying the Baker-Campbell-Hausdorff formula (\ref{BCH}), along 
with the commutation relations (\ref{KHPlambda1}), one easily obtains
$$e^{\psi K}e^{x_0 H}e^{-\psi K}=1+\left[C(x_0;\epsilon_1)-1\right]E_{23}+
S(x_0;\epsilon_1)H$$ $$+\psi\left[C(x_0;\epsilon_1)-1\right]K-\psi S(x_0;
\epsilon_1)P.$$
Because of the relations (\ref{EHPK}), the right-hand-side is the same as 
$$e^{x_0H}e^{-\psi S(x_0;\epsilon_1)P}e^{\psi [C(x_0;\epsilon_1)-1]K}.$$
But $e^{\psi K}e^{x P}e^{-\psi K}=e^{x P}$, as $P$ and $K$ do commute, and
therefore,
$$e^{\psi K}e^{x_0H}e^{x P}e^{-\psi K}=e^{\psi K}e^{x_0H}e^{-\psi K}\,
e^{\psi K}e^{x P}e^{-\psi K}$$ $$=e^{x_0H}e^{[x-\psi S(x_0;\epsilon_1)]P}
e^{\psi [C(x_0;\epsilon_1)-1]K}$$
which has the same equivalence class as $e^{x_0H}e^{[x-\psi S(x_0;
\epsilon_1)]P}$. Therefore, the transformation law is
\begin{eqnarray} & &
x_0^\prime=x_0,\nonumber \\ & & x^\prime=x-\psi\, S(x_0;\epsilon_1).
\label{e2translaw}
\end{eqnarray}
For $\epsilon_1=0$, we get Galilean transformations with $x_0=t$ and $\psi=V$,
which are the natural choice of group parameter dimensions for the Galilei 
kinematics.

For Newton-Hook kinematics we have a characteristic time $\tau$ and under the 
natural choice of group parameters $\epsilon_1=1$, $x_0=t/\tau$, $\psi=V\tau$
the transformation law (\ref{e2translaw}) takes the form
\begin{eqnarray} & &
x_0^\prime=x_0,\nonumber \\ & & x^\prime=x-V\tau\, \sinh{\frac{t}{\tau}}.
\label{NHtranslaw}
\end{eqnarray} 
In contrast to the Poincar\'{e} or Galilean case, a particle at rest $x=0$
experiences exponentially accelerated inertial motion $$x^\prime=-V\tau\, 
\sinh{\frac{t^\prime}{\tau}}$$ if subjected to a pure inertial transformation.
Therefore, Newton-Hook kinematics corresponds to expanding universe \cite{73}.

In the case of anti Newton-Hook kinematics, $\epsilon_1=-1$ and we have 
oscillating universe with inertial transformations
\begin{eqnarray} & &
x_0^\prime=x_0,\nonumber \\ & & x^\prime=x-V\tau\, \sin{\frac{t}{\tau}}.
\label{ANHtranslaw}
\end{eqnarray}

Would Minkowski follow the logic of his staircase-wit from the ``{\it Raum 
und Zeit}'' to the end, he would reveal that the Poincar\'{e} group is 
mathematically less intelligible than the de Sitter or anti de Sitter groups, 
whose limiting cases it is. Therefore, he missed an opportunity to predict 
the expanding universe already in 1908 \cite{40}.

However, as the existence of the Newton-Hook groups indicate, the Newtonian 
($\epsilon_2=0$) relativity is sufficient to predict the expanding universe 
and it is not well-known that the expanding universe was indeed proposed
by Edgar Allan Poe in 1848, many years before Friedmann and Lema\^{i}tre!
Poe's {\it Eureka}, from which we borrowed the preface, is an extended version
of the lecture Poe gave at the Society Library of New York. The lecture was 
entitled ``On the Cosmogony of the Universe'', and despite its quite naive 
and metaphysical premises this bizarre mixture of metaphysics, philosophy, 
poetry, and science contains several brilliant ideas central in modern-day 
cosmology, including a version of the Big Bang and evolving universe with 
inflation of the primordial atom at the start, resolution of the Olbers' 
paradox (why sky is dark at night), an application of the Anthropic 
Cosmological Principle to explain why the universe is so immensely large, a  
suggestion of the multiverse with many causally disjoint universes, each with
its own set of physical laws \cite{82,83,84}. Although this is quite 
fascinating, it seems Poe was driven mainly by his poetic aesthetics in 
producing these ideas rather than by scientific logic \cite{85}. As a result, 
these ideas, being far ahead of the time, remained obscure for contemporaries 
and have not played any significant role in the historical development of 
cosmology. Curiously, as witnessed by his biographers, Poe was Friedmann's 
favorite writer. ``Did  Friedmann read {\it Eureka}? It would be not serious 
to push this game too far'' \cite{82}.   

\section{Once more about mass}
It seems worthwhile to return to ``the virus of relativistic mass'' \cite{86}
and inspect the concept of mass from the different viewpoint provided by
the quantum theory. After all, the creation of quantum mechanics was another
and more profound conceptual revolution in physics. Unfortunately, the 
corresponding dramatic change of our perspective of reality is usually 
ignored in teaching relativity. 

The states of a quantum-mechanical system are described by vectors in 
a Hilbert space. At that, vectors which differ only in phase represent the
same state. That is, the quantum-mechanical state is represented by a ray
\begin{equation}
e^{i\alpha}|\Psi\left > \right .,
\label{ray}
\end{equation}
where $\alpha$ is an arbitrary phase, rather than by a single vector 
$|\Psi\left > \right .$. This gauge freedom has some interesting consequences 
for a discussion of symmetries in quantum case.

Symmetries of a quantum system are represented by unitary (or anti-unitary, if
time reversal is involved) operators in the Hilbert space. Because of the
gauge freedom (\ref{ray}), these unitary operators are also defined only up 
to phase factors. Let us take a closer look at these things assuming that
under the symmetry transformation $g\in G$ the space-time point 
$x=(t,\vec{x})$ transforms into $x^\prime=(t^\prime,\vec{x}^{\,\prime})$. 
For the short-hand notation we will write $x^\prime=g(x)$. Let 
$|x\left >\equiv|\vec{x},t\right >$ be the basis vectors of the 
$x$-representation, that is, of the representation  where the coordinate 
operator is diagonal. The symmetry $g$ is represented in the Hilbert space by 
a unitary operator $U(g)$ and, obviously, $|x^\prime\left > \right .$ and 
$U(g)|x\left > \right .$ should represent the same state, that is, they can 
differ only by a phase factor:
\begin{equation}
|x^\prime\left > \right .=e^{i\alpha(x;\,g)}U(g)|x\left > \right .,\;\;\;\; 
\left < x\right .|U(g)=e^{i\alpha_1(x;\,g)}\left <\, g\right . ^{-1}(x)|.
\label{alpha1}
\end{equation}
Here the second identity follows from the first one when we take into 
account that operators $U^+(g)=U^{-1}(g)$ and $U(g^{-1})$ can differ from each
other only by a phase factor. For an arbitrary state vector $|\Psi\left > 
\right .$ we have the transformation law
$$|\Psi^\prime\left>=U(g)|\Psi\right>.$$
Using (\ref{alpha1}), one has for the wave function
$$\Psi^\prime(\vec{x},t)=\left <x|\Psi^\prime\right >=\left <x|U(g)|\Psi
\right>=e^{i\alpha_1(x;\,g)}\left <x^{\prime\prime}|\psi\right >=
e^{i\alpha_1(x;\,g)}\Psi(\vec{x}^{\,\prime\prime},t^{\prime\prime}),$$
where $x^{\prime\prime}=g^{-1}(x)$. Therefore,
\begin{equation}
\Psi^\prime (x)=e^{i\alpha_1(x;\,g)}\Psi(g^{-1}(x)).
\label{Psiprime}
\end{equation}
Now let us compare $\left <x|U(g_1)U(g_2)|\Psi\right >$ and 
$\left <x|U(g_1\cdot g_2)|\Psi\right >$ 
\cite{87}. Using (\ref{alpha1}), we get
$$\left <x|U(g_1)U(g_2)|\Psi\right >=e^{i\alpha_1(x;\,g_1)}\left <x_1|U(g_2)|
\Psi\right >=e^{i\alpha_1(x;\,g_1)} e^{i\alpha_1(x_1;\,g_2)} \left <x_{12}|
\Psi\right >,$$
where $x_1=g_1^{-1}(x)$ and $x_{12}=(g_1\cdot g_2)^{-1}(x)$. On the other 
hand,
$$\left <x|U(g_1\cdot g_2)|\Psi\right >=e^{i\alpha_1(x;\,g_1\cdot g_2)}
\left <x_{12}|\Psi\right >.$$
Therefore, 
\begin{equation}
\left <x|U(g_1)U(g_2)|\Psi\right >=e^{i\alpha_2(x;\,g_1,g_2)}
\left <x|U(g_1\cdot g_2)|\Psi\right >
\label{U(g1g2)}
\end{equation}
for all space-time points $x$ and for all state vectors 
$|\Psi\left > \right .$. Here
$$\alpha_2(x;g_1,g_2)=\alpha_1(g_1^{-1}(x);g_2)-\alpha_1(x;g_1\cdot g_2)+
\alpha_1(x;g_1)$$
\begin{equation}
=(\delta \alpha_1)(x;g_1,g_2).
\label{alpha2}
\end{equation}
It will be useful to use elementary cohomology terminology \cite{88}, 
although we will not go into any depth in this high-brow theory. Any real 
function $\alpha_n(x;g_1,g_2,\ldots,g_n)$ will be called a cochain. The action
of the coboundary operator $\delta$ on this cochain is determined as follows:
\begin{eqnarray} && \hspace*{35mm}
(\delta \alpha_n)(x;g_1,g_2,\ldots,g_n,g_{n+1})= \nonumber \\ && 
\alpha_n(g_1^{-1}(x);g_2,g_3,\ldots,g_n,g_{n+1})-\alpha_n(x;g_1\cdot g_2,
g_3,\ldots,g_n,g_{n+1})+ \nonumber \\ &&
\alpha_n(x;g_1,g_2\cdot g_3,g_4,\ldots,g_n,g_{n+1})-
\alpha_n(x;g_1,g_2,g_3\cdot g_4,g_5,\ldots,g_n,g_{n+1})+\nonumber \\ && 
\cdots+(-1)^{n}\alpha_n(x;g_1,g_2,\ldots,g_n\cdot g_{n+1})+(-1)^{n+1}
\alpha_n(x;g_1,g_2,\ldots,g_n).
\label{coboundary}
\end{eqnarray}
The coboundary operator has the following fundamental property
\begin{equation}
\delta^2=0.
\label{delta2}
\end{equation}
A cochain with zero coboundary is called a cocycle. Because of (\ref{delta2}),
every coboundary $\alpha_n=\delta \alpha_{n-1}$ is a cocycle. However, not 
all cocycles can be represented as coboundaries. Such cocycles will be called 
nontrivial.

Low dimensional cocycles play an important role in the theory of 
representations of the symmetry group $G$ \cite{88}. For example, if 
$\alpha_1(x;g)$ is a cocycle, so that $\alpha_2(x;g_1,g_2)$ vanishes, then
(\ref{U(g1g2)}) indicates that 
\begin{equation}
U(g_1\cdot g_2)=U(g_1)U(g_2)
\label{Urep}
\end{equation}
and the unitary operators $U(g)$ realize a representation of the group $G$.

However, because of the gauge freedom related to the phase ambiguity, it is 
not mandatory that the unitary operators $U(g)$ satisfy the representation 
property (\ref{Urep}). It will be sufficient to have a projective (or ray) 
representation \cite{89}
\begin{equation}
U(g_1\cdot g_2)=e^{-i\xi(g_1,g_2)}U(g_1)U(g_2).
\label{UProjrep}
\end{equation}
The so-called local exponent $\xi(g_1,g_2)$ can not be quite arbitrary. 
In particular, the associativity property 
$(g_1\cdot g_2)\cdot g_3=g_1\cdot (g_2\cdot g_3)$ implies that
\begin{equation}
\xi(g_1,g_2)+\xi(g_1\cdot g_2, g_3)=\xi(g_2,g_3)+\xi(g_1,g_2\cdot g_3),
\label{cocyclecond}
\end{equation}
and, therefore, $\xi(g_1,g_2)$ is a global (not-dependent on the 
space-time point $x$) cocycle:
$$(\delta \xi)(g_1,g_2,g_3)=\xi(g_2,g_3)-\xi(g_1\cdot g_2,g_3)+
\xi(g_1,g_2\cdot g_3)-\xi(g_1,g_2)=0.$$
If the cocycle (\ref{alpha2}) does not depend on $x$, one can identify
$$\xi(g_1,g_2)=\alpha_2(x;g_1,g_2).$$
At that $\alpha_2(x;g_1,g_2)$ is a trivial local cocycle, as (\ref{alpha2})
indicates, but globally it is not necessarily trivial, that is, representable 
as the coboundary of a global cochain.

If unitary operators $U(g)$ constitute a projective representation of the 
symmetry group $G$, then the correspondence $(\theta,g)\to e^{i\theta}U(g)$,
where $\theta$ is a real number and $g\in G$, gives an ordinary representation 
of the slightly enlarged group $\tilde G$ consisting of all pairs 
$(\theta,g)$. A group structure on $\tilde G$ is given by the multiplication 
law
\begin{equation}
(\theta_1,g_1)\cdot (\theta_2,g_2)=(\theta_1+\theta_2+\xi(g_1,g_2),\;g_1\cdot 
g_2).
\label {Gtildemult}
\end{equation}
Indeed, the cocycle condition (\ref{cocyclecond}) ensures that the 
multiplication law (\ref{Gtildemult}) is associative. If we assume, as is
usually done, that $U(e)=1$, where $e$ is the unit element of the group $G$,
then (\ref{UProjrep}) indicates that $\xi(e,e)=0$. By setting, respectively,
$g_2=g_3=e$, $g_1=g_2=e$ and $g_1=g_3=g, g_2=g^{-1}$ in (\ref{cocyclecond}),
we get
\begin{equation}
\xi(e,g)=\xi(g,e)=0,\;\;\;\xi(g,g^{-1})=\xi(g^{-1},g),
\label{xiinval}
\end{equation}
for every element $g\in G$. However, it is evident then that $(0,e)$ 
constitutes the unit element of the extended group $\tilde G$ and the inverse 
element of $(\theta, g)$ is given by
$$(\theta, g)^{-1}=(-\theta-\xi(g,g^{-1}),\;g^{-1}).$$ 
Elements of the form $(\theta, e)$ commute with all elements of $\tilde G$,
that is, they belong to the center of $\tilde G$, and respectively $\tilde G$ 
is called a central extension of $G$.

The structure of the group $G$ and the functional relation (\ref{cocyclecond})
greatly constrain the possible forms of admissible two-cocycles
$\xi(g_1,g_2)$ \cite{89,90}. We will not reproduce here the relevant 
mathematics with somewhat involved technical details, but instead clarify 
the physical meaning behind this mathematical construction \cite{87,89,91,92}. 

Let $\Psi(\vec{x},t)$ be a wave function of a free non-relativistic particle 
of mass $m$. Then  $\Psi(\vec{x},t)$ satisfies the Schr\"{o}dinger equation
\begin{equation}
i\frac{\partial \Psi(\vec{x},t)}{\partial t}=-\frac{1}{2\mu}\Delta 
\Psi(\vec{x},t),
\label{Schrodinger}
\end{equation}
where $\mu=m/\hbar$. 

The Galilean invariance ensures that  $\Psi^\prime(\vec{x},t)$ is also  a 
solution of the same Schr\"{o}dinger equation (\ref{Schrodinger}), because 
$\Psi^\prime$ is just the same wave function in another inertial reference 
frame. But according to (\ref{Psiprime}) 
$$\Psi(\vec{x},t)=\exp{[-i\alpha_1(\vec{x}^{\,\prime},t^\prime;g)]}\;
\Psi^\prime(\vec{x}^{\,\prime},t^\prime),$$
where $x^\prime =g(x)$, or
\begin{eqnarray} &&
x^\prime_i=R_{ij}x_j-V_i t +a_i, \nonumber \\ &&
t^\prime=t+b,
\label{Gtransf}
\end{eqnarray}
$R$ being the spatial rotation matrix: $R R^T=R^T R=1$. From (\ref{Gtransf})
we get
$$\frac{\partial}{\partial t}=\frac{\partial t^\prime}{\partial t}\,
\frac{\partial}{\partial t^\prime}+\frac{\partial x_i^\prime}{\partial t}\,
\frac{\partial}{\partial x_i^\prime}=\frac{\partial}{\partial t^\prime}-
\vec{V}\cdot \nabla^{\,\prime}$$
and
$$\nabla_i=\frac{\partial x_j^\prime}{\partial x_i}\,
\frac{\partial}{\partial x_j^\prime}=R_{ji}\nabla^{\,\prime}_j, \;\;\;
\Delta=\Delta^\prime.$$

Therefore the Schr\"{o}dinger equation (\ref{Schrodinger}) can be rewritten 
as follows (we have dropped the primes except in $\Psi^\prime$) 
\begin{equation}
\left . \left . \left (i\,\frac{\partial}{\partial t}-i\,\vec{V}\cdot \nabla+
\frac{1}{2\mu}\Delta \right )\right \{\exp{[-i\alpha_1(\vec{x},t;g)]}
\Psi^\prime(\vec{x},t)\right \}=0.
\label{Schrodinger1}
\end{equation}
But
$$\frac{\partial}{\partial t}\left (e^{-i\alpha_1}\Psi^\prime \right )=
e^{-i\alpha_1}\left [
-i\,\frac{\partial \alpha_1}{\partial t}\;\Psi^\prime+
\frac{\partial \Psi^\prime}{\partial t}\right ],$$
$$\nabla (e^{-i\alpha_1}\Psi^\prime)=e^{-i\alpha_1}\left [
-i\,(\nabla \alpha_1)\;\Psi^\prime+\nabla \Psi^\prime\right ],$$
and
$$\Delta (e^{-i\alpha_1}\Psi^\prime)=e^{-i\alpha_1}\left [
-i\,(\Delta \alpha_1)\Psi^\prime - (\nabla \alpha_1)^2\Psi^\prime -2i\,
(\nabla \alpha_1)\cdot (\nabla \Psi^\prime)+\Delta \Psi^\prime\right ].$$
Therefore, (\ref{Schrodinger1}) takes the form
$$\Psi^\prime \left [ \frac{\partial \alpha_1}{\partial t}-
\vec{V}\cdot \nabla \alpha_1-\frac{i}{2\mu}\Delta\alpha_1-\frac{1}{2\mu}
(\nabla \alpha_1)^2
\right ]$$ $$+\left (i\frac{\partial}{\partial t}+\frac{1}{2\mu}
\Delta \right )\Psi^\prime-i\left (\vec{V}+\frac{1}{\mu}
\nabla \alpha_1\right )\cdot \nabla \Psi^\prime=0.$$
As $\Psi^\prime$ satisfies the Schr\"{o}dinger equation (\ref{Schrodinger}),
we are left with two more conditions on the $\alpha_1$ cochain:
\begin{equation}
\vec{V}+\frac{1}{\mu}\nabla \alpha_1=0
\label{constr1}
\end{equation}
and 
\begin{equation}
\frac{\partial \alpha_1}{\partial t}-\vec{V}\cdot \nabla \alpha_1-
\frac{i}{2\mu}\Delta \alpha_1 -\frac{1}{2\mu}(\nabla \alpha_1)^2=0.
\label{constr2}
\end{equation}
From (\ref{constr1}), we get
$$\alpha_1(\vec{x},t)=-\mu\vec{V}\cdot\vec{x}+f(t)$$
and (\ref{constr2}) takes the form
$$\frac{df}{dt}+\frac{1}{2}\mu V^2=0.$$
Therefore,
$$f(t)=-\frac{1}{2}\mu V^2 t+C_g,$$
where $C_g$ is an integration constant that can depend on $g$, that is, on
$(b,\vec{a},\vec{V},R)$. A convenient choice \cite{92} is
$$C_g=\frac{1}{2}\mu V^2 b+\mu \vec{V}\cdot \vec{a}.$$
Then $\alpha_1(\vec{x},t;g)$ cochain takes the form
\begin{equation}
\alpha_1(\vec{x},t;g)=-\mu\vec{V}\cdot(\vec{x}-\vec{a})-\frac{1}{2}\mu V^2
(t-b).
\label{a1cochain}
\end{equation}
Substituting (\ref{a1cochain}) into (\ref{alpha2}), one can check that
$\xi(g_1,g_2)=\alpha_2(x;g_1,g_2)$ is indeed a global cocycle (does not 
depend on $x$):
\begin{equation}
\xi(g_1,g_2)=\frac{1}{2}\mu V_1^2\,b_2-\mu \vec{V}_1\cdot (R_1\vec{a}_2),
\label{BargmannCocycle}
\end{equation}  
where $\vec{V}_1\cdot (R_1\vec{a}_2)=(R^{-1}_1\vec{V}_1)\cdot\vec{a}_2=
V_{1i}R_{1ij}a_{2j}$. Note that (\ref{BargmannCocycle}) differs from the 
Bargmann's result \cite{89} by a coboundary $C_{g_2}-C_{g_1\cdot g_2}+
C_{g_1}$ (note the overall sign difference in $\alpha_1$, $\alpha_2$, as well 
as the sign difference of the velocity term in (\ref{Gtransf}) from the 
conventions of \cite{89}). Cocycles which differ by a coboundary should be 
considered as equivalent because unitary operators $U(g)$ are determined only 
up to a phase. If we change representatives of the projective representation 
rays as follows $$U(g)\to e^{i\phi(g)}U(g),$$
the two-cocycle $\xi(g_1,g_2)$ will be changed by a coboundary
$$\xi(g_1,g_2)\to\xi(g_1,g_2)+\phi(g_2)-\phi(g_1\cdot g_2)+\phi(g_1).$$
In particular, different choices of the integration constant $C_g$ produce
equivalent cocycles.

The Bargmann cocycle (\ref{BargmannCocycle}) is nontrivial. This can be shown
as follows \cite{89,92}. The multiplication table of the Galilei group is 
given by 
\begin{eqnarray} &&
(b_1,\vec{a}_1,\vec{V}_1,R_1)\cdot (b_2,\vec{a}_2,\vec{V}_2,R_2) \nonumber
\\ && = (b_1+b_2,\,\vec{a}_1+R_1\vec{a}_2-b_2\vec{V}_1\,,\vec{V}_1+
R_1\vec{V}_2,\,R_1R_2),
\nonumber \\ &&
(b,\vec{a},\vec{V},R)^{-1}=(-b,\,-R^{-1}\vec{a}-bR^{-1}\vec{V},\,
-R^{-1}\vec{V},\,R^{-1}),
\label{Gmult}
\end{eqnarray}
where $R\vec{a}$ denotes the action of the rotation $R$ on the vector 
$\vec{a}$, that is $(R\vec{a})_i=R_{ij}a_j$. We have already used 
(\ref{Gmult}) while deriving (\ref{BargmannCocycle}). 

It follows from this multiplication table that elements of the form \linebreak
$(0,\vec{a},\vec{V},1)$ (space translations and boosts) form an Abelian
subgroup:  
$$(0,\vec{a}_1,\vec{V}_1,1)\cdot (0,\vec{a}_2,\vec{V}_2,1)=
(0,\vec{a}_1+\vec{a}_2,\vec{V}_1+\vec{V}_2,1),$$ $$
(0,\vec{a}_1,\vec{V}_1,1)^{-1}=(0,-\vec{a},-\vec{V},1).$$
every trivial cocycle, having the form $\phi(g_2)-\phi(g_1\cdot g_2)+
\phi(g_1)$, is necessarily symmetric in $g_1,g_2$ on Abelian subgroups. 
However, the Bargmann cocycle remains asymmetric on the Abelian subgroup of 
space translations and boosts:
$$\xi(g_1,g_2)=-\mu \vec{V}_1\cdot \vec{a}_2,\;\; 
g_1=(0,\vec{a}_1,\vec{V}_1,1),\;\; g_2=(0,\vec{a}_2,\vec{V}_2,1).$$
Therefore, it can not be a trivial cocycle.

The presence of a nontrivial two-cocycle in quantum field theory usually 
signifies the existence of anomalous Schwinger terms in current 
commutators \cite{93,94}. Remarkably,  in Galilean quantum mechanics mass 
plays the role of the Schwinger term, as we demonstrate below \cite{87}.

Let us take $g_1=(0,\vec{0},\vec{V},1)$ and $g_2=(0,\vec{a},\vec{0},1)$. Then 
$\xi(g_1,g_2)=-\mu \vec{V}\cdot\vec{a}$ and $\xi(g_2,g_1)=0$. Therefore,
\begin{equation}
U(g_1)U(g_2)=e^{-i\mu \vec{V}\cdot\vec{a}}U(g_1\cdot g_2)=
e^{-i\mu \vec{V}\cdot\vec{a}}U(g_2)U(g_1).
\label{UUcom}
\end{equation}
In terms of the (anti-Hermitian) generators $K_i$ and $P_i$ of the Galilean 
group, we have
$$U(g_1)=e^{\vec{K}\cdot \vec{V}},\;\;\;\; U(g_2)=e^{\vec{P}\cdot \vec{a}},$$
and (\ref{UUcom}) takes the form
\begin{equation}
e^{\vec{K}\cdot \vec{V}}e^{\vec{P}\cdot \vec{a}}
e^{-\vec{K}\cdot \vec{V}}e^{-\vec{P}\cdot \vec{a}}=
e^{-i\mu \vec{V}\cdot\vec{a}}.
\label{UUcom1}
\end{equation}
Expanding up to second-order terms in the  infinitesimals $\vec{V}$ and 
$\vec{a}$, we get from (\ref{UUcom1})
$$[\vec{K}\cdot \vec{V},\,\vec{P}\cdot \vec{a}]=-i\mu \vec{V}\cdot\vec{a},$$
or
\begin{equation}
[P_i,K_j]=i\mu\,\delta_{ij}.
\label{PKancom}
\end{equation}
The commutator (\ref{PKancom}) is anomalous because space translations and 
boosts commute on the group level while their generators in the 
representation $U(g)$ do not.

The analogous commutator in the Poincar\'{e} group Lie algebra 
\begin{equation}
[P_i,K_j]=\delta_{ij}\, H
\label{PKcom}
\end{equation}
is not anomalous, as the following argument shows \cite{87}. Let $g_1$ 
be a pure Lorentz transformation
\begin{eqnarray} &&
t^\prime=\gamma \left ( t-\frac{\vec{V}\cdot\vec{x}}{c^2}\right ), \nonumber
\\ && \vec{x}^{\,\prime}=\vec{x}-\gamma\vec{V}t+\frac{\gamma^2}{1+\gamma}\,
\frac{\vec{V}\cdot\vec{x}}{c^2}\,\vec{V},\nonumber
\end{eqnarray}
and $g_2$, a space translation
\begin{eqnarray} &&
t^\prime=t, \nonumber
\\ && \vec{x}^{\,\prime}=\vec{x}+\vec{a}.\nonumber
\end{eqnarray}
Then for infinitesimal $\vec{V}$ and $\vec{a}$ we get that $g_1\cdot g_2
\cdot g_1^{-1}\cdot g_2^{-1}$ is a pure time translation:
\begin{equation}
\left (g_1\cdot g_2\cdot g_1^{-1}\cdot g_2^{-1}\right )\left ( 
\begin{array}{c} t \\ \\ \vec{x}\end{array}\right )= \left ( 
\begin{array}{c} t- \gamma\,\frac{\vec{V}\cdot\vec{a}}{c^2} \\ \\
\vec{x}+\frac{\gamma^2}{1+\gamma}\,\frac{\vec{V}\cdot\vec{a}}{c^2}\,
\vec{V}\end{array}\right )
\approx \left ( \begin{array}{c} t-\frac{\vec{V}\cdot\vec{a}}{c^2} \\  
\\ \vec{x} \end{array}\right ),
\label{timetrans}
\end{equation}
up to second-order terms. However, this is exactly  what is expected from 
(\ref{PKcom}). For infinitesimal $\vec{V}$, the Lorentz transformation 
$g_1$ is represented by the unitary operator
$$U(g)=\exp{\left (\psi\,\frac{\vec{K}\cdot\vec{\beta}}{\beta}\right )}\approx 
\exp{\left ( \frac{\vec{K}\cdot\vec{V}}{c}\right )},$$
where $\psi\approx \tanh{\psi}=\beta$. Therefore,
$$U(g_1)U(g_2)U^{-1}(g_1)U^{-1}(g_2)\approx 1+\left [\frac{\vec{K}\cdot\vec{V}
}{c},\,\vec{P}\cdot\vec{a}\right ]=1-\frac{\vec{V}\cdot\vec{a}}{c}\,H.$$
While the time translation (\ref{timetrans}) is represented by
$$e^{\Delta x_0\, H}=\exp{\left (-\frac{\vec{V}\cdot\vec{a}}{c} H\right )}
\approx 1-\frac{\vec{V}\cdot\vec{a}}{c}\,H,$$
where $x_0=ct$.

The above argument indicates that the Poincar\'{e} group does not admit
nontrivial cocycles. This can be confirmed by the following observation. 
Instead of the Schr\"{o}dinger equation (\ref{Schrodinger}), in the 
relativistic case we will have the Klein-Gordon equation
$$(\Box +\mu^2 c^2)\Psi(x)=0,$$
where $\Box=\partial_\mu \partial^\mu=\Box^\prime$ is now invariant under the
action of the symmetry (Poincar\'{e}) group and, therefore, we will have,
instead of (\ref{Schrodinger1}), the following equation 
$$(\Box +\mu^2 c^2)[e^{-i\alpha_1}\Psi]$$ $$=e^{-i\alpha_1}\left \{
(\Box +\mu^2 c^2)\Psi-2i(\partial_\mu \alpha_1)(\partial^\mu \Psi)-
[(\partial_\mu \alpha_1)(\partial^\mu \alpha_1)+i\Box\alpha_1]\Psi 
\right \}=0.$$
It follows then that
$$\partial^\mu \alpha_1=0,$$
which implies the independence of $\alpha_1$ from the space-time point $x$,
$$\alpha_1(x;g)=C_g,$$
and therefore $\xi(g_1,g_2)=(\delta \alpha_1)(g_1,g_2)$ is a globally trivial
two-cocycle.

It is an interesting question how the nontrivial cocycle of the Galilei group
arises in the process of the Poincar\'{e} group contraction. This can be 
clarified as follows \cite{95}. The general Poincar\'{e} transformations have
the form \cite{96}
\begin{eqnarray} &&
t^\prime=\gamma \left (t-\frac{\vec{V}\cdot R\vec{x}}{c^2}\right ) +b,
\nonumber \\ &&
\vec{x}^{\,\prime}=R\vec{x}-\gamma\vec{V}t+\frac{\gamma^2}{1+\gamma}
\frac{\vec{V}\cdot R\vec{x}}{c^2}\,\vec{V}+\vec{a}.
\label{Ptrasf}
\end{eqnarray}
Let us take
$$C_g=-\mu c^2 b,$$
so that $\alpha_1(g)=C_g$ diverges as $c^2\to\infty$. Nevertheless, its 
coboundary $\xi(g_1,g_2)=(\delta \alpha_1)(g_1,g_2)$ does have  a well 
defined limit under the Poincar\'{e} group contraction. Indeed, it follows 
from (\ref{Ptrasf}) that
$$b(g_1\cdot g_2)=b_1+\gamma_1 b_2-\gamma_1\frac{\vec{V}_1\cdot R_1\vec{a}_2}
{c^2}.$$
Therefore,
$$(\delta \alpha_1)(g_1,g_2)=\alpha_1(g_2)-\alpha_1(g_1\cdot g_2)+\alpha_1
(g_1)=\mu c^2 (\gamma_1-1)b_2-\mu\gamma_1\,\vec{V}_1\cdot R_1\vec{a}_2,$$
and this expression converges to the Bargmann cocycle (\ref{BargmannCocycle})
as $c^2\to\infty$:
$$\mu c^2 (\gamma_1-1)b_2-\mu\gamma_1\vec{V}_1\cdot R_1\vec{a}_2
\to \frac{1}{2}\mu V_1^2 b_2-\mu \vec{V}_1\cdot R_1\vec{a}_2.$$ 

These considerations indicate that mass plays somewhat different 
conceptual roles in relativistic and non-relativistic quantum theories. In 
Galilei invariant theory mass has a cohomological origin and appears as 
a Schwinger term (central extension parameter). Both in relativistic and 
non-relativistic quantum theories, mass plays the role of a label (together 
with spin) distinguishing different irreducible representations of the 
symmetry group and, therefore, different elementary quantum systems. Of 
course, this new quantum facets of mass are a far cry from what is usually 
assumed in the Newtonian concept of mass (a measure of inertia and a source 
of gravity). Nevertheless, I believe they are more profound and fundamental, 
while the Newtonian facets of mass are just emerging concepts valid only in 
a restricted class of circumstances (in non-relativistic situations and in 
the classical limit). 

Considering mass as a Schwinger term has an important physical
consequence because it leads to the mass superselection rule \cite{89,91}.
Let us return to (\ref{UUcom1}) which indicates that the identity 
transformation $g_1\cdot g_2\cdot g_1^{-1}\cdot g_2^{-1}$ of the Galilei 
group is represented by the phase factor $e^{-i\mu\vec{V}\cdot\vec{a}}$ in 
the Hilbert space. This is fine, except for coherent superpositions of 
different mass states for which (\ref{UUcom1}) leads to an immediate trouble. 
For example, if $|\Psi_1\left > \right .$ and $|\Psi_2\left > \right .$ are 
two states with different masses $m_1\neq m_2$, then
$$\left . \left . U(g_1)U(g_2)U^{-1}(g_1)U^{-1}(g_2)\,\right (|\Psi_1\left >
\right . +\;|\Psi_2\left > \right . \right )$$
$$=e^{-i\mu_1\vec{V}\cdot\vec{a}}\left (|\Psi_1\left > \right . +\; 
e^{i(\mu_1-\mu_2)\vec{V}\cdot\vec{a}}\,|\Psi_2\left > \right . \right )$$
which is physically different from $|\Psi_1\left > \right . +\;|\Psi_2\left >
\right .$.

Therefore, quantum states with different masses should belong to different 
superselection sectors and can not be coherently superposed. In particular, 
there are no neutrino oscillations in Galilei invariant theory \cite{97}. In
the relativistic case we have no such prohibition of coherent superpositions
of different energy states which must often also be considered as 
superpositions of different mass states. Experimental observation of neutrino 
oscillations, therefore, directly indicates that we do not live in a Galilei 
invariant world.

Besides, we see once again that the Galilei group is a rather singular and
subtle limit of the Poincar\'{e} group, and not always correctly describes
the non-relativistic limit of the Poincar\'{e} invariant quantum theory. 
There are real physical phenomena (like neutrino oscillations), persistent in
the non-relativistic limit, which Galilei invariant quantum theory fails to 
describe \cite{98}. One may expect, therefore, that the inverse road from 
Galilean world to the relativistic one can have some pitfalls (like 
relativistic mass). It is correct that historically just this adventurous 
road was used to reach the relativistic land. ``This was so because Einstein 
was going from a Galilean universe to an as yet unknown one. So, he had to use
Galilean concepts in his approach to a new theory'' \cite{81}. However, after 
a hundred years of persistent investigation of this new land maybe it will be 
wiser not to use the Newtonian road to the relativistic world any more 
and avoid pitfalls by embarking on safer pathways.  

\section{The return of {\ae}ther?}
A by-product of special relativity was that {\ae}ther became a banished word 
in physics, given on worry to crackpots. However, the concept of {\ae}ther 
has too venerable a history \cite{99} and after a century of banishment we 
may ask whether it is reasonable to give up the term. After all we are still 
using the word `atom' without attaching to it the same meaning as given 
initially by ancient Greeks.

It was Einstein himself who tried to resurrect {\ae}ther as the general 
relativistic space-time in his 1920 inaugural lecture at the University of 
Leiden \cite{100}:
   
``Recapitulating: we may say that according to the general theory of 
relativity space is endowed with physical qualities; in this sense, therefore,
there exists {\ae}ther. According to the general theory of relativity space
without {\ae}ther is unthinkable; for in such space there not only would be no 
propagation of light, but also no possibility of existence for standards of 
measuring rods and clocks, nor therefore any space-time intervals in the
physical sense. But this {\ae}ther may not be thought of as endowed with the 
quality characteristic of ponderable media, as consisting of parts which may 
be tracked through time. The idea of motion may not be applied to it.''

However, Einstein was more successful in eliminating {\ae}ther than in giving
it new life later. Actually all classical {\ae}ther theories had got a death 
blow  and became doomed after special relativity. But quantum mechanics added
a new twist in the story. This was first realized (to my knowledge) by another 
great physicist Paul Dirac \cite{101}. His argument goes as follows.

Usually it is supposed that {\ae}ther is inconsistent with special relativity 
because it defines a preferred inertial reference frame -- where the {\ae}ther 
is at rest. In other reference frames the {\ae}ther moves with some velocity 
and this velocity vector provides a preferred direction in space-time which 
should show itself in suitably designed experiments.

But in quantum mechanics the velocity of the {\ae}ther is subject to 
uncertainty relations and usually it is not a well-defined quantity but
distributed over a range of possible values according to the probabilities 
dictated by {\ae}ther's wave function. One can envisage a wave function
(although not normalizable and hence describing an idealized state which
can be approached indefinitely close but never actually realized) which makes
all values of the {\ae}ther's velocity equally probable.

"We can now see that we may very well have an {\ae}ther, subject to quantum 
mechanics and conforming to relativity, provided we are willing to consider 
the perfect vacuum as an idealized state, not attainable in practice. From the
experimental point of view, there does not seem to be any objection to this.
We must take some profound alterations in our theoretical ideas of the vacuum.
It is no longer a trivial state, but needs elaborate mathematics for its 
description" \cite{101}. 

The subsequent development of quantum field theory completely confirmed
Dirac's prophecy about the quantum vacuum. According to our modern 
perspective, the quantum vacuum is  seething with activity of creating and 
destroying virtual quanta of various fields if probed locally. Therefore, it 
is much more like {\ae}ther than empty space. Nevertheless, this new
quantum {\ae}ther is Lorentz invariant: it looks alike from all inertial 
reference frames. How is this possible? The following example of 
quasiclassical quantum {\ae}ther of the electromagnetic field demonstrates 
the main points \cite{102}. 

In the quasiclassical approximation the electromagnetic quantum {\ae}ther can
be viewed as space filled with a fluctuating electromagnetic field which by 
itself can be represented as a superposition of the transverse plane waves
\begin{eqnarray} &&
\vec{E}(\vec{x},t)=\sum\limits_{\lambda=1}^2\int d\vec{k}\,f(\omega)
\cos{\left (\omega t-\vec{k}\cdot\vec{x}-\theta(\vec{k},\lambda)\right )}
\,\vec{\epsilon}\,(\vec{k},\lambda),
\nonumber \\ &&
\vec{B}(\vec{x},t)=\sum\limits_{\lambda=1}^2\int d\vec{k}\,f(\omega)
\cos{\left (\omega t-\vec{k}\cdot\vec{x}-\theta(\vec{k},\lambda)\right )}
\,\frac{\vec{k}\times \vec{\epsilon}\,(\vec{k},\lambda)}{k},
\label{ZPE}
\end{eqnarray}
where $\lambda$ labels different polarizations, the frequency $\omega$ and
wave vector $\vec{k}$ are related  by the relation $\omega=ck,\;k=|\vec{k}|$,
and $\vec{\epsilon}\,(\vec{k},\lambda)$ are unit polarization vectors:
$$\vec{\epsilon}\,(\vec{k},\lambda)\cdot \vec{k}=0,\;\;\;\;
\vec{\epsilon}\,(\vec{k},\lambda_1)\cdot \vec{\epsilon}\,(\vec{k},\lambda_2)
=\delta_{\lambda_1 \lambda_2}.$$
The fluctuating character of the electromagnetic field is indicated by 
introducing the uniformly distributed random phase $\theta(\vec{k},\lambda)$
for which the following averages hold
\begin{eqnarray} &&
\left <\cos{\theta(\vec{k}_1,\lambda_1)}\,\cos{\theta(\vec{k}_2,\lambda_2)}
\right >=\frac{1}{2}\,\delta(\vec{k}_1-\vec{k}_2)\,
\delta_{\lambda_1 \lambda_2},\nonumber \\ &&
\left <\sin{\theta(\vec{k}_1,\lambda_1)}\,\sin{\theta(\vec{k}_2,\lambda_2)}
\right >=\frac{1}{2}\,\delta(\vec{k}_1-\vec{k}_2)\,
\delta_{\lambda_1 \lambda_2},\nonumber \\ &&
\left <\cos{\theta(\vec{k}_1,\lambda_1)}\,\sin{\theta(\vec{k}_2,\lambda_2)}
\right >=0.
\label{averaging}
\end{eqnarray}
For quantum {\ae}ther to be Lorentz invariant, the weight-function $f(\omega)$ 
must have a special form which we will now find.

The electric and magnetic fields contribute equally to the energy density
$$u=\frac{1}{8\pi}\left <E^2+B^2\,\right >=\frac{1}{4\pi}\left <E^2\,
\right >.$$
Substituting (\ref{ZPE}) and using (\ref{averaging}) along with the 
decomposition
$$\cos{\left (\omega t-\vec{k}\cdot\vec{x}-\theta(\vec{k},\lambda)\right )}
$$ $$=\cos{\left (\omega t-\vec{k}\cdot\vec{x}\right )}
\cos{\theta(\vec{k},\lambda)}+\sin{\left (\omega t-\vec{k}\cdot\vec{x}
\right )}\sin{\theta(\vec{k},\lambda)},$$ 
we obtain
$$u=\frac{1}{4\pi}\int d\vec{k} f^2(\omega)=\int\limits_0^\infty \rho(\omega)
\,d\omega,$$
where the spectral energy-density function is
$$\rho(\omega)=\frac{\omega^2}{c^3}f^2(\omega).$$

Under a Lorentz transformation along the $x$ axis, the $\vec{E}$ and $\vec{B}$
fields mix up. In particular, 
\begin{equation}
E^\prime_x=E_x,\;\;\;E^\prime_y=\gamma\left (E_y-\beta B_z\right ),\;\;\;
E^\prime_z=\gamma\left (E_z+\beta B_y\right ).
\label{EBtrans}
\end{equation}
At that, transverse plane waves go into transverse plane waves with 
transformed frequencies and wave vectors
\begin{equation}
\vec{E}^{\,\prime}(\vec{x}^{\,\prime},t^\prime)=\sum\limits_{\lambda=1}^2\int 
d\vec{k}\,f(\omega)\cos{\left (\omega^\prime t^\prime-\vec{k}^\prime\cdot
\vec{x}^{\,\prime}-\theta(\vec{k},\lambda)\right )}
\,\vec{\epsilon}^{\;\prime}\,(\vec{k},\lambda),
\label{Eprime}
\end{equation}
where
\begin{equation}
\omega^\prime=\gamma \left (\omega-Vk_x\right ),\;\;\;
k^\prime_x=\gamma \left (k_x-\frac{V}{c^2}\,\omega\right ),\;\;\;
k^\prime_y=k_y,\;\;\;k^\prime_z=k_z.
\label{ktrans}
\end{equation}
Taking into account (\ref{ZPE}) and (\ref{EBtrans}), we get for the primed 
polarization vectors
\begin{equation}
\epsilon_x^\prime=\epsilon_x,\;\;\;\epsilon_y^\prime=\gamma\left (
\epsilon_y-\frac{\beta}{k}\left (\vec{k}\times \vec{\epsilon}\right )_z
\,\right ),\;\;\; \epsilon_z^\prime=\gamma\left (\epsilon_z+\frac{\beta}{k}
\left (\vec{k}\times \vec{\epsilon}\right )_y\,\right ),
\label{epstrans}
\end{equation}
or in the vector form
\begin{equation}
\vec{\epsilon}^{\;\prime}(\vec{k},\lambda)=\gamma\vec{\epsilon}\left (1-\beta
\,\frac{k_x}{k}\right )+\gamma\beta\,\frac{\epsilon_x}{k}\,\vec{k}+(1-\gamma)
\epsilon_x\,\vec{i}.
\label{epsprime}
\end{equation}
It is not evident that the primed polarization vectors are transverse but this
can be checked by an explicit calculation:
$$\vec{\epsilon}^{\;\prime}(\vec{k},\lambda)\cdot \vec{k}^{\,\prime}=
\gamma\,\vec{\epsilon}\cdot\vec{k}-\gamma\beta\,\epsilon_x\left [ k-
\frac{k_y^2}{k}-\frac{k_z^2}{k}\right ]-\gamma\beta\,\frac{k_x}{k}\left 
(\epsilon_y k_y+\epsilon_z k_z\right )$$ $$=\gamma\,\vec{\epsilon}\,(\vec{k},
\lambda)\cdot\vec{k}\left ( 1-\beta\,\frac{k_x}{k}\right)=0.$$
It follows from (\ref{epsprime}) that
$$\vec{\epsilon}^{\;\prime}(\vec{k},\lambda)\cdot \vec{\epsilon}^{\;\prime}
(\vec{k},\lambda)=\gamma^2\left (1-\beta\frac{k_x}{k}\right )^2$$ $$+
\epsilon_x^2\left [ \gamma^2\beta^2+(1-\gamma)^2+2\beta\gamma\frac{k_x}{k}
(1-\gamma)+2\gamma(1-\gamma)\left (1-\beta\frac{k_x}{k}\right )\right ].$$
However,
$$\gamma^2\beta^2+(1-\gamma)^2+2\gamma(1-\gamma)=0,$$
and we get
\begin{equation}
\vec{\epsilon}^{\;\prime}(\vec{k},\lambda)\cdot \vec{\epsilon}^{\;\prime}
(\vec{k},\lambda)=\gamma^2\left (1-\beta\,\frac{k_x}{k}\right )^2.
\label{epsprime2}
\end{equation}
It follows then that the energy density of the zero-point field in the primed 
frame is
\begin{equation}
u^\prime=\left <E^{\,\prime 2}\,\right >=\frac{1}{4\pi}\int d\vec{k}\,\gamma^2
\left (1-\beta\,\frac{k_x}{k}\right )^2\, f^2(\omega).
\label{uprime1}
\end{equation}
But from (\ref{ktrans})
$$d\vec{k}^{\,\prime}=\gamma \left (1-\beta\,\frac{k_x}{k}\right )\,d\vec{k},
\;\;\;\; \gamma \left (1-\beta\,\frac{k_x}{k}\right )=\frac{\omega^\prime}
{\omega},$$
and (\ref{uprime1}) can be rewritten as follows
$$u^\prime=\frac{1}{4\pi}\int d\vec{k}^{\,\prime}\,\frac{\omega^\prime}
{\omega}\,f^2(\omega).$$
The Lorentz invariance demands that
$$u^\prime=u=\frac{1}{4\pi}\int d\vec{k}\,f^2(\omega)=
\frac{1}{4\pi}\int d\vec{k}^{\,\prime}\,f^2(\omega^\prime),$$
where the last equation follows from the dummy character of the integration
variable. Therefore,
$$\frac{f^2(\omega^\prime)}{\omega^\prime}=\frac{f^2(\omega)}{\omega}=
\alpha,$$
where $\alpha$ is some constant. 

As we see, the Lorentz invariant quantum {\ae}ther is ensured by the following 
spectral energy-density function
$$\rho(\omega)=\alpha\,\frac{\omega^3}{c^3}.$$
The constant $\alpha$ is not fixed by the Lorentz invariance alone but it can
be determined from the elementary quantum theory which predicts 
$\frac{1}{2}\hbar\omega$ zero-point energy per normal mode. The number of 
normal modes of the electromagnetic field with two independent transverse 
polarizations is $2\frac{1}{(2\pi)^3}$ per momentum interval $d\vec{k}=
4\pi\frac{\omega^2 d\omega}{c^3}$ and, therefore, the quantum theory predicts
the spectral energy density of the quantum vacuum
\begin{equation}
\rho(\omega)=\frac{4\pi\omega^2}{c^3}\,\frac{2}{(2\pi)^3}\,\frac{1}{2}\,
\hbar\omega=\frac{\hbar\omega^3}{2\pi^2 c^3}  
\label{rhodensity}
\end{equation}
which fixes $\alpha$ at
$$\alpha=\frac{\hbar}{2\pi^2}.$$

The above derivation of the spectral energy density (\ref{rhodensity}) of
the Lorentz invariant quantum {\ae}ther emphasizes the wave character of the
electromagnetic field \cite{102,103}. Alternatively one can emphasize the 
photonic picture by considering the quantum electromagnetic {\ae}ther as
consisting of photons of all frequencies moving in all directions \cite{104}.
The result is the same: for radiation to be Lorentz invariant its intensity 
at every frequency must be proportional to the cube of that frequency.

It is instructive to show by explicit calculations that the Lorentz invariant
zero-point {\ae}ther does not lead to any drag force for bodies moving through
it. We will demonstrate this in the framework of the Einstein-Hopf model.
Einstein and Hopf  showed in 1910 that in general there is a drag force 
slowing down a particle moving through stochastic electromagnetic background
\cite{105}.

Let us consider a particle of mass $m$ and charge $e$ harmonically attached to
another particle of mass $M\gg m$ and opposite charge $-e$. We assume 
following Einstein and Hopf that this dipole oscillator is immersed in a 
fluctuating electromagnetic field of the form (\ref{ZPE}), moves in the 
$x$-direction with velocity $V\ll c$, and is oriented so that oscillations 
are possible only along the $z$-axis.

In the rest frame of the oscillator, the equation of motion of mass $m$ looks
like 
\begin{equation}
\ddot{\,z^{\,\prime}}-\Gamma\,\dddot{\,z^{\,\prime}}+\omega_0^{\,\prime\, 2}
z^{\,\prime}=\frac{e}{m}E^{\,\prime}_z(\vec{x}^{\,\prime},t^\prime)
\approx \frac{e}{m}E^{\,\prime}_z(\vec{0},t^\prime),
\label{zequation}
\end{equation}
where we have assumed that the mass $M$ is located at the origin and the 
oscillation amplitude is small. In (\ref{zequation}) the Abraham-Lorentz 
radiation reaction force is included with the well-known radiation damping 
constant \cite{106}
$$\Gamma=\frac{2}{3}\frac{e^2}{mc^3}.$$
The electric field $\vec{E}^{\,\prime}(\vec{x}^{\,\prime},t^\prime)$ in the
primed frame is given by (\ref{Eprime}), and therefore, the steady-state 
solution of (\ref{zequation}) should have the form
\begin{equation}
z^{\,\prime}(t^\prime)=\sum\limits_{\lambda=1}^2\int d\vec{k}\, \left [
a(\omega^{\,\prime})e^{i(\omega^\prime t^\prime-\theta(\vec{k},\lambda))}+
a^*(\omega^{\,\prime})e^{-i(\omega^\prime t^\prime-\theta(\vec{k},\lambda))}
\right ]\epsilon^{\,\prime}_z(\vec{k},\lambda).
\label{zprime}
\end{equation}
Substituting (\ref{zprime}) in (\ref{zequation}), we get
\begin{equation}
a(\omega^{\,\prime})=\frac{e}{m}\,\frac{f(\omega)}{2}\,
\frac{1}{\omega_0^{\,\prime\, 2}+i\Gamma\, \omega^{\,\prime\, 3}-
\omega^{\,\prime\, 2}}.
\label{aomega}
\end{equation}
In the $x$-direction the electromagnetic field exerts ponderomotive force on 
the dipole
\begin{equation}
F^{\,\prime}_x=e z^{\,\prime}\,\frac{\partial E^{\,\prime}_x}{\partial 
z^{\,\prime}}-\frac{e}{c}\,\dot{\,z^{\,\prime}}B^{\,\prime}_y,
\label{dragFprime}
\end{equation}
here $\frac{\partial E^{\,\prime}_x}{\partial z^{\,\prime}}$ and 
$B^{\,\prime}_y$ are evaluated at the origin $\vec{x}^{\,\prime}=0$ thanks to 
the assumed smallness of the oscillation amplitude.

Now it is straightforward, although rather lengthy, to calculate the average 
value of (\ref{dragFprime}) by using
\begin{eqnarray} &&
\left <e^{i\theta(\vec{k}_1,\lambda_1)}\,e^{i\theta(\vec{k}_2,\lambda_2)}
\right > = \left <e^{-i\theta(\vec{k}_1,\lambda_1)}\,e^{-i\theta(\vec{k}_2,
\lambda_2)}\right > =0, \nonumber \\ &&
\left <e^{i\theta(\vec{k}_1,\lambda_1)}\,e^{-i\theta(\vec{k}_2,\lambda_2)}
\right > = \left <e^{-i\theta(\vec{k}_1,\lambda_1)}\,e^{i\theta(\vec{k}_2,
\lambda_2)}\right > =\delta_{\lambda_1 \lambda_2}\,\delta(\vec{k}_1-
\vec{k}_2), \nonumber
\end{eqnarray}
which follow from (\ref{averaging}). As a result, we get
$$\left <e z^{\,\prime}\,\frac{\partial E^{\,\prime}_x}{\partial 
z^{\,\prime}}\right > = \sum\limits_{\lambda=1}^2\int d\vec{k}\;\frac{e
f(\omega)}{2}\;\epsilon^{\,\prime}_z(\vec{k},\lambda)\,\epsilon^{\,\prime}_x
(\vec{k},\lambda)\,k^{\,\prime}_z\,i\left [ a(\omega^{\,\prime})-
a^*(\,\omega^{\prime})\right ].$$
The polarization sum can be performed with the help of
$$\sum\limits_{\lambda=1}^2\,\epsilon^{\,\prime}_i(\vec{k},\lambda)\,
\epsilon^{\,\prime}_j(\vec{k},\lambda)=\left (\frac{\omega^{\,\prime}}
{\omega}\right )^2\left (\delta_{ij}-\frac{k^{\,\prime}_i\,k^{\,\prime}_j}
{k^{\,\prime\,2}}\right ),$$
which follows from (\ref{ktrans}), (\ref{epstrans}), and
$$\sum\limits_{\lambda=1}^2\,\epsilon_i(\vec{k},\lambda)\,
\epsilon_j(\vec{k},\lambda)=\delta_{ij}-\frac{k_i\,k_j}{k^2}.$$
Besides,
$$i\left [ a(\,\omega^{\prime})-a^*(\omega^{\,\prime})\right ]=\frac{e}{m}
\,f(\omega)\,\frac{\Gamma \omega^{\,\prime\,3}}{\left (\omega_0^{\,\prime\,2}
-\omega^{\,\prime\,2}\right )^2+\Gamma^2\omega^{\,\prime\,6}},$$
and we get
$$\left <e z^{\,\prime}\,\frac{\partial E^{\,\prime}_x}{\partial z^{\,\prime}}
\right > = \int d\vec{k}\;\frac{e^2}{2m}\,f^2(\omega) \left (\frac{\omega^{\,
\prime}}{\omega}\right )^2\left (-\frac{k_z^{\,\prime\,2}\,k_x^{\,\prime}}
{k^{\,\prime\,2}}\right )\,\frac{\Gamma \omega^{\,\prime\,3}}{\left 
(\omega_0^{\,\prime\,2}-\omega^{\,\prime\,2}\right )^2+\Gamma^2\omega^{\,
\prime\,6}}.$$
However,
$$f^2(\omega)=\frac{c^3}{\omega^2}\,\rho(\omega),\;\;\;\frac{e^2}{m}=\frac{3}
{2}\Gamma c^3,\;\;\; \frac{\omega^{\,\prime}}{\omega}\,d\vec{k}=
d\vec{k}^{\,\prime},$$
and, therefore,
$$
\left <e z^{\,\prime}\,\frac{\partial E^{\,\prime}_x}{\partial z^{\,\prime}}
\right > = \int d\vec{k}^{\,\prime}\;\frac{3}{4}\,\frac{c^3\omega^{\,\prime}}
{\omega^3}\,\rho(\omega)\,\left (-\frac{k_z^{\,\prime\,2}\,k_x^{\,\prime}}
{k^{\,\prime\,2}}\right )\,\frac{\Gamma^2 c^3 \omega^{\,\prime\,3}}{\left 
(\omega_0^{\,\prime\,2}-\omega^{\,\prime\,2}\right )^2+\Gamma^2\omega^{\,
\prime\,6}}.$$
In order to change $\rho(\omega)/\omega^3$ over to a function of 
$\omega^{\,\prime}$, we expand up to the first order in $\beta$ \cite{102}:
\begin{eqnarray} &&
\rho(\omega)\approx \rho(\omega^{\,\prime})+\frac{\partial \rho(\omega^{\,
\prime})}{\partial \omega^{\,\prime}}\,(\omega-\omega^{\,\prime})\approx 
\rho(\omega^{\,\prime})+\beta\,\frac{k_x^{\,\prime}}{k^{\,\prime}}\,
\omega^{\,\prime} \frac{\partial \rho(\omega^{\,\prime})}{\partial 
\omega^{\,\prime}},\nonumber \\ &&
\frac{1}{\omega^3}\approx \frac{1}{\omega^{\,\prime\,3}}-3\frac{1}{\omega^
{\,\prime\,4}}\,(\omega-\omega^{\,\prime})\approx \frac{1}{\omega^{\,
\prime\,3}}\left ( 1-3\beta\,\frac{k_x^{\,\prime}}{k^{\,\prime}}\right ).
\nonumber
\end{eqnarray}
Hence, up to the first order in $\beta$,
$$\left <e z^{\,\prime}\,\frac{\partial E^{\,\prime}_x}{\partial z^{\,\prime}}
\right > = \int d\vec{k}^{\,\prime}\,\frac{3}{4}\,\frac{c^3}
{\omega^{\,\prime\,2}}\,\frac{\Gamma^2 c^3 \omega^{\,\prime\,3}}{\left 
(\omega_0^{\,\prime\,2}-\omega^{\,\prime\,2}\right )^2+\Gamma^2\omega^{\,
\prime\,6}} $$ $$\times
\left [\rho(\omega^{\,\prime})+\beta\,\frac{k_x^{\,\prime}}
{k^{\,\prime}}\,\omega^{\,\prime}\, \frac{\partial \rho(\omega^{\,\prime})}
{\partial \omega^{\,\prime}}\right ]\left ( 3\beta\,\frac{k_z^{\,\prime\,2}
k_x^{\,\prime\,2}}{k^{\,\prime\,3}}-\frac{k_z^{\,\prime\,2}\,k_x^{\,\prime}}
{k^{\,\prime\,2}}\right ).$$
Analogously, we get
$$\left <-\frac{e}{c}\,\dot{\,z^{\,\prime}}B^{\,\prime}_y \right > =
-\sum\limits_{\lambda=1}^2\int d\vec{k}\;\frac{e f(\omega)}{2c}\,
\omega^{\,\prime}\;\epsilon^{\,\prime}_z\,\frac{\left (\vec{k}^{\,\prime}
\times \vec{\epsilon}^{\;\prime}\right )_y}{k^{\,\prime}}\;i\left [ 
a(\omega^{\,\prime})-a^*(\omega^{\,\prime})\right ],$$
because $B^{\,\prime}_y=\gamma (B_y+\beta E_z)$, and one can check the 
identity
$$\gamma \beta \,\epsilon_z+\gamma\, \frac{\left (\vec{k}\times\vec{\epsilon}
\right )_y}{k}=\frac{\left (\vec{k}^{\,\prime}
\times \vec{\epsilon}^{\;\prime}\right )_y}{k^{\,\prime}}.$$
After performing the polarization sum, we are left with the expression
$$\left <-\frac{e}{c}\,\dot{\,z^{\,\prime}}B^{\,\prime}_y \right > =
\int d\vec{k}^{\,\prime}\;\frac{3}{4}\,\frac{c^3\omega^{\,\prime}}
{\omega^3}\,\rho(\omega)\,k^{\,\prime}_x \,\frac{\Gamma^2 c^3 \omega^{\,
\prime\,3}}{\left (\omega_0^{\,\prime\,2}-\omega^{\,\prime\,2}\right )^2+
\Gamma^2\omega^{\,\prime\,6}},$$
or up to the first order in $\beta$:
$$ \left <-\frac{e}{c}\,\dot{\,z^{\,\prime}}B^{\,\prime}_y \right > =
\int d\vec{k}^{\,\prime}\,\frac{3}{4}\,\frac{c^3}{\omega^{\,\prime\,2}}\,
\frac{\Gamma^2 c^3 \omega^{\,\prime\,3}}{\left (\omega_0^{\,\prime\,2}-
\omega^{\,\prime\,2}\right )^2+\Gamma^2\omega^{\,\prime\,6}} $$ $$\times
\left [\rho(\omega^{\,\prime})+\beta\,\frac{k_x^{\,\prime}}
{k^{\,\prime}}\,\omega^{\,\prime}\, \frac{\partial \rho(\omega^{\,\prime})}
{\partial \omega^{\,\prime}}\right ]\left ( k^{\,\prime}_x-
3\beta\,\frac{k^{\,\prime\,2}_x}{k^{\,\prime}}\right ).$$
Angular integrations, assumed in the decomposition $d\vec{k}^{\,\prime}=
\frac{\omega^{\,\prime\,2}d\omega^{\,\prime}}{c^3}\,d\Omega^{\,\prime}$, are
straightforward if the change to the polar coordinates is made and give
\begin{eqnarray} &&
\int\left (k_x^{\,\prime}-3\beta\,\frac{k_x^{\,\prime\,2}}{k^{\,\prime}}+
3\beta\,\frac{k_x^{\,\prime\,2}k_z^{\,\prime\,2}}{k^{\,\prime\,3}}-
\frac{k_x^{\,\prime}k_z^{\,\prime\,2}}{k^{\,\prime\,2}}\right )
d\Omega^{\,\prime}=-\frac{16\pi}{5}\,\beta\,k^{\,\prime},\nonumber \\ &&
\int\left (k_x^{\,\prime}-3\beta\,
\frac{k_x^{\,\prime\,2}}{k^{\,\prime}}+3\beta\,\frac{k_x^{\,\prime\,2}
k_z^{\,\prime\,2}}{k^{\,\prime\,3}}-\frac{k_x^{\,\prime}k_z^{\,\prime\,2}}
{k^{\,\prime\,2}}\right )\frac{k_x^{\,\prime}}{k^{\,\prime}}\,
d\Omega^{\,\prime}=\frac{16\pi}{15}\,k^{\,\prime}. \nonumber
\end{eqnarray}
Therefore,
\begin{equation}
\left < F_x^{\,\prime} \right > = -\frac{12\pi}{5}Vc\int\limits_0^\infty
\frac{\Gamma^2 \omega^{\,\prime\,4}}{\left (\omega_0^{\,\prime\,2}-
\omega^{\,\prime\,2}\right )^2+\Gamma^2\omega^{\,\prime\,6}}
\left [\rho(\omega^{\,\prime})-\frac{1}{3}\,\omega^{\,\prime}\, 
\frac{\partial \rho(\omega^{\,\prime})}{\partial \omega^{\,\prime}}\right ]
d\omega^{\,\prime}.
\label{Fxprime}
\end{equation}
We assume that $\Gamma \omega_0^{\,\prime}\ll 1$. But in the limit $\Gamma 
\omega_0^{\,\prime}\to 0$ one has
$$\frac{\Gamma \omega^{\,\prime\,4}}{\left (\omega_0^{\,\prime\,2}-
\omega^{\,\prime\,2}\right )^2+\Gamma^2\omega^{\,\prime\,6}} \to 
\frac{\Gamma \omega_0^{\,\prime\,4}}{\left (\omega_0^{\,\prime\,2}-
\omega^{\,\prime\,2}\right )^2+\Gamma^2\omega_0^{\,\prime\,6}} $$
and
$$
\frac{\Gamma \omega_0^{\,\prime\,3}}{\left (\omega_0^{\,\prime\,2}-
\omega^{\,\prime\,2}\right )^2+\Gamma^2\omega_0^{\,\prime\,6}}\to
\pi\delta\left (\omega^{\,\prime\,2} - \omega_0^{\,\prime\,2}\right )$$
$$=\frac{\pi}{2\omega_0^{\,\prime}}\left [\delta(\omega^{\,\prime}-
\omega_0^{\,\prime})+\delta(\omega^{\,\prime}+\omega_0^{\,\prime})\right ].$$
This makes the integration in (\ref{Fxprime}) trivial, and if we drop primes 
in the result, because we are interested only in the lowest order terms, we 
finally get the expression \cite{102,105}
\begin{equation}
\left < F_x \right > = -\frac{6\pi^2}{5}Vc\,\Gamma
\left [\rho(\omega_0)-\frac{1}{3}\,\omega_0\, 
\frac{\partial \rho(\omega_0)}{\partial \omega_0}\right ],
\label{EHdragforce}
\end{equation}
for the Einstein-Hopf drag force in the laboratory frame.
As expected, this drag force disappears for the Lorentz invariant {\ae}ther
due to the cubic dependence of its spectral energy density on the frequency 
(\ref{rhodensity}). Alternatively one can consider the above calculations
as yet another derivation of the spectral energy density which ensures the
Lorentz invariance.

Interestingly, the Einstein-Hopf drag force does not vanish for the cosmic
microwave background radiation with its black body spectrum
$$\rho(\omega, T)=\frac{\hbar \omega^3}{\pi^2 c^3}\,\frac{1}
{e^{\hbar \omega/k T}-1}.$$
Therefore, one has a curious situation that formally the Aristotelian view of 
motion is realized instead of Newtonian one: a body begins to slow down
with respect to a reference frame linked to the isotropic microwave background 
radiation if no force is applied to it \cite{107}. This example shows clearly
that it is futile to ascribe absolute truth to physical laws. Any valid 
physical theory is just some idealization of reality based on concepts which 
are completely sound and useful within the realm of their applicability but 
which can go completely astray if pushed outside of this domain.

Quantum theory inevitably leads to fluctuating quantum vacuum which can be 
considered as a Lorentz invariant {\ae}ther. The Lorentz invariance ensures
that this {\ae}ther does not define a preferred inertial frame and therefore
the Principle of Relativity is not violated. Nevertheless, this quantum 
{\ae}ther is completely real as it leads to experimentally observable physical
effects such as mass and charge renormalizations, Casimir effect, Lamb shift,
Van der Waals forces and fundamental linewidth of a laser \cite{108}. 
A coherent description of these effects is provided by quantum 
electrodynamics. However, many aspects can qualitatively be understood even 
at the classical level if the existence of Lorentz invariant fluctuating 
{\ae}ther is postulated. Probably it will come as a surprise for many 
physicists that the resulting classical theory, the so-called stochastic 
electrodynamics pioneered by Marshall \cite{103,109} and Boyer \cite{102,110},
provides a classical foundation for key quantum concepts \cite{111}. Of course 
stochastic electrodynamics can not be considered as a full-fledged substitute
for the quantum theory, but it is truly remarkable that the introduction of 
Lorentz invariant fluctuating {\ae}ther is sufficient to grasp the essence of 
many concepts thought to be completely quantum. The stochastic electrodynamics
offers a new and useful viewpoint narrowing a gap between quantum weirdness 
and our classical intuition.

The electromagnetic field in the quantum {\ae}ther is bound to fluctuate 
around the zero mean value in order to preserve the Lorentz invariance. There
is no such constraint for scalar fields (elementary or composite) and they 
can develop non-zero vacuum expectation values. The corresponding vacuum 
condensates represent another example of quantum mechanical {\ae}ther. 
These {\ae}ther states play an important role in modern elementary particle 
theory as they lead to the phenomena of spontaneous symmetry breaking and 
generation of mass via the Higgs mechanism \cite{112}.

In 1993, the UK Science Minister, William Waldegrave, challenged physicists 
to produce a one-page answer to the question `What is the Higgs boson, and 
why do we want to find it?' David Miller from University College London won a 
bottle of champagne for a very picturesque description of the Higgs mechanism 
\cite{113} reproduced with slight modifications in \cite{114}.

Imagine a conference hall crowded by physicists. The physicists represent 
a non-trivial medium ({\ae}ther) permeating the space. A gorgeous woman 
enters the hall and tries to find her way through the crowd. A cluster
of her admirers is immediately formed around her slowing her scientific 
progress. At the same time the cluster gives her more momentum for the same 
speed of movement across the room and once moving she is harder to stop. 
Therefore this ephemeral creature acquires much greater effective mass. This 
is the Higgs mechanism. When the woman leaves the hall, a gossip about her 
propagates in the opposite direction bringing an excitement in the crowd. This 
excitement of the medium also propagates in the form of a scandalmongers 
cluster. This is the Higgs boson which will be hunted at LHC \cite{114}.

According to Matvei Bronstein, {\it each epoch in the history of physics has 
its own specific {\ae}ther} \cite{115}. ``The {\ae}ther of the 21-st century 
is the quantum vacuum. The quantum {\ae}ther is a new form of matter. This 
substance has very peculiar properties strikingly different from the other 
forms of matter (solids, liquids, gases, plasmas, Bose condensates, radiation,
etc.) and from all the old {\ae}thers'' \cite{115}.

However, there is a serious problem with the Lorentz invariant quasiclassical 
{\ae}ther with the spectral energy density (\ref{rhodensity}) because the 
integral
\begin{equation}
u=\int\limits_0^\infty \rho(\omega)\,d\omega
\label{vacuumenergy}
\end{equation}
severely diverges. This problem is not cured by the full machinery of quantum 
field theory. It just hides this and some other infinities in several 
phenomenological constants, if the theory is renormalizable (like quantum 
electrodynamics). 

The problem is possibly caused by our ignorance of the true physics at very 
small scales (or, what is the same, at very high energies). A natural 
ultraviolet cut-off in (\ref{vacuumenergy}) is provided by the Planck 
frequency 
$$\omega_p=\frac{m_pc^2}{\hbar},\;\;\;\; m_p=\sqrt{\frac{\hbar c}{G}},$$
$G$ being the Newtonian gravitational constant. Ignoring the factors of $2\pi$ 
and the like, the particle Compton wavelength $\hbar c/E_p$ becomes equal 
to its gravitational radius $G/E_p c^4$ at the Planck energy $E_p=
\hbar \omega_p\approx 1.22 \times 10^{19}~\mathrm{GeV}$. Hence the particle
becomes trapped in its own gravitational field and cardinal alteration of our
notions of space-time is expected  \cite{116,117}. 

If we accept the Planck frequency as the ultraviolet cut-off in 
(\ref{vacuumenergy}), the vacuum energy density becomes (ignoring again some
numerical factors which are not relevant for the following)
\begin{equation}
u\sim \frac{E_p^4}{\hbar^3 c^3}=\frac{c^7}{\hbar G^2}.
\label{vacuumenergy1}
\end{equation}
And now we have a big problem: (\ref{vacuumenergy1}) implies the cosmological
constant which is fantastically too large (123 orders of magnitude!) compared
to the experimental value inferred from the cosmological observations 
\cite{115, 118}.

Of course one can remember Dirac's negative energy sea of the fermionic 
quantum fields and try to compensate (\ref{vacuumenergy1}). We could even 
succeed in this, thanks to supersymmetry. However, the supersymmetry, if it 
exists at all, is badly broken in our low energy world. Therefore it can 
reduce (\ref{vacuumenergy1}) somewhat but not by 123 orders of magnitude. 
Anyway it is much harder to naturally explain so small non-zero vacuum energy 
density than to make it exactly zero. This is the notorious cosmological 
constant problem. Among suggested solutions of this problem \cite{119} the 
most interesting, in my opinion, is the one based on the analogy with 
condensed matter physics \cite{118}. 

``Is it really surprising in our century that semiconductors and cosmology
have something in common? Not at all, the gap between these two subjects 
practically disappeared. The same quantum field theory describes our Universe 
with its quantum vacuum and the many-body system of electrons in 
semiconductors. The difference between the two systems is quantitative, rather
than qualitative: the Universe is bigger in size, and has an extremely low 
temperature when compared to the corresponding characteristic energy scale''
\cite{115}.

The temperature that corresponds to the Planck energy is indeed very high in
ordinary units $T_p=E_p/k_B\approx 1.4 \times 10^{32}~\mathrm{K}$. The natural
question is then why all degrees of freedom are not frozen out at the 
temperature $T=300~\mathrm{K}$ at which we live \cite{115}. Be the Universe
like ordinary semiconductors or insulators, the expected equilibrium density
of excitations (elementary particles) at our living temperature $T$ would be
suppressed by the extremely large factor $e^{T_p/T}=e^{10^{30}}$. That we 
survive such a freezer as our Universe indicates that the Universe is more
like metals with a Fermi surface and gapless electron spectra, or, to be 
more precise, is like special condensed matter systems with 
topologically-protected Fermi points \cite{115}. There are only a very few 
such systems like superfluid phases of liquid $^3\mathrm{He}$ and 
semiconductors of a special type which can be used to model cosmological 
phenomena \cite{120,121}. At that, in the condensed matter system the role of 
the Planck scale is played by the atomic scale. At energies much lower than 
the atomic scale, all condensed matter systems of Fermi point universality 
class exhibit a universal generic behavior and such ingredients of the 
Standard Model as relativistic Weyl fermions, gauge fields and effective 
gravity naturally emerge \cite{120,121}. Amusingly, if the Fermi point 
topology is the main reason why the elementary particles are not frozen out 
at temperatures $T\ll T_p$, then we owe our own existence to the hairy ball 
theorem of algebraic topology which says that {\it one cannot comb the hair 
on a ball smooth}, because Fermi point is the hedgehog in momentum space and 
its stability is ensured by just that theorem \cite{115}. 

Therefore, if the condensed matter analogy is really telltaling, we are left
with the exciting possibility that physics probably does not end even at the 
Planck scale and we have every reason to restore the word '{\ae}ther' in the 
physics vocabulary. At the present moment we can not even guess what the 
physics of this trans-Planckian {\ae}ther is like because our familiar 
physical laws, as emergent low energy phenomena, do not depend much on the 
fine details of the trans-Planckian world, being determined only by the 
universality class, which the whole system belongs to. ``The smaller is our 
energy, the thinner is our memory on the underlying high-energy 
trans-Planckian world of the quantum {\ae}ther where we all originated from. 
However, earlier or later we shall try to refresh our memory and concentrate 
our efforts on the investigation of this form of matter'' \cite{115}.

\section{Concluding remarks}
Although the idea of a four-vector can be traced down to Poincar\'{e} 
\cite{122}, it was Minkowski who gave the formulation of special relativity 
as a four-dimensional non-Euclidean geometry of space-time. Very few 
physicists were well trained in pure mathematics in general, and in 
non-Euclidean geometry in particular, at that time. Minkowski himself was 
partly motivated in his refinement of Einstein's work, which led him to now 
standard four-dimensional formalism, by his doubts in Einstein's skills in 
mathematics \cite{122}. When Minkowski presented his mathematical elaboration
of special relativity at a session of the Gottingen Mathematical Society, he 
praised the greatness of Einstein's scientific achievement, but added that 
``the mathematical education of the young physicist was not very solid which 
I'm in a good position to evaluate since he obtained it from me in Zurich 
some time ago'' \cite{123}.

In light of Minkowski's assessment of Einstein's skills in mathematics, it is 
fair to say that ``in the broader context of education in German-speaking 
Europe at the end of the nineteenth century Einstein received excellent 
preparation for his future career'' \cite{124}. Neither was he dull in 
mathematics as a secondary school pupil. For example, at the final examination
at Aarau trade-school Einstein gave a very original solution of the suggested  
geometrical problem by using a general identity for the three angles of a
triangle (which I'm not sure is known to every physicist)
$$\sin^2{\frac{\alpha}{2}}+\sin^2{\frac{\beta}{2}}+\sin^2{\frac{\gamma}{2}}+
2\,\sin{\frac{\alpha}{2}}\,\sin{\frac{\beta}{2}}\,\sin{\frac{\gamma}{2}}=1$$
and by solving a cubic equation which he got from this equality by 
substitutions. ``Although it depended on instant recall of  complicated 
mathematical formulas, Einstein's solution was the very opposite of one based
on brute-force calculations. He was careful to arrive at numerical values only
after having made general observations on, among other things, the rationality
of the roots of the cubic equation and on the geometrical requirements that a
solution would have to satisfy'' \cite{124}.

Anyway, after Minkowski's sudden death from appendicitis at age 44, shortly 
after his seminal Cologne lecture, neither Einstein nor any other physicist
at that time was in a position to duly appreciate the non-Euclidean readings
of special relativity. Although the space-time formalism, energetically 
promoted by Sommerfeld, quickly became a standard tool in special relativity,
its non-Euclidean facets remained virtually unnoticed, with just one exception.

Inspired by  Sommerfeld's interpretation of the relativistic velocity
addition as a trigonometry on an imaginary sphere, the Croatian mathematician 
Vladimir Vari\v{c}ak established that relativistic velocity space possessed 
a natural hyperbolic (Lobachevsky) geometry \cite{125,126}.

One can check the hyperbolic character of metric in the relativistic velocity
space as follows. The natural distance between two (dimensionless) velocities
$\vec{\beta}_1$ and $\vec{\beta}_2$ in the velocity space is the relative
velocity \cite{127,128}
$$\beta_{rel}=\frac{\sqrt{\left (\vec{\beta}_1-\vec{\beta}_2\right )^2-
\left (\vec{\beta}_1\times \vec{\beta}_2\right )^2}}{1-\vec{\beta}_1\cdot
\vec{\beta}_2}.$$
Taking $\vec{\beta}_1=\vec{\beta}$ and $\vec{\beta}_2=\vec{\beta}+
d\vec{\beta}$, one gets for the line element in the velocity space 
\cite{127,128}
\begin{equation}
ds^2=\frac{\left (d\vec{\beta}\right)^2-\left (\vec{\beta}\times d\vec{\beta}
\right)^2}{(1-\beta^2)^2}.
\label{ds2vel}
\end{equation}
To make contact with previous formulas, we will assume two-dimensional 
velocity space and change to the new coordinates \cite{129}
$$\beta_x=\frac{2x}{1+x^2+y^2},\;\;\;\;\beta_y=\frac{2y}{1+x^2+y^2}.$$
Then
\begin{eqnarray} &&
\vec{\beta}=\frac{2x}{1+x^2+y^2}\,\vec{i}+\frac{2y}{1+x^2+y^2}\,\vec{j}=
f(x,y)\,\vec{i}+g(x,y)\,\vec{j},\nonumber \\ &&
d\vec{\beta}=\left( \frac{\partial f}{\partial x}dx+
\frac{\partial f}{\partial y}dy\right )\,\vec{i}+
\left( \frac{\partial g}{\partial x}dx+
\frac{\partial g}{\partial y}dy\right )\,\vec{j},\nonumber
\end{eqnarray}
and substituting this into (\ref{ds2vel}), we get after some algebra
$$ds^2=\frac{4(dx^2+dy^2)}{(1-x^2-y^2)^2}.$$
However, this is nothing but the line element of the hyperbolic geometry 
from (\ref{CKds2}), with $\epsilon_1=1,\,\epsilon_2=-1$.

The non-Euclidean style in relativity was pursued mainly by mathematicians
and led to very limited physical insights, if any, in the relativity 
mainstream \cite{125}. In a sense, non-Euclidean readings of relativity were
ahead of time as illustrated by the unrecognized discovery of the Thomas 
precession in 1913 by the famous French mathematician \'{E}mile Borel, 
a former doctoral student of Poincar\'{e}.

In Borel's non-Euclidean explanation, the Thomas precession, which is usually
regarded as an obscure relativistic effect, gets a very transparent meaning.
If a vector is transported parallel to itself along a closed path on the 
surface of a sphere, then its orientation undergoes a change proportional to 
the enclosed area. The Lobachevsky space is a space of constant negative 
curvature and, as Borel remarked, the similar phenomenon should take place: 
if a velocity vector circumscribes a closed path under parallel transport in 
the kinematical space, it will undergo precession with magnitude proportional 
to the enclosed area. Borel ``was careful to point out that the effect is a 
direct consequence of the structure of the Lorentz transformations'' 
\cite{125}. This remarkable discovery of Borel, however, was of limited 
historical value because its physical significance was not recognized (I 
doubt it could have been done before the advent of quantum mechanics).

The hyperbolic geometry of the relativistic velocity space is an interesting 
but only a particular aspect of special relativity. A more important and 
really non-Euclidean reading of Minkowski identifies the Minkowski geometry 
itself as a kind of non-Euclidean geometry, as just one representative of the
whole family of non-Euclidean Cayley-Klein geometries. Placing special 
relativity in such a broader context, we see instantly (see Fig.\ref{Kgroups})
that it is not a fundamental theory but a limiting case of more general 
theory. A slight generalization of the Relativity Principle (in the sense of 
Bacry and L\'{e}vy-Leblond \cite{73}) leads to eleven different relativity 
theories which all are various limiting cases of the two really fundamental 
homogeneous space-times -- de Sitter and anti de Sitter spaces.

It is not surprising, therefore, that astrophysical data indicate the non-zero
cosmological constant, which means that the correct asymptotic (vacuum) 
space-time in our Universe is the de Sitter space-time, not Minkowski. What
is surprising is the incredible smallness of the cosmological constant which
makes the special relativity valid for all practical purposes.

As yet we do not know the resolution of the cosmological constant problem.
Maybe this enigma leads to a trans-Planckian {\ae}ther with yet unknown 
physics, as the condensed matter analogy \cite{115} indicates. Anyway Einstein 
was lucky enough to be born in the Universe with nearly vanishing cosmological 
constant and nowadays we formulate all our theories of fundamental interactions
in the Minkowskian background space-time.

The case of gravity needs some comment because it is usually assumed that
special relativity is no longer correct in the presence of gravity, and 
should be replaced by general relativity. However, the interpretation of 
gravity as a pseudo-Riemannian metric of curved space-time is not the only 
possible interpretation, nor is it always the best one, because this 
interpretation sets gravity apart from other interactions --- ``\ldots too 
great an emphasis on geometry can only obscure the deep connections between 
gravitation and the rest of physics'' \cite{130}.

Is it possible to develop a theory of gravity as a quantum theory of massless
spin-two field in the flat Minkowski space-time by analogy with other 
interactions? It seems Robert Kraichnan, the only post-doctoral student that 
Einstein ever had, was the first who initiated the study of this question. 
``He recalls that, though he received some encouragement from Bryce DeWitt, 
very few of his colleagues supported his efforts. This certainly included 
Einstein himself, who was appalled by an approach to gravitation that rejected
Einstein's own hard-won geometrical insights'' \cite{131}. Maybe just because 
of this lack of support, Kraichnan left the Institute of Advanced Study 
(Princeton) and the field of gravitation in 1950, to establish himself in 
years that followed as one of the world's leading turbulence theorists.

Kraichnan published his work \cite{132} only in 1955, after Gupta's paper 
\cite{133} on the similar subject appeared. Since then the approach was 
developed in a number of publications by various authors. It turned out that 
the flat Minkowski metric is actually unobservable and  the geometrical 
interpretation  of gravity as a curved and dynamical effective metric arises 
at the end all the same. ``The fact is that a spin-two field has this 
geometrical interpretation: this is not something readily explainable -- it is 
just marvelous. The geometrical interpretation is not really necessary or 
essential to physics. It might be that the whole coincidence might be 
understood as representing some kind of invariance'' \cite{134}. 

Although the field theory approach to gravity has an obvious pedagogical
advantages \cite{134,135}, especially for high-energy physics students, we can
not expect from it the same success as from quantum electrodynamics, because
the theory is non-renormalizable. Besides the condensed matter analogy shows
\cite{115,136} that gravity might be really somewhat different from other
interactions, and in a sense more classical, because in such interpretation
gravity is a kind of elasticity of quantum vacuum (trans-Planckian {\ae}ther) 
--- the idea that dates back to Sakharov \cite{137}.

An interesting hint that gravity might be an emergent macroscopic phenomenon
with some underlying microscopic quantum theory is the fact that general 
relativity allows the existence of space-time horizons  with well defined 
notions of temperature and entropy which leads to an intriguing analogy 
between the gravitational dynamics of the horizons and thermodynamics. 
It is even possible to obtain the Einstein equations from the proportionality 
of entropy to the horizon area together with the fundamental thermodynamical 
relation between heat, entropy, and temperature. ``Viewed in this way, the 
Einstein equation is an equation of state'' \cite{138}. This perspective has 
important consequences for quantization of gravity: ``it may be no more 
appropriate to canonically quantize the Einstein equation than it would be to 
quantize the wave equation for sound in air'' \cite{138}.

All these considerations indicate that Einstein was essentially right and
the Riemannian trend in geometry, which at first sight appears entirely 
different from Klein's Erlangen program, is crucial for describing gravity,
although this geometry might be a kind of low energy macroscopic emergent 
illusion and not the fundamental property of trans-Planckian {\ae}ther.

Anyway, it seems reasonable to explore also another way to unify gravity 
with other interactions: the geometrization of these interactions. From
the beginning we have to overcome a crucial obstacle on this way: for 
interactions other than gravity the Kleinian geometry seems to be of much 
more importance than the Riemannian geometry of curved space-time. ``There 
is hardly any doubt that for physics special relativity theory is of much 
greater consequence than the general theory. The reverse situation prevails 
with respect to mathematics: there special relativity theory had 
comparatively little, general relativity theory very considerable, influence,
above all upon the development of a general scheme for differential 
geometry'' \cite{139}. 

Just this development of mathematics, spurred by general relativity, led 
finally to the resolution of the dilemma we are facing in our attempts to find 
a common geometrical foundation for all interactions. This dilemma was 
formulated by Cartan as follows \cite{49,140}:

``The principle of general relativity brought into physics and philosophy the
antagonism between the two leading principles of geometry due to Riemann
and Klein respectively. The space-time manifold of classical mechanics and
of the principle of special relativity is of the Klein type, and the one
associated with the principle of general relativity is Riemannian. The fact
that almost all phenomena studied by science for many centuries could be
equally well explained from either viewpoint was very significant and 
persistently called for a synthesis that would unify the two antagonistic 
principles''.

The crucial observation which enables the synthesis is that ``most 
properties of Riemannian geometry derive from its Levi-Civita parallelism'' 
\cite{47}. Let us imagine a surface in Euclidean space. A tangent vector at 
some point of the surface can be transferred to the nearby surface point by 
parallel displacement in the ambient Euclidean space, but at that, in general,
it will cease to be tangent to the surface. However, we can split the new 
vector into its tangential and (infinitesimal) normal components and throw 
away the latter. This process defines the Levi-Civita connection which may be 
viewed as a rule for parallel transport of tangent vectors on the surface 
\cite{139}.

There is a more picturesque way to explain how Levi-Civita parallel transport 
reveals the intrinsic geometry of the surface \cite{141}. Imagine a hamster
closed inside a ``hamster ball''. When hamster moves inside the ball, the 
latter rolls on a lumpy surface. We can say that the hamster studies the 
intrinsic geometry of the lumpy surface by rolling more simple model surface 
(the sphere) on it. Rolling the ball without slipping or twisting along two 
different paths connecting given points on the surface will give, in general, 
results differing by some $SO(3)$ rotation, an element of the 
principal group of the model geometry, which encodes information about the 
intrinsic geometry of the surface. An infinitesimal $SO(3)$ rotation of the 
hamster ball, as it begins to move along some path,  breaks up into a part 
which describes the $SO(2)$ rotation of the sphere around the axis through
the point of tangency and into a part which describes an infinitesimal 
translation  of the point of tangency. The $SO(3)$ connection (the analog of
the  Levi-Civita connection for this example) interrelates these two parts 
and thus defines a method of rolling the tangent sphere along the surface 
\cite{141}.

\begin{figure}[htb]
 \begin{center}
    \mbox{\includegraphics[scale=0.43]{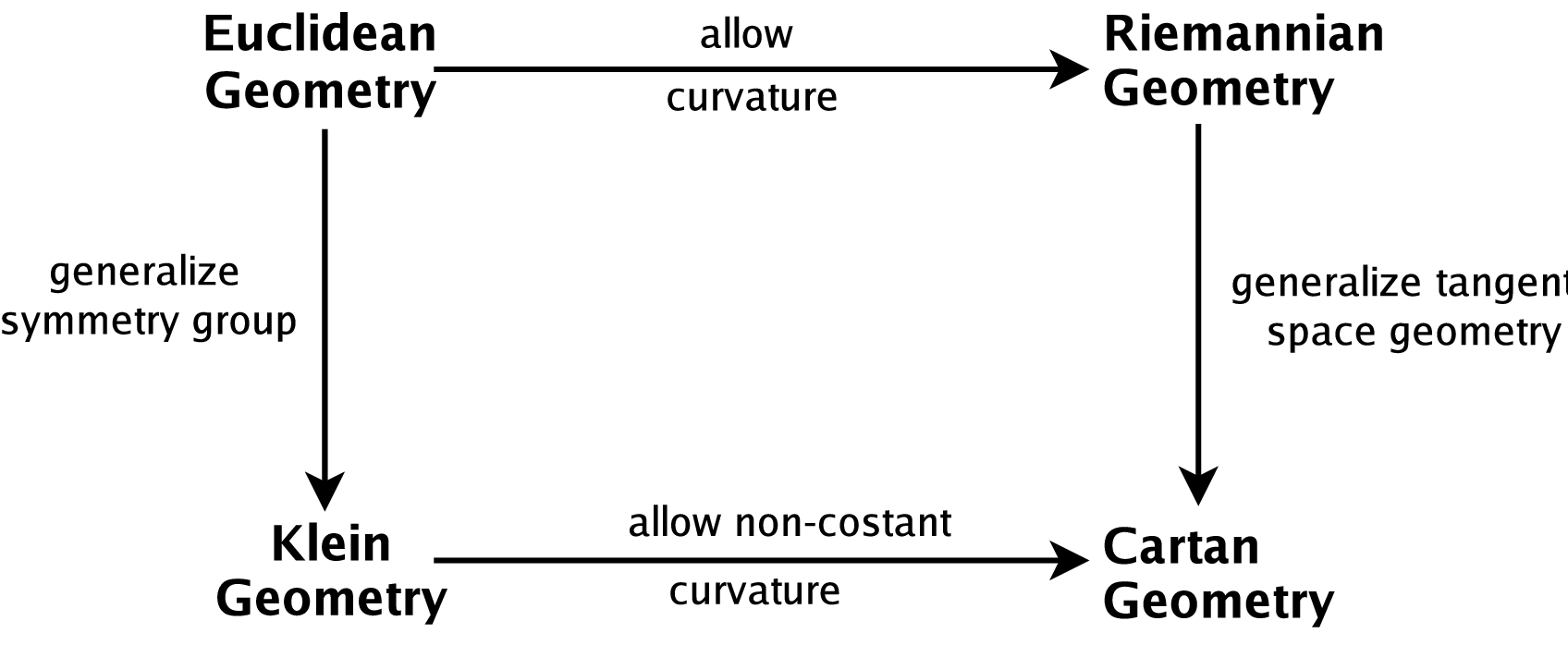}}
  \end{center}
\caption {The unification of Riemannian and Klein geometries into the Cartan
geometry \cite{141}.}
\label{Cartan}
\end{figure}

The far reaching generalization due to \'{E}lie Cartan is now obvious: we can
take any homogeneous Klein space as a model geometry and try to roll it on the
space the geometry of which we would like to study. Information about the
geometry of the space under study will then be encoded in the Cartan 
connection which gives a method of the rolling. Figure \ref{Cartan} (from 
\cite{141}) shows how the Cartan geometry unifies both Kleinian and 
Riemannian trends in geometry.  

There is no conceptual difference between gravity and other physical 
interactions as far as the geometry is involved because the Cartan geometry 
provides the fully satisfying way in which gauge theories can be truly 
regarded as geometry \cite{142}.

Therefore, we see that special relativity and principles it rests upon are
really fundamental, at the base of modern physics with its gauge theories
and curved space-times. I believe its teaching should include all the beauty
and richness behind it which was revealed by modern physics, and avoid 
historical artifacts like the second postulate and relativistic mass. I hope 
this article shows that such a presentation requires some quite elementary 
knowledge of basic facts of modern mathematics and quantum theory. ``The 
presentation of scientific notions as they unfolded historically is not the 
only one, nor even the best one. Alternative arguments and novel derivations 
should be pursued and developed, not necessarily to replace, but at least to 
supplement the standard ones'' \cite{18}.

The game of abstraction outlined above is not over. For example, one hardly
questions our {\it a~priory} assumption that physical quantities are 
real-valued. But why? Maybe the root of this belief lies in the pre-digital 
age assumption that physical quantities are defined operationally in terms of 
measurements with classical rulers and pointers that exist in the classical
continuum physical space \cite{143}. But quantum mechanics teaches us that
the Schr\"{o}dinger's cat is most likely object-oriented \cite{144}. That 
suggests somewhat Platonic view of reality that physics is a concrete 
realization, in the realm of some Topos, of abstract logical relations 
among elements of reality. ``Topos'' is a concept proposed by Alexander 
Grothendieck, one of the most brilliant mathematicians of the twentieth 
century, as the ultimate generalization of the concept of space. Our notion
of a smooth space-time manifold, upon which the real-valuedness of the 
physical quantities rests, is, most likely, an emergent low energy concept 
and we do not know what kind of abstractions will be required when we begin 
``to refresh our memory'' about primordial trans-Planckian {\ae}ther 
\cite{143,145}. 
 
Grothendieck abruptly ended his academic career at the age of 42  and 
``withdrew more and more into his own tent''. If his last time visitors can be
believed, ``he is obsessed with the Devil, whom he sees at work everywhere 
in the world, destroying the divine harmony and replacing 300,000 km/sec by 
299,887 km/sec as the speed of light!'' \cite{146}. I'm afraid a rather 
radical break of the traditional teaching tradition of special relativity is
required to restore its full glory and divine harmony.

\section*{Acknowledgments}
The author is grateful to Alexander Ulyanov for his help with the 
manu\-script. The work is supported in part by grants 
Sci.School-905.2006.2 and RFBR 06-02-16192-a.

\end{document}